\documentclass[aos]{imsart}

\RequirePackage{amsthm,amsmath,amsfonts,amssymb}
\RequirePackage[authoryear]{natbib}
\RequirePackage[colorlinks,citecolor=blue,urlcolor=blue]{hyperref}
\RequirePackage{graphicx}
\RequirePackage{comment}
\DeclareMathOperator*{\argmin}{arg\,min}

\startlocaldefs
\theoremstyle{plain}

\newtheorem{theorem}{Theorem}[section]

\newtheorem{lemma}{Lemma}[section]

\theoremstyle{definition}
\newtheorem{remark}{Remark}[section]

\newtheorem{assumption}{Assumption}[section]
\numberwithin{equation}{section}

\theoremstyle{remark}

\endlocaldefs

\begin{document}
	
	\begin{frontmatter}
		\title{A Two-step Estimating Approach for Heavy-tailed AR Models with Non-zero Median GARCH-type Noises}
		\runtitle{A Two-step Estimating Approach}
		
		\begin{aug}
			\author[A]{\fnms{Rui}~\snm{She}\ead[label=e1]{rshe@swufe.edu.cn}},
			\author[A]{\fnms{Linlin}~\snm{Dai}\ead[label=e2]{ldaiab@swufe.edu.cn}}
			\and
			\author[B]{\fnms{Shiqing}~\snm{Ling}\ead[label=e3]{maling@ust.hk}}
			\address[A]{School of Statistics,
				The Southwestern University of Finance and Economics\printead[presep={,\ }]{e1,e2}}
			
			\address[B]{Department of Mathematics, The Hong Kong University of Science and Technology\printead[presep={,\ }]{e3}}
		\end{aug}
		
		\begin{abstract}
			This paper develops a novel two-step estimating procedure
			for heavy-tailed AR models with non-zero median GARCH-type noises, allowing for time-varying volatility. We first establish the self-weighted quantile regression estimator (SQE) across all quantile levels $\tau\in (0,1)$  for the AR parameters  $\theta_{0}$. We show that the SQE, less a bias, converges weakly to a Gaussian process at a rate of $n^{-1/2}$. The bias is zero if and only if $\tau$   equals   $\tau_{0}$, the probability that the noise is less than zero. Based on the SQE, we propose an  approach to estimate $\tau_{0}$ in the second step and {feed the estimated $\tau_0$ back into the SQE to estimate $\theta_0$.} Both the estimated $\tau_{0}$ and $\theta_{0}$ are shown to be consistent and asymptotically normal.  A random weighting bootstrap method is  developed to approximate the complex distribution.
			The problem we study is non-standard because   $\tau_{0}$ may not be identifiable in conventional quantile regression, and the usual methods cannot verify the existence of the SQE bias.
			Unlike existing procedures for heavy-tailed time series, our method does not require prior information about the symmetry, tail index, or the parametric form of the noise, nor does it require classical identification conditions, such as zero-mean or zero-median.
		\end{abstract}
		
		\begin{keyword}[class=MSC]
			\kwd{62F12}
			\kwd{62F35}
			\kwd{62M10}
		\end{keyword}
		
		\begin{keyword}
			\kwd{GARCH}
			\kwd{heavy-tailed}
			\kwd{self-weighted}
			\kwd{quantile}
			\kwd{nonstationary}
		\end{keyword}
		
	\end{frontmatter}
	
	\section{Introduction}
	
Many economic and financial datasets often exhibit characteristics such as skewness, heavy tails, temporal dependence and persistent volatility. Many models have been proposed to address these features in the field of time series, of which  the most popular model often used as a   benchmark is   the AR  model with GARCH noise, defined as
\begin{align}\label{eq_1_1}
	y_t=\mu_0+\sum_{j=1}^{p}{\phi_{j0} y_{t-j}}+\varepsilon_{t} \mbox{ and }\varepsilon_{t}=\eta_{t}\sigma_{t},
\end{align}
where the innovation $\{\eta_{t}\}$ is a sequence of independent and identically distributed (i.i.d.) random variables, and the volatility $\sigma_{t} \in \mathcal{F}_{t-1}=(\eta_{t-1},\eta_{t-2},\cdots)$ is strictly positive. This framework includes   the standard GARCH model \citep{engle1982}, the absolute value GARCH model \citep{taylor1986}, the GJR model \citep{gjr1993}, the nonlinear GARCH model \citep{engle1990}, among others; see, e.g., \citet{Fanyao2003} or \citet{francq2010} for a comprehensive overview.
The conventional inference methods for this model assume that $\varepsilon_{t}$ has a  light-tail, see \citet{weiss1986}, \citet{ling1997}, and \citet{francq2004} for stationary $\{\sigma_{t}\}$, and   \citet{xu2006}, \citet{xu2008}, \citet{zhu2019}, and \citet{robinson2020} for time-varying $\{\sigma_{t}\}$.

However,  heavy-tailed phenomena have been  demonstrated  widely in financial markets \citep{koed,francq,yang2017}. Moreover,  GARCH-type noise  always exhibits heavy-tailed behavior with a tail index less than 4, even when $\eta_{t}$ follows a Gaussian distribution \citep{basrak2002, zhang2015a,she2023}.
The presence of heavy-tailed GARCH noises would mislead practitioners because  conventional
time series analysis procedures will result in the estimated parameters  having a slower
convergence rate than $\sqrt{n}$ with non-pivotal asymptotic distributions, see  \citet{zhang2015a} and \citet{zhangx2022}.

When  the volatility $\sigma_{t}$ is completely unspecified, the self-weighted  least absolute deviation estimator (LADE)   proposed by \citet{zhu2015} is  the only feasible estimator in the literature that achieves asymptotic normality without imposing moment assumptions on  $\eta_t$ and $\varepsilon_t$.  In line with typical LAD-type estimation methods,
\citet{zhu2015} assume that the conditional median of $\varepsilon_{t}$  is zero, i.e.,
\begin{equation}\label{zero-median}
	P(\eta_{t}\le 0)=\frac{1}{2}.
\end{equation}
However, it is often observed that the zero-median assumption does not hold in financial and economic data because of  factors such as  the leverage effect. {Evidence} for this is given in asset pricing   \citep{harvey2000,patton2004} and  exchange rates \citep{francq2012,ma2022}. With a nonzero-median of $\eta_t$, we have
\begin{align}
	\label{nonzero-median}
	P(\eta_t\leq 0)=\tau_0 \mbox{ for some  } \tau_0\in (0,1).
\end{align}
If  $\tau_0$ in (\ref{nonzero-median}) is known, { paralleling} \citet{zhu2015}, a {standard} statistical inference procedure based on the $\tau_0$-th quantile estimator can be developed. However,   $\tau_0$ is typically unknown in practice.   One potential solution is to transform the innovation by setting $\eta^{*}_{t}=\eta_{t}-\mbox{median}(\eta_{t})$ to satisfy the zero-median assumption. Unfortunately, this would disrupt the AR-GARCH structure \citep{ma2022}. For more on the effects of transformation on skewed data, see \citet{fan2014}.
Another possible solution is to  parameterize the  volatility $\sigma_{t}$; see,  e.g., \citet{ling2007a,zhu2011,ling2007b,jiang2020,zhuq2018,zhu2021}.
Such fully parametric frameworks    carry the risk of model misspecification.
Thus, it remains a challenging open problem to  carry out inference  for model (\ref{eq_1_1}) with   unspecified volatility $\sigma_{t}$ and the unknown $\tau_0$.

This paper is to build  a  two-step procedure to solve this problem. The main innovations of this paper are three folds:
\begin{itemize}
	\item {\it Determine the identification condition} under which  the level $\tau_{0}$ as defined in (\ref{nonzero-median}) and the AR parameter $\theta_{0} = (\mu_0,\phi_{10},\cdots, \phi_{p0})^\top$ are inherently identifiable without requiring any additional conditions, such as the zero-median assumption. Specifically, $\tau_{0}$ is the unique quantile level for which the corresponding conditional quantile of $y_{t}$ can be expressed as a linear combination of the covariates, see Lemma \ref{lemma_3_1}.
 This condition is necessary  to estimate  $\tau_{0}$ and $ \theta_{0}$ consistently, see Remarks \ref{remark_3_1}--\ref{remark_3_1_2} for further discussion.


	\item  {\it Justify  the existence of the bias function}, denoted by $\delta_{0}(\tau)$, of  the self-weighted $\tau$-th quantile regression estimator (SQE), where $\tau \in (0,1)$ denotes the unknown value of $\tau_{0}$; see Theorem \ref{theorem_3_1}.
	This result is  crucial to establish our inference procedure. Almost all existing research assumes that the quantile regression estimator (QRE) has a unique minimizer \(\theta_0(\tau) := \theta_0 + \delta_0(\tau)\) when \(\tau \neq \tau_0\), and that this minimizer lies within a given compact subset \citep{white2003, angrist2006}. Alternatively, some studies assume that \(\tau = \tau_0 + O(n^{-1/2})\) \citep{escan2010,escan2014}.
 Theorem \ref{theorem_3_1} is the first to show that the conventional QRE has a unique global minimizer when \(\tau \neq \tau_{0}\) (i.e., under model misspecification).
	
	\item {\it  Offer a two-step procedure to estimate $(\tau_{0}, \theta_{0})$.}
	We first derive the SQE for  all $\tau\in (0,1)$
	and show that, after subtracting a bias, it  converges weakly to a Gaussian process at the rate of  $n^{-1/2}$, which involves    a new functional central limit theorem for the nonstationary time series in Lemma \ref{theorem_3_2}. In the second step, we estimate $\tau_{0}$ using the SQE and then use  the SQE corresponding to the estimated $\tau_{0}$ to serve as the estimator for $\theta_{0}$. Both the estimated $\tau_{0}$ and $\theta_{0}$ are consistent and asymptotically normal with the  rate $n^{-1/2}$.
	Furthermore, we develop a random weighting approach to estimate the asymptotic distributions of  the estimated $\tau_{0}$ and $\theta_{0}$, which can be used to construct  inference for $\theta_{0}$ and $\tau_{0}$. This novel method offers a feasible inference for the AR model with general GARCH-type noise, without imposing any restrictions on parametric form, symmetry, or tail index.
\end{itemize}
	The problem we study is non-standard because   $\tau_{0}$ may not be identifiable in conventional quantile regression, and the usual methods cannot verify the existence of the SQE bias.
	The estimators based on the two-step procedure demand innovative and complex techniques to develop their asymptotic theories, which are outlined in  Remarks \ref{remark_3_0_1}, \ref{remark_5_1} and \ref{remark_7_1}.

The remainder of the paper proceeds as follows. Section \ref{section2} gives the models and  assumptions. Section  \ref{section3} presents our two-step estimation procedure. Section \ref{section4} investigates the bias of the SQE and Section \ref{section5} gives its limiting property. Section \ref{section6} presents the asymptotic normality of the two-step SQE estimator. A random weighting  bootstrap approximation of the limiting distributions is discussed in Section \ref{section7}. Section \ref{section8} conducts simulation experiments to assess the finite sample performance of our procedure and one real example is illustrated in Section \ref{section9}. Proofs of Lemma \ref{lemma_3_1}, Theorem \ref{theorem_3_1} and Theorems \ref{theorem_4_1}-\ref{theorem_4_2} are relegated to the Appendix. Some additional simulation results, technical lemmas, the proofs for Sections \ref{section5} and \ref{section7}, as well as potential extensions, are
provided in the supplementary material.

Throughout the paper, \(\overset{p}{\longrightarrow}\) denotes convergence in probability, and \(\overset{d}{\longrightarrow}\) denotes convergence in distribution. The notation \(|A|\) represents the usual Euclidean norm, and \(A^{\top}\) denotes the transpose. For a random vector \(X\) and \(\kappa > 0\), \(\|X\|_{\kappa} = (E|X|^{\kappa})^{1/\kappa}\). In particular, \(\|X\|\) means \(\|X\|_2\).

\section{Models and Assumptions}\label{section2}
We consider  model (\ref{eq_1_1}) with  time-varying  heteroscedastic volatility:
\begin{align}\label{eq_1_2}
	\sigma_{t}:=\sigma(t/n, \mathcal{F}_{t-1}),
\end{align}
where $\sigma(\cdot): [0,1]\times \mathbb{R}^{\infty} \to \mathbb{R}$ is a positive measurable function.
When the function $\sigma(\cdot)$ is independent of the time ratio $t/n$, the framework in (\ref{eq_1_2}) simplifies to a stationary time series, otherwise, it is non-stationary. Crucially, including $t/n$ in (\ref{eq_1_2}) allows the noise to vary smoothly, enabling the modeling of a wide range of non-stationary error patterns. For instance, consider a continuous and positive function $v_0(x)$ and a measurable function $\sigma_0(\cdot)$. We can  define two typical error structures:
\begin{align}
	\mbox{(i)}\, \varepsilon_t = \eta_{t} \times v_0(t/n)\sigma_0(\mathcal{F}_{t-1});\quad  \mbox{(ii)}\, \varepsilon_t =\eta_{t}\times  [\sigma_0(\mathcal{F}_{t-1})]^{v_0(t/n)}.\nonumber
\end{align}
The multiplier $v_0(\cdot)$ in (i) captures the non-stationarity of variance in light-tailed cases (e.g., \citet{rho2015}, \citet{zhu2019}, and \citet{jiang2021}), while the exponent $v_0(\cdot)$ in (ii) reflects the smoothly varying tail index in heavy-tailed errors as those in \citet{dehaan2021} and \citet{D2023}.

Now, we present some fundamental assumptions for model (\ref{eq_1_1}) with (\ref{eq_1_2}).
\begin{assumption}\label{assumption_2_1}
	$\phi_0(z)=1-\sum_{j=1}^{p}{\phi_{j0}z^j}$ has all roots outside the unit circle.
\end{assumption}
\begin{assumption}\label{assumption_2_2}
	There exists $\alpha_0 \in (0,1]$ such that $\Vert \sup_{x\in [0,1]}{[\sigma_{t}(x)]}\Vert_{\alpha_0}<\infty$, $\Vert \eta_{t} \Vert_{\alpha_0}<\infty$, and $\Vert Y_p \Vert_{\alpha_0}<\infty$, where $\sigma_{t}(x)=\sigma(x,\mathcal{F}_{t-1})$ and $Y_p=(y_p,\cdots,y_1)^{\top}$.
\end{assumption}
\begin{assumption}\label{assumption_2_3}
	$\sup\limits_{\substack{x_1 \neq x_2, x_1, x_2 \in [0,1]}}{\{ \Vert \sigma_{t}(x_1)-\sigma_{t}(x_2) \Vert_{\alpha_0}/\vert x_1-x_2\vert \}}<\infty$.
\end{assumption}
Assumption \ref{assumption_2_1} is a standard assumption for AR($p$) models, implying existence of $\pi_0(z)=\phi^{-1}_0(z)=\sum_{j=0}^{\infty}{\pi_{j0}z^j}$ with $\pi_{j0}=O(c^j_0)$ for some $c_0 \in (0,1)$. For any $x \in [0,1]$, denote
\begin{align}\label{eq_epsilon}
	\varepsilon_t(x)=\eta_t \sigma_{t}(x),
\end{align}
and then Assumption \ref{assumption_2_2} indicates that $\Vert \sup_{x \in [0,1]}{\vert \varepsilon_{t}(x) \vert}\Vert_{\alpha_0}<\infty$. Furthermore, by the monotone convergence theorem, we have that the summation
$$\vert \pi_0(1)\mu_0\vert^{\alpha_0}+\sum_{j=0}^{\infty}{\big \{\vert \pi_{j0}\vert^{\alpha_0} \sup_{x\in[0,1]}{\vert \varepsilon_{t-j}(x)\vert^{\alpha_0}}\big \}}$$ converges almost surely. Given that  $(a+b)^{\alpha_0}\leq a^{\alpha_0}+b^{\alpha_0}$ for $a,b\geq 0$ and $\alpha_0\in (0,1]$, the stationary functional time series is well defined almost surely:
\begin{align}\label{eq_2_1}
y_t(x)=\pi_0(1)\mu_0+\sum_{j=0}^{\infty}{\pi_{j0} \varepsilon_{t-j}(x)}, \mbox{ for } x \in [0,1].
\end{align}
Note that we only require a finite $\alpha_0$-fractional moment for the time series, allowing for noise with infinite variance. Assumption \ref{assumption_2_3}  guarantees that the nonstationary time series $\{y_t\}$ is well approximated by its stationary counterpart $\{y_t(t/n)\}$, as in Lemma S.1  of \citet{she2022}; see also (\ref{eq_a_2})-(\ref{eq_a_5}) in the Appendix.

Let $f(z)$ be the density of $\eta_{t}$ with $f^{(1)}(z)$ as its first derivative. We now introduce two additional assumptions, which are pivotal to the validity of the two-step estimation procedure.

\begin{assumption}\label{assumption_2_4}
	(i) There exists a constant $\sigma_0>0$ such that $\sigma_{t}(x)\geq \sigma_0$ for any $x\in [0,1]$; (ii) $\sigma_{t}(x)$ is a non-degenerate random variable for any $x\in[0,1]$.
\end{assumption}
\begin{assumption}\label{assumption_2_5}
	(i) The continuous density $f(z)$ is positive for any $z\in \mathbb{R}$ and uniformly bounded; (ii) $f^{(1)}(z)$ is Lipschitz continuous and uniformly bounded.
\end{assumption}

Assumption \ref{assumption_2_4}  is used to ensure the identifiability of $\tau_{0}$, as given in Section \ref{section3_1}, and is applicable to nearly all GARCH-type structures.  Assumption \ref{assumption_2_5}  is necessary for  technical proofs, see Remark \ref{remark_5_1}. It is clear that (\ref{nonzero-median}) must hold for some $\tau_{0}\in (0,1)$, under which the $\tau_0$-th conditional quantile of $y_t$ given the past information $\mathcal{F}_{t-1}$ has the following expression:
\begin{align}\label{eq_1_5}
Q_{y_t}(\tau_0\vert\mathcal{F}_{t-1})=Z^{\top}_{t-1}\theta_0,
\end{align}
where $Z_{t-1}=(1,Y^{\top}_{t-1})^{\top}$ and $Y_{t-1}=(y_{t-1},\cdots,y_{t-p})^{\top}$. It is natural to use the $\tau_0$-th linear quantile {regression} approach to estimate  $\theta_{0}$.
 However,  $\tau_{0}$ is unknown in practice.


	
\section{Feasible Estimation Strategies}\label{section3}
  We first  need to study the identification and estimability of   $(\tau_{0},\theta_{0})$ {in the sense of the existence of its consistent estimators}, which is a crucial and  highly non-standard problem.

\subsection{The intrinsic identification of $\tau_0$}\label{section3_1}
Denote the unknown value of $\tau_{0}$ to be $\tau\in (0,1)$.  Then the form of the $\tau$-th conditional quantile of $y_t$ is
\begin{align}\label{eq_2_4}
	Q_{y_t}(\tau\vert\mathcal{F}_{t-1})=Z^{\top}_{t-1}\theta_0+F^{-1}(\tau)\sigma_{t},
\end{align}
where $F^{-1}(\cdot)$ is the inverse of $F(\cdot)$, the distribution function of $\eta_{t}$. Clearly,  equation (\ref{eq_1_5}) is a special case of (\ref{eq_2_4}) with $F^{-1}(\tau_{0})=0$. Given the non-degeneracy of $\sigma_{t}$ in Assumption \ref{assumption_2_4} (ii), we can derive a significant result as follows.
\begin{lemma}\label{lemma_3_1}
	Given Assumptions \ref{assumption_2_4} and \ref{assumption_2_5} (i), it follows that
	(i) $P(Z^{\top}_{t-1}\delta=\sigma_{t})<1$ for any $\delta\in \mathbb{R}^{p+1}$; (ii) $P(Z^{\top}_{t-1}\delta=0)=1$ if and only if $\delta=0$.
\end{lemma}
\noindent
Since $F^{-1}(\tau)=0$ if and only if $\tau=\tau_{0}$ by Assumption \ref{assumption_2_5} (i),
Lemma \ref{lemma_3_1} implies that
\begin{align}\label{identify}
	Q_{y_t}(\tau\vert\mathcal{F}_{t-1})=Z^{\top}_{t-1}\theta \,\, \mbox{if and only if }\,\, \tau=\tau_{0}\,\,\mbox{and}\,\,\theta=\theta_{0},
\end{align}
that is, it  rules out the possibility of two quantile levels
	$\tau_{1}<\tau_{2}$ such that $Q_{y_t}(\tau_1\vert\mathcal{F}_{t-1})=Z^{\top}_{t-1}\theta_1$ and $Q_{y_t}(\tau_2\vert\mathcal{F}_{t-1})=Z^{\top}_{t-1}\theta_2$ for some $\theta_1,\theta_2\in \mathbb{R}^{p+1}$.
It is equivalent to
\begin{align}\label{identify3}
	\begin{cases}
		& (i) \mbox{ when }\tau=\tau_{0},\,\, E[ \psi_{\tau_0}(y_t-Z^{\top}_{t-1}{\theta})\vert\mathcal{F}_{t-1}]=0 \mbox{ only for }\theta=\theta_{0},\\
		& (ii) \mbox{ when }\tau\neq\tau_{0},\,\, E[ \psi_{\tau}(y_t-Z^{\top}_{t-1}{\theta})\vert\mathcal{F}_{t-1}]\neq 0 \mbox{ for any }\theta \in \mathbb{R}^{p+1},
	\end{cases}
\end{align}
where $\psi_{\tau}(x)=\tau-I(x\leq 0)$.
 Therefore, the value $\tau_{0}$ is identifiable in the sense that  it is the unique quantile level such that the related conditional quantile of $y_{t}$ is  some linear combination of the covariate $Z_{t-1}$.
  This non-standard identifiability condition is our motivation  to construct a consistent estimator of $\tau_{0}$ in Section \ref{section3.2}.

\begin{remark}\label{remark_3_1} When the noise is i.i.d. (i.e., \(\sigma_{t} \equiv 1\)), Lemma \ref{lemma_3_1} (i) no longer holds because  Assumption \ref{assumption_2_4} (ii) fails and (\ref{eq_2_4}) can be rewritten as follows:
	\begin{align*}
	Q_{y_t}(\tau \mid \mathcal{F}_{t-1})= Z^{\top}_{t-1} \theta_0 + F^{-1}(\tau)= (\mu_0 + F^{-1}(\tau), \phi_{10}, \cdots, \phi_{p0}) Z_{t-1}.
	\end{align*}
	So for any $\tau\in(0,1)$, there exists some $\theta$ such that $E[\psi_{\tau}(y_t-Z^{\top}_{t-1}{\theta})\vert\mathcal{F}_{t-1}]=0$ and hence $\tau_{0}$ and $\theta_{0}$ are not identifiable.
 Thus, the GARCH structure is essential for the identifiability of \(\tau_{0}\) and our two-step estimation procedure. The locally time-varying volatility, as described in (\ref{eq_1_2}), represents an important class of time series models and is  attracting increasing attention in the literature; see \cite{xu2006}. Under this general setting, our two-step estimation procedure will have a much broader scope of application.

\end{remark}

\begin{remark}\label{remark_3_1_2}
	The GARCH-type noise  can be further relaxed to accommodate a more general noise process \(\varepsilon_t\).  Specifically, let \(\varepsilon_t \in \mathcal{F}_t\), and denote its conditional distribution function given   \(\mathcal{F}_{t-1}\) as \(F_t(\cdot)\) and its density as \(f_t(\cdot)\). (\ref{identify}) and (\ref{identify3}) still hold under the following assumptions:
	(i) \(f_t(z)\) is continuous and positive on the real line almost surely;
	(ii) there exists a \(\tau_0 \in (0,1)\) such that \(F_t^{-1}(\tau_0) = 0\) almost surely for all $t$;
	(iii) for any \(\tau \neq \tau_0\), \(F_t^{-1}(\tau)\) is a non-degenerate random variable.
	With these relaxed assumptions, the proposed two-step estimation procedure remains applicable to AR models with general noise  \(\varepsilon_t\).
 Detailed theoretical results for more general noise structures can be found in the supplementary material.
\end{remark}


\subsection{Two-step estimation procedure}\label{section3.2}
{Given the identifiability condition (\ref{identify3}), we first apply linear quantile regression to estimate the conditional $\tau$-th quantile of $y_t$ for any $\tau \in (0,1)$, with the resulting
	estimator denoted as $\hat{\theta}_n(\tau)$ and then  use   (\ref{identify3}) to construct the  estimator of $\tau_{0}$.
To be precise, in the first step, we  use $Z^{\top}_{t-1}\theta$ to fit the $\tau$-th conditional quantile of $y_{t}$. The linear quantile regression loss function is defined as:
\begin{align}\label{loss_function}
	\tilde{L}_n(\theta,\tau)=n^{-1}\sum_{t=p+1}^{n}{w_t \rho_{\tau}(y_t-Z^{\top}_{t-1}\theta)},
\end{align}
where $w_t=w(t/n,Y_{t-1})$ for some positive measurable function $w(\cdot)$, and  $\rho_{\tau}(x)=x\psi_{\tau}(x)$.
As in \citet{ling2005},  $w_t$ is to downweight large values  in $Z_{t-1}$.
Furthermore,  because of  the nonstationarity of $y_t$, the following assumption is required:
\begin{assumption}\label{assumption_2_7}
	The weight $w_t=w(t/n, Y^{\top}_{t-1})$ for some function $w(x_0,x_1,\cdots,x_p)$, and $\{w(x_0,x_1,\cdots,x_p)x^{l_i}_{i}x^{l_j}_j x^{l_k}_k x^{l_m}_m \}$ are bounded and Lipschitz continuous on $[0,1]\times \mathbb{R}^p$, where $i,j,k,m\in \{1,\cdots,p\}$ and $l_i, l_j, l_k,l_m \in \{0,1\}$.
\end{assumption}
A  method commonly used to select the weight function $w_t$ is the data-driven approach proposed by \citet{ling2005}. However, as noted by \citet{zhou2009}, the sample quantile of ${y^2_t}$ typically converges to a time-varying curve as a result of  the non-stationarity of the noise. Therefore, we prefer to use a weight with a pre-determined form, such as:
$$ w_t=\prod_{i=1}^{p}{(1+\vert y_{t-i}\vert^k)^{-1}}\,\,\mbox{or}\,\,w_t=\prod_{i=1}^{p}{({1+e^{\vert y_{t-i}\vert^k}})^{-1}}.\,\,$$
For  a fixed $\tau$, the minimizer of $\tilde{L}_n(\theta,\tau)$,  referred to as the self-weighted quantile regression estimator (SQE), is denoted by $\hat{\theta}_{n}(\tau)$; that is
\begin{align}\label{eq_3_0}
	\hat{\theta}_{n}(\tau)=\argmin_{\theta \in \mathbb{R}^{p+1}}{ \tilde{L}_n(\theta,\tau) }.
\end{align}
Under  mild conditions in Sections \ref{section4}--\ref{section5}, we will show that there always exists a bias curve denoted by $\delta_0(\tau)$, such that
\begin{align}\label{eq_3_6_1}
	\sup_{\tau\in\mathcal{T}}{\vert \hat{\theta}_n(\tau)-\theta_{0}-\delta_{0}(\tau) \vert}=O_p(n^{-1/2}),
\end{align}
where $\mathcal{T}=[\epsilon,1-\epsilon]$ is any closed interval in $(0,1)$ such that $\tau_{0}$ is an interior point. In particular, $\delta_{0}(\tau)\neq 0$ if and only if $\tau\neq\tau_{0}$.

Denote $\theta_{0}(\tau)=\theta_{0}+\delta_{0}(\tau)$, then $\theta_{0}(\tau)=\theta_{0}$ if and only $\tau=\tau_{0}$. In the second step, we use the following identifiability condition from (\ref{identify3}) to estimate $\tau_{0}$:
\begin{align}\label{eq_4_1}
	E[ \psi_{\tau}(y_t-Z^{\top}_{t-1}{\theta}_0(\tau)) \vert \mathcal{F}_{t-1} ]=0 \mbox{ if and only if }\tau=\tau_{0}.
\end{align}
It is well-known that for any random variable $Y$ and the $\sigma$-{algebra} $\mathcal{F}$, $E[Y\vert \mathcal{F}]=0$ {is equivalent to} $E[XY]= 0$ for any bounded random variable $X\in \mathcal{F}$.  Motivated by the work of \citet{newey1985} and \citet{bierens1982},  we use a series of bounded functions
$\{\tilde{w}_{lt}=\tilde{w}_l(t/n,\tilde{Y}_{t-1})\}^{L}_{l=1}$  to capture the  information from the $\sigma$-{algebra} $\mathcal{F}_{t-1}$, where   $\tilde{Y}_{t-1}=(y_{t-1},\cdots,y_{t-\tilde{p}})$ for some $L$ and $\tilde{p}$.
Then, {subject to the regularity conditions} in Section \ref{section6},
the conditional moment condition in (\ref{eq_4_1}) can be replaced by
\[
{E[ \tilde{w}_{lt} \psi_{\tau}(y_t-Z^{\top}_{t-1}{\theta}_0(\tau)) ]}=0 \mbox{ for any }l\leq L \,\,\mbox{ if and only if }\tau=\tau_{0}.
\]
Note that $ \hat{\theta}_n(\tau)$ is a consistent estimator of  ${\theta}_0(\tau)$.
Thus, our second step is to estimate $\tau_0$ by $\hat{\tau}_n$  as follows:
\begin{align}\label{eq_4_2}
	\hat{\tau}_n=\argmin_{\tau \in \mathcal{T}}{\sum_{l=1}^{L}{ \Big\{n^{-1}\sum_{t=p+1}^{n}{r_t(\tau;\tilde{w}_l)}\Big\}^2}},
\end{align}
where $r_t(\tau;\tilde{w}_l)=\tilde{w}_{lt} \psi_{\tau}(y_t-Z^{\top}_{t-1}\hat{\theta}_n(\tau))$. Then, a {feasible} estimator for $\theta_{0}$ is
\begin{eqnarray}\label{feasible_theta}
	\hat{\theta}_{n}= \hat{\theta}_{n}(\hat{\tau}_n).
\end{eqnarray}
The asymptotic normality of $\hat{\tau}_{n}$ and $\hat{\theta}_{n}$ will be given in Section \ref{section6}.

Based on the theoretical analysis in Section \ref{subsection_6_2}, we can choose
 the following form for ${\tilde{w}_{lt}}$  in practical applications:
\begin{align}\label{weight_practical}
	\Bigg\{\tilde{w}_0(t/n,{Y}^{\top}_{t-1})\Big(\frac{y_{t-1}}{\sqrt{1+y^2_{t-1}}}\Big)^{d_1} \cdots \Big(\frac{y_{t-p}}{\sqrt{1+y^2_{t-p}}}\Big)^{d_p}\Bigg\},
\end{align}
where $\tilde{w}_0(\cdot)$ satisfies Assumption \ref{assumption_2_7} and $0\leq d_i\leq d_0$ for some given integer $d_0$. As shown in Section \ref{section8}, this setting can yield a satisfactory finite-sample performance for $\tau_{0}$ and $\theta_{0}$, even when $d_0$ is small.

\begin{remark} Intuitively, one possible   estimator of   $\tau_{0}$ is
	\begin{align*}
		\check{\tau}_{n}=\argmin_{\tau\in\mathcal{T}}{\tilde{L}_n(\hat{\theta}_n(\tau),\tau)}.
	\end{align*}
	Since $\rho_{\tau}(x)$ is monotonic with respect to $\tau$ for every fixed $x$ because $\partial \rho_{\tau}(x)/\partial \tau=x$,
then	using Theorem \ref{theorem_3_1} in the next section,  we can show that
 $\tau_{0}$ is  not the minimizer of $E\tilde{L}_n(\theta_{0}(\tau),\tau)$ on the set $\mathcal{T}$ and hence $\check{\tau}_{n}$ will  not be  a consistent estimator of $\tau_{0}$.
\end{remark}

\section{Existence and Uniqueness of  Bias}\label{section4}
{This section discusses the bias $\delta_{0}(\tau)$ when $\tau\neq\tau_{0}$.
The loss $\tilde{L}_n(\theta,\tau)$ in (\ref{loss_function}) is the empirical version of the function $$E\tilde{L}_n(\theta,\tau)=n^{-1}{ \sum_{t=p+1}^{n}{E[w_t \rho_{\tau}(y_t-Z^{\top}_{t-1}\theta)]} }.$$
Since $E\tilde{L}_n(\theta,\tau)$ is convex with respect to $\theta$, the global minimizer of $E\tilde{L}_n(\theta,\tau)$ must be the zero point of  the following score function (i.e., the left derivative of $E\tilde{L}_n(\theta,\tau)$):
\begin{align}\label{eq_score}
	\frac{\partial E\tilde{L}_n(\theta,\tau)}{\partial \theta}=-n^{-1}\sum_{t=p+1}^{n}{E[w_tZ_{t-1}\psi_{\tau}(y_{t}-Z^{\top}_{t-1}\theta) ]}.
\end{align}
Denote $Z_{t-1}(s)=(1,Y^{\top}_{t-1}(s))^{\top}$ and $Y_{t-1}(s)=(y_{t-1}(s),\cdots,y_{t-p}(s))^{\top}$. Recall that $\mathcal{T}=[\epsilon,1-\epsilon]$ is any closed interval in $(0,1)$ such that $\tau_{0}$ is an interior point.
Now, we derive the asymptotic form of the score function and prove the existence of the zero point for (\ref{eq_score}).}
\begin{theorem}\label{theorem_3_1}
	Suppose that Assumptions \ref{assumption_2_1}--\ref{assumption_2_3}, Assumption \ref{assumption_2_4} (i), Assumption \ref{assumption_2_5} (i), and Assumption \ref{assumption_2_7} hold. Then it follows that
	
	(i) for any bounded set $\Theta \subset \mathbb{R}^{p+1}$, we have
	\begin{align*}
		\sup_{\tau \in \mathcal{T},\theta \in \Theta}{\Big \vert   \frac{\partial E\tilde{L}_n(\theta,\tau)}{\partial \theta}-g(\theta-\theta_{0},\tau)  \Big \vert}=O(n^{-\alpha_0}),
	\end{align*}
	where $g(x,\tau)=-\int_{0}^{1}{E\Big\{w(s,Y^{\top}_{t-1}(s))Z_{t-1}(s)\big[\tau-F\big(\frac{x^{\top}Z_{t-1}(s)}{\sigma_{t}(s)}\big) \big]   \Big\} ds}$.
	
	(ii) for any $\tau \in \mathcal{T}$, in the whole space $\mathbb{R}^{p+1}$, there exists a unique $\delta_0(\tau)$ such that $g(\delta_0(\tau),\tau)=0$ and  $\delta_0(\tau)$ has the continuous derivative with respect to $\tau$ with
\begin{align*}
\frac{\partial\delta_0(\tau_{0})}{\partial \tau} & =\Big\{\int_{0}^{1}{E\Big[w(s,Y^{\top}_{t-1}(s))Z_{t-1}(s)Z^{\top}_{t-1}(s)\sigma^{-1}_{t}(s)  \Big]ds}\Big\}^{-1}\\
& \qquad \quad \times f^{-1}(0) \int_{0}^{1}{E[w(s,Y^{\top}_{t-1}(s))Z_{t-1}(s)]ds}.
\end{align*}
\end{theorem}

We will see that  $\delta_0(\tau)$ is exactly the bias of 	$\hat{\theta}_{n}(\tau)$ in the next section.
The statement (i) shows that, for any $\tau\in \mathcal{T}$, the score function ${\partial E\tilde{L}_n(\theta,\tau)}/{\partial \theta}$ is equivalent to the function $g(\theta-\theta_{0},\tau)$ and the statement (ii) proves that there exists a unique $$\theta_{0}(\tau):=\theta_{0}+\delta_0(\tau)$$ such that $g(\theta_{0}(\tau)-\theta_{0},\tau)=0$. In other words, for any $\tau \in \mathcal{T}$,  $E\tilde{L}_n(\theta,\tau)$ can attain its minimum at the point  $\theta=\theta_{0}(\tau)$.
By the definition of $\delta_{0}(\tau)$ and $F^{-1}(\tau_{0})=0$, we can further derive that
	\begin{align}
	\int_{0}^{1}{E\Big\{w(s,Y^{\top}_{t-1}(s))Z_{t-1}(s)   \Big\} ds}=0 & \Longleftrightarrow \delta_{0}(\tau)=0 \mbox{ for any }\tau \in (0,1),\nonumber\\
	\int_{0}^{1}{E\Big\{w(s,Y^{\top}_{t-1}(s))Z_{t-1}(s)   \Big\} ds}\neq 0 & \Longleftrightarrow  \delta_{0}(\tau_0)=0\mbox{ and }\delta_{0}(\tau) \mbox{ is injective.}\label{conclusion2}
	\end{align}
	As a result, the constant in $Z_{t-1}(s)$ ensures (\ref{conclusion2}) and thus  $\delta_0(\tau)\neq 0$ if and only if $\tau\neq \tau_{0}$.


\begin{remark}\label{remark_3_0_1}

	The proof of Theorem \ref{theorem_3_1} (ii) presents significant challenges. For any fixed \(\tau\), \(g(x,\tau)\) is a multivariate vector-valued function from \(\mathbb{R}^{p+1}\) to \(\mathbb{R}^{p+1}\), and \(\delta_{0}(\tau)\) must be chosen such that all components of \(g(\delta_0(\tau),\tau)\) are simultaneously zero. Directly solving for \(\delta_{0}(\tau)\) from \(g(\delta_{0}(\tau),\tau)=0\) is infeasible because  the distribution function \(F(\cdot)\) is unknown. Meanwhile, classical zero-point existence theorems, such as the intermediate value theorem or the Poincar\'e-Miranda theorem, are not applicable here because they require a compact domain, whereas our setting involves the entire space \(\mathbb{R}^{p+1}\).
	To address these challenges, we adopt an innovative proof strategy. Starting with \(g(0,\tau_{0})=0\), we prove that \(g(x,\tau)\) is continuously differentiable at this point and apply the implicit function theorem to extend the zero point from \(\tau_{0}\) to its neighborhood. The most difficult part is  extending this neighborhood to the entire interval \((0,1)\). We first construct two boundaries:
	\begin{align*}
	\tilde{\tau}_{l} &= \inf\{\tau_1: g(\delta_0(\tau),\tau)=0,\,\forall \tau \in [\tau_1,\tau_0)\}, \\
	\tilde{\tau}_{u} &= \sup\{\tau_2: g(\delta_0(\tau),\tau)=0,\,\forall \tau \in (\tau_0,\tau_2]\}.
	\end{align*}
and then show that \(\tilde{\tau}_l=0\) and \(\tilde{\tau}_{u}=1\). To this end, we introduce \(\tilde{g}(r,\phi,\tau)\), a new real-valued function obtained by transforming \(g(x,\tau)\) via polar coordinates. Using the properties of \(\tilde{g}\) at the boundaries, we achieve the desired conclusion; see (\ref{eq_a_9})--(\ref{eq_a_19}) for details.
\end{remark}

\section{Asymptotic Theory of the SQE $\hat{\theta}_n(\tau)$}\label{section5}
In this section,  we investigate  the  uniform properties of $\hat{\theta}_n(\tau)$ in (\ref{eq_3_0}). Inspired by Theorem \ref{theorem_3_1}, we reparameterize the loss (\ref{loss_function}) with $u=\sqrt{n}[\theta-\theta_{0}(\tau)]$ and let
\begin{align*}
	L_n(u,\tau) & ={n}\Big\{\tilde{L}_n\Big(\frac{u}{\sqrt{n}}+\theta_{0}(\tau),\tau\Big)-\tilde{L}_n(\theta_{0}(\tau),\tau)\Big\},
\end{align*}
where $\theta_{0}(\tau)$ is defined as in Section \ref{section4}.
It is straightforward to see that for any $\tau$, $\hat{u}_n(\tau)=\sqrt{n}(\hat{\theta}_n(\tau)-\theta_{0}(\tau))$ is the minimizer of $L_n(u,\tau)$. We further rewrite  $L_n(u,\tau)$ as follows:
\begin{align*}
	L_n(u,\tau) 		& =-u^{\top} T_n(\tau)+\sum_{t=p+1}^{n}{E[\zeta_t(u,\tau)\vert \mathcal{F}_{t-1}]}+\alpha_n(u,\tau),
\end{align*}
where $T_n(\tau)=\sum_{t=p+1}^{n}{\xi_t(\tau)}$, $\alpha_n(u,\tau)=\sum_{t=p+1}^{n}{\Big\{ \zeta_t(u,\tau)-E[\zeta_t(u,\tau)\vert \mathcal{F}_{t-1}] \Big\}}$ and
\begin{align}
	\xi_t(\tau) & =\frac{1}{\sqrt{n}}w_t Z_{t-1} \psi_{\tau}(\varepsilon_{t}-\delta^{\top}_0(\tau)Z_{t-1}),\label{eq_3_2}\\
	\zeta_t(u,\tau) & =w_t \int_{0}^{\frac{u^{\top}}{\sqrt{n}}Z_{t-1}}{[I(\varepsilon_{t}-\delta^{\top}_0(\tau)Z_{t-1}\leq s)-I(\varepsilon_{t}-\delta^{\top}_0(\tau)Z_{t-1}\leq 0)]ds}.\label{eq_3_3}
\end{align}		
Note that $ET_n(\tau)=-\sqrt{n}{\partial {E\tilde{L}_n(\theta_{0}(\tau),\tau)}}/{\partial \theta}$, by Theorem \ref{theorem_3_1}, we have
\begin{align*}
	\sup_{\tau \in \mathcal{T}}{\Big \vert   ET_n(\tau)\Big \vert}=O(n^{1/2-\alpha_0}).
\end{align*}
As a result,  when $\alpha_0>1/2$,   $ET_n(\tau)$ is asymptotically eliminated.

Now, we focus on the dominating term $\{T_n(\tau)-ET_n(\tau)\}$. For its uniform property on the set $\mathcal{T}$, it is necessary to characterize the more accurate temporal dependence of $\{\varepsilon_{t}\}$ and $\{y_t\}$. Let $\mathcal{F}^{*(t-m)}_{t}=(\eta_t,\cdots,\eta_{t-m+1},\eta^{*}_{t-m},\eta_{t-m-1},\cdots)$, where $\{\eta^{*}_{t}\}$ is an independent copy of $\{\eta_t\}$. For any random variable $X_t=G(\mathcal{F}_t)$, denote $X^{*(t-m)}_t=G(\mathcal{F}^{*(t-m)}_{t})$. Then we proceed with the following assumption.
\begin{assumption}\label{assumption_3_1}
	There exists a $\kappa_0 \in (0,1)$ such that, for any $m \geq 0$,
	\begin{align*}
	\Vert Y_p-Y^{*(p-m)}_p\Vert_{\alpha_0}+\sup_{x \in [0,1]}{\Vert \sigma_t(x)-\sigma^{*(t-m)}_t(x) \Vert_{\alpha_0}}=O(\kappa_0^m).
	\end{align*}
\end{assumption}
Assumption \ref{assumption_3_1} is  sufficiently general to cover a wide range of GARCH-type models that are commonly used in the literature; see \citet{wu2005bahadur} for details. In that Assumption \ref{assumption_3_1} allows for $\alpha_0<1$,  it is a weaker restriction than those imposed on the predictors and noise in quantile regression \citep{zhou2017, zhu2019}.
Let $l^{\infty}(\mathcal{T})$ denote the space of all bounded functions on the set $\mathcal{T}=[\epsilon,1-\epsilon]$ equipped with the uniform topology.
The weak convergence of the centered process is presented as below.
\begin{lemma}\label{theorem_3_2}
	Suppose that Assumptions \ref{assumption_2_1}-\ref{assumption_2_4} (i), Assumption \ref{assumption_2_5}, Assumption \ref{assumption_2_7}, and Assumption \ref{assumption_3_1} hold. Then
	\begin{align*}
		\{T_n(\tau)-ET_n(\tau)\}_{\tau \in \mathcal{T}} \rightsquigarrow T_0(\tau),
	\end{align*}
	as $n \to \infty$, where $\rightsquigarrow$ denotes weak convergence on the space $l^{\infty}(\mathcal{T})$, and $T_0(\tau)$ is a centered Gaussian process with $Cov(T_0(\tau_{1}),T_0(\tau_{2}))=\sum_{k\in \mathbb{Z}}{\int_{0}^{1}{\mbox{Cov}(\xi_{0}(\tau_{1},s),\xi_{k}(\tau_{2},s)) ds}}$,  where $\tau_{1},\tau_{2} \in \mathcal{T}$
	and $\xi_{t}(\tau,s)=w(s,Y_{t-1}(s)) Z_{t-1}(s)$ $\psi_{\tau}(\varepsilon_{t}(s)-\delta^{\top}_0(\tau)Z_{t-1}(s))$.
\end{lemma}

 \begin{remark}\label{remark_5_1}
 	Lemma 5.1 essentially  establishes a functional central limit theorem for an  empirical process of the non-stationary time series $\{(Y_{t}, \varepsilon_{t})\}$ with $w_{t}$ and $\delta_{0}(\tau)$. The existing results, such as those in \citet{zhou2020} and \citet{phan2022}, are only for  cases with  $w_t=1$, $Z_{t-1}=1$ and $\delta_{0}(\tau)=\tau$, and hence  cannot be applied to this process.
			The proof of Lemma 5.1 is to show the   tightness and  convergence of finite-dimensional distributions of the process $\{T_n(\tau)-ET_n(\tau)\}$. Following   \citet{wu2008}, we  decompose the process  into a martingale component $M_n(\tau)$ in (B.1) and a remainder $N_n(\tau)$ in (B.2) and show the tightness of each part.
Since $N_n(\tau)$ is non-zero when $\tau\neq \tau_0$,
			 it is essential to account for the  temporal dependencies of $w_tZ_{t-1}$, $f(\delta^{\top}_{0}(\tau)Z_{t-1}/\sigma_{t})$ and $f^{(1)}(\delta^{\top}_{0}(\tau)Z_{t-1}/\sigma_{t})$
 to establish its tightness; see (C.20)--(C.28) in the supplementary material.
 Furthermore, because of  the presence of the term \(N_n(\tau)\), the martingale CLT cannot be applied directly. Instead, a martingale approximation, as in \citet{wu2007} and \citet{zhang2012}, is required to show the convergence of  finite-dimensional distributions.
\end{remark}

Using Theorem \ref{theorem_3_1} and Lemma \ref{theorem_3_2} and by the standard convex argument in \citet{kato2009}, the asymptotic properties of $\hat{\theta}_n(\tau)$ can be derived as follows.

\begin{theorem}\label{theorem_3_3}
	Suppose that all the conditions in Lemma \ref{theorem_3_2} hold. If $\alpha_0>1/2$, then
	\begin{align*}
		\sqrt{n}(\hat{\theta}_n(\tau)-\theta_0(\tau)) \rightsquigarrow \Sigma^{-1}(\tau) T_0(\tau),
	\end{align*}
	where  $\Sigma(\tau)=\int_{0}^{1}{E\big[{w(s,Y^{\top}_{t-1}(s))}Z_{t-1}(s)Z^{\top}_{t-1}(s)f(\frac{\delta^{\top}_0(\tau)Z_{t-1}(s)}{\sigma_{t}(s)})\sigma^{-1}_{t}(s)  \big]ds}$.
\end{theorem}
Theorem \ref{theorem_3_3} shows that the SQE,  $\hat{\theta}_n(\tau)$,  converges {uniformly} to $\theta_0(\tau)$ in $\mathcal{T}$ at a rate of  $\sqrt{n}$, and $\delta_0(\tau)$ is its bias when $\tau\neq\tau_0$. These findings are corroborated by the simulation results provided in the supplementary material.
{The condition $\alpha_0>1/2$ is primarily introduced  to mitigate the estimation effect arising from non-stationarity.} When the noise is strictly stationary, this requirement {becomes} unnecessary.

\section{Asymptotic Theory of $\hat{\tau}_n$ and $\hat{\theta}_n$}\label{section6}
\subsection{The Asymptotic Normality}
In this section, we first establish the limiting theory of $\hat{\tau}_n$ defined in (\ref{eq_4_2}).
Let  $r_{t0}(\tau;\tilde{w}_l)=\tilde{w}_{lt} \psi_{\tau}(y_t-Z^{\top}_{t-1}{\theta}_0(\tau))$ and $R_t(\tau;\tilde{w}_l)=r_t(\tau;\tilde{w}_l)-r_{t0}(\tau;\tilde{w}_l)$. 	
We introduce an assumption and a lemma to describe the  estimation error in $R_t(\tau;\tilde{w}_l)$ as follows.

\begin{assumption}\label{assumption_4_1}
	 $\{\tilde{w}_l(x_0,x_1,\cdots,x_{\tilde{p}})x^{l_i}_{i}x^{l_j}_j x^{l_k}_k x^{l_m}_m \}$ are bounded and Lipschitz continuous on $[0,1]\times \mathbb{R}^{\tilde{p}\vee p}$, where $i,j,k,m\in \{1,\cdots,p\}$ and $l_i, l_j, l_k,l_m \in \{0,1\}$.
\end{assumption}

\begin{lemma}\label{lemma_4_1}
	Suppose that all the conditions in Theorem \ref{theorem_3_3} and Assumption \ref{assumption_4_1} hold. Then for any $l\in \{1,\cdots,L\}$, we have
	\begin{align*}
		& \sup_{\tau \in \mathcal{T}} \Big \vert  \frac{1}{\sqrt{n}}\sum_{t=p+1}^{n}{R_t(\tau;\tilde{w}_l)}+b^{\top}(\tau;\tilde{w}_l)\sqrt{n}(\hat{\theta}_n(\tau)-\theta_0(\tau)) \Big \vert=o_p(1),\\
		& \sup_{\tau\in \mathcal{T}} \Big \vert \frac{1}{\sqrt{n}}\sum_{t=p+1}^{n}{E[r_{t0}(\tau;\tilde{w}_l)]}-\sqrt{n}\tilde{g}(\tau;\tilde{w}_l)\Big \vert=o(1),
	\end{align*}
	as $n \to \infty$, where $b(\tau;\tilde{w}_l)=\int_{0}^{1}$ $ E\big[ {\tilde{w}_l(s,\tilde{Y}^{\top}_{t-1}(s))}Z_{t-1}(s)f(\frac {\delta^{\top}_0(\tau)Z_{t-1}(s)}{\sigma_{t}(s)})\sigma^{-1}_{t}(s)   \big] ds$ and $\tilde{g}(\tau;\tilde{w}_l)=\int_{0}^{1}{E\big\{\tilde{w}_l(s,\tilde{Y}^{\top}_{t-1}(s))[\tau-F\big( \frac{\delta_0^{\top}(\tau)Z_{t-1}(s)}{\sigma_{t}(s)} \big) ]   \big\} ds}$ .
\end{lemma}
  Like Assumption \ref{assumption_2_7}, Assumption \ref{assumption_4_1} plays a key role in coping with the heavy tail of $Z_{t-1}$. Now, an additional assumption is required for the asymptotic normality of $\hat{\tau}_n$.
\begin{assumption}\label{assumption_4_2}
	(i). $\sum_{l=1}^{L}{\tilde{g}^2(\tau;\tilde{w}_l)}>0$ for any $\tau \neq \tau_{0}$; (ii). $\partial \tilde{g}(\tau_{0};\tilde{w}_l)/\partial\tau \neq 0$ for some $l\leq L$.
\end{assumption}

 Assumption \ref{assumption_4_2} (i) ensures that $\sum_{l=1}^{L}{\tilde{g}^2(\tau;\tilde{w}_l)}$ achieves its unique minimizer  at $\tau_{0}$ since $\tilde{g}(\tau_{0};\tilde{w}_l)=0$.
From    (\ref{eq_a_40}),  we  can see that
\begin{align*}
\sum_{l=1}^{L}{ \big\{n^{-1}\sum_{t=p+1}^{n}{r_t(\tau;\tilde{w}_l)}\big\}^2} & =\sum_{l=1}^{L}{\tilde{g}^2(\tau;\tilde{w}_l)}+o_p(1),
\end{align*}
 and hence  Assumption \ref{assumption_4_2} (i) is   necessary for the  consistency of $\hat{\tau}_{n}$.
 From   (\ref{eq_a_47}),  we can see that  Assumption \ref{assumption_4_2} (ii) is essential to establish asymptotic normality. In fact,
  Assumption \ref{assumption_4_2} is the usual  condition for $M$-estimation; see Section 3.2 in \citet{van1996}.

Define $a_l=-[\partial \tilde{g}(\tau_{0};\tilde{w}_l)/\partial\tau]/\sum_{l\in \mathcal{A}}{[\partial \tilde{g}(\tau_{0};\tilde{w}_l)/\partial\tau]^2}$ and
$$\tilde{h}_{t}(s)=\sum_{l\in \mathcal{A}}{a_l[\tilde{w}_l(s,\tilde{Y}^{\top}_{t-1}(s))-b^{\top}(\tau_{0};\tilde{w}_l)\Sigma^{-1}(\tau_{0}) w(s,Y^{\top}_{t-1}(s))Z_{t-1}(s)  ]},$$
where  $\mathcal{A}=\{l: \partial \tilde{g}(\tau_{0};\tilde{w}_l)/\partial\tau \neq  0\}$. The asymptotic normality of $\hat{\tau}_n$ is shown as follows.
\begin{theorem}\label{theorem_4_1}
	Suppose that Assumptions \ref{assumption_2_1}--\ref{assumption_2_5}, Assumption \ref{assumption_2_7}, Assumption \ref{assumption_3_1} and Assumptions \ref{assumption_4_1}-\ref{assumption_4_2} hold and $\alpha_0>1/2$. Then, it follows that
	\begin{align*}
		\sqrt{n}(\hat{\tau}_n-\tau_{0}) \overset{d}{\longrightarrow} N(0,\gamma^2_1),
	\end{align*}
	as $n \to \infty$, where $\gamma^2_1=\tau_{0}(1-\tau_{0})\int_{0}^{1}{E\tilde{h}^2_t(s)ds   }$.
\end{theorem}

We will show how to choose $\{\tilde{w}_{lt}\}$ such that Assumptions \ref{assumption_4_1}--\ref{assumption_4_2} are satisfied in the next subsection.
Further, from (\ref{eq_a_49}), we can  show that
	          \begin{align*}
	          \sqrt{n}(\hat{\theta}_n(\hat{\tau}_n)-\theta_0)=\sqrt{n}(\hat{\theta}_n(\tau_{0})-\theta_0(\tau_{0}))+\frac{\partial \delta_0(\tau_{0})}{\partial \tau}\sqrt{n}(\hat{\tau}_n-\tau_{0})+o_p(1),
	          \end{align*}
in which the second term demonstrates the effect of replacing $\tau_{0}$ in the  oracle estimator $\hat{\theta}_n(\tau_{0})$ asymptotically by
the estimator $\hat{\tau}_{n}$.
Since the existence of the constant term in $Z_{t-1}(s)$ such that the vector $\partial \delta_{0}(\tau_{0})/\partial \tau$ is non-zero from Theorem \ref{theorem_3_1} (ii),  this effect  cannot be eliminated.
Using Theorems \ref{theorem_4_1} and \ref{theorem_3_3}, the limiting property of the feasible estimator $\hat{\theta}_n$, as defined in (\ref{feasible_theta}), is derived as below.

\begin{theorem}\label{theorem_4_2}
	Suppose that the conditions in Theorem \ref{theorem_4_1} hold. Then we have
	\begin{align*}
		\sqrt{n}(\hat{\theta}_n-\theta_0)\overset{d}{\longrightarrow} N(0, \Gamma_1),
	\end{align*}
	as $n \to \infty$, where $\Gamma_1=\tau_{0}(1-\tau_{0})\int_{0}^{1}{E\Big\{\tilde{H}_t(s)\tilde{H}^{\top}_t(s)\Big\}ds}$ with $$\tilde{H}_t(s)=\Sigma^{-1}(\tau_{0})w(s,Y^{\top}_{t-1}(s))Z_{t-1}(s)+\frac{\partial \delta_0(\tau_{0})}{\partial \tau}\tilde{h}_t(s).$$
\end{theorem}

Theorem \ref{theorem_4_2} demonstrates that the proposed two-step estimator $\hat{\theta}_n$ is consistent and weakly converges to a normal distribution, without imposing any restrictions on the structure of heteroscedasticity, the tail index or the symmetry of the noises. Notably, even in the case of light-tailed distributions, the usual condition $E[\varepsilon_t\vert \mathcal{F}_{t-1}]=0$ is not required. This makes the proposed procedure highly attractive for practical applications, with significant potential in finance and econometrics.
 When the second term $\partial \delta_0(\tau_{0})/\partial \tau \tilde{h}_t(s)$ in $\tilde{H}_t(s)$ is removed, the matrix $\Gamma_1$  reduces to
$$\Gamma_{10}:=\tau_{0}(1-\tau_{0})\Sigma^{-1}(\tau_{0}) \Big\{\int_{0}^{1}{E[w^2(s,Y^{\top}_{t-1}(s))Z_{t-1}(s)Z^{\top}_{t-1}(s)]ds} \Big\} \Sigma^{-1}(\tau_{0}),$$
which is exactly the asymptotic matrix of $\sqrt{n}(\hat{\theta}_n(\tau_{0})-\theta_{0})$ in Theorem \ref{theorem_3_3}.

\begin{remark}
	It is interesting to see whether  the previous two-step estimating procedure can be extended for
	the AR($\infty$) model.  We study the following AR($\infty$) model:
	\begin{align}\label{high_model}
		y_t = \mu_0 + \sum_{j=1}^{\infty} \phi_{j0} y_{t-j} + \varepsilon_t \mbox{ and }\varepsilon_t=\eta_{t} \sigma_{t},
	\end{align}
	where $\{\sigma_{t}\}$ is strictly stationary and $\phi_{j0}=O(c^{j}_0)$ for some $c_0 \in (0,1)$.
	Let $p=M_{0}\log (n)$ be the truncated order with a constant $M_{0}>0$ and denote $\theta_0=(\mu_0,\phi_{10},\cdots,\phi_{p0})^{\top}$. Some further results are presented in the supplementary material, in which
	we show that the bias $\delta_{0}(\tau)$ exists  uniquely  in  Theorem G.1 and
	under certain conditions, there exists a constant $\beta_{0}$ such that
	\begin{eqnarray}\label{neweq}
		\sup_{\tau \in \mathcal{T}}{\vert \hat{\theta}_n(\tau)-\theta_{0}(\tau) \vert}=O_p(p^{\beta_0+1/2}n^{-1/2}),
	\end{eqnarray}
	in Theorem G.4.
	As observed in (\ref{neweq}), the convergence rate of the SQE is slightly slower than in the case where $p$ is fixed. We can also show that \(\hat{\tau}_n\) will converge to \(\tau_0\) at a similar rate. As a result,  we cannot obtain the limiting distributions like those in Theorems \ref{theorem_3_3}, \ref{theorem_4_1}, and \ref{theorem_4_2} in Sections \ref{section5}--\ref{section6} and hence, our two-step inference approach is not applicable to model (\ref{high_model}). We leave the inference problem for
	AR$(\infty)$ to future research.
\end{remark}

\subsection{The selection of  $\{\tilde{w}_l\}$}\label{subsection_6_2}
Here, we provide a general way to  determine a class of  the weight functions $\{\tilde{w}_l\}_{l=1}^L$ suitable for verifying the conditional moment condition in  (\ref{eq_4_1}). This method yields weight functions with a straightforward form that also satisfy Assumptions \ref{assumption_4_1}--\ref{assumption_4_2}.

Let $\tilde{w}_{0}(\cdot)$ be any positive function on $[0,1]\times \mathbb{R}^{p+1}$ that satisfies Assumption \ref{assumption_2_7}. By the law of total expectation, the nonstationary version of the condition (\ref{eq_4_1}), and the fact that \[E[\psi_{\tau}(y_t(s)-Z^{\top}_{t-1}(s){\theta}_0(\tau))\vert \mathcal{F}_{t-1}]=\big[\tau-F({\delta_0^{\top}(\tau)Z_{t-1}(s)}/\sigma_{t}(s) ) \big],\]  it is clear to see that
\begin{align}\label{eq_4_4}
	\int_{0}^{1}{E\Big\{\tilde{w}_0(s,{Y}^{\top}_{t-1}(s))\big[\tau-F({\delta_0^{\top}(\tau)Z_{t-1}(s)}/\sigma_{t}(s)) \big] \psi_{\tau}(y_t(s)-Z^{\top}_{t-1}(s){\theta}_0(\tau))  \Big\} ds}\geq 0,
\end{align}
and the equality holds if and only if $\tau=\tau_{0}$.

For many parametric models, such as  those in \citet{engle1982} and \citet{taylor1986}, the innovations ${\eta_{t}}$ can be  approximated accurately by the noise ${\varepsilon_t(s)}$. Therefore, by (\ref{eq_4_4}) and  $\varepsilon_{t}(s)=y_t(s)-\theta^{\top}_0Z_{t-1}(s)$,
for any $\tau\neq \tau_{0}$, there exists some ${q}(\tau),\tilde{p}(\tau)$ such that
\begin{align*}
	\int_{0}^{1}{E\Big\{\tilde{w}_0(s,{Y}^{\top}_{t-1}(s))G_{q(\tau)}(\tau,y_{t-1}(s),\cdots,y_{t-\tilde{p}(\tau)}(s)) \psi_{\tau}(y_t(s)-Z^{\top}_{t-1}(s){\theta}_0(\tau))  \Big\} ds}> 0,
\end{align*}
where $G_{q(\tau)}(\tau,y_{t-1}(s),\cdots,y_{t-\tilde{p}(\tau)}(s))$ is the approximation of $\big[\tau-F({\delta_0^{\top}(\tau)Z_{t-1}(s)}/\sigma_{t}(s) ) \big]$. This inequality reveals that for   any $\tau\neq \tau_{0}$,
$$E[\tilde{w}_0(s,{Y}^{\top}_{t-1}(s))\psi_{\tau}(y_t(s)-Z^{\top}_{t-1}(s){\theta}_0(\tau))\vert \{y_{t-1}(s),\cdots,y_{t-\tilde{p}(\tau)}(s)\}]\neq 0.$$

Furthermore, let $t(\cdot)$ be a bounded, Lipschitz continuous and one-to-one function. By Theorem 2 in \citet{bierens1982}, for any $\tau \neq \tau_{0}$, we can find  power series of $\{t(y_{t-1}(s)),\cdots,$ $ t(y_{t-\tilde{p}(\tau)}(s))\}$ such that
\begin{align}\label{eq_4_6}
	\int_{0}^{1} E\Big\{\tilde{w}_{1t}(\tau,s)\psi_{\tau}(y_t(s)-Z^{\top}_{t-1}(s){\theta}_0(\tau))  \Big\} ds\neq 0,
\end{align}
where $\tilde{w}_{1t}(\tau,s)=\tilde{w}_0(s,{Y}^{\top}_{t-1}(s))\times t(y_{t-1}(s))^{d_1(\tau)}\cdots t(y_{t-\tilde{p}(\tau)}(s))^{d_{\tilde{p}(\tau)}(\tau)}.$
By (\ref{eq_4_6}) and the continuity, for any $\tau_{i} \neq \tau_{0}$, there exists an open ball $B_{r_i}(\tau_{i})$ with the radius $r_i>0$ such that for any $\tau\in B_{r_i}(\tau_{i})$, we have
\begin{align}\label{eq_4_8}
	\int_{0}^{1}{E\Big\{\tilde{w}_{1t}(\tau_i,s) \psi_{\tau}(y_t(s)-Z^{\top}_{t-1}(s){\theta}_0(\tau))  \Big\} ds}\neq 0.
\end{align}
Write $\delta'_0(\tau_{0}) = \partial \delta_0(\tau_{0})/\partial \tau$. Following a similar derivation as in (\ref{eq_4_6}), we obtain
\begin{align}\label{eq_4_7}
	\int_{0}^{1}{E\Big\{\tilde{w}_{1t}(\tau_{0},s) \Big[1-f(0){Z^{\top}_{t-1}(s)\delta'_0(\tau_{0})}/\sigma_{t}(s)\Big]  \Big\} ds}\neq  0, \mbox{ for } \tau=\tau_0,
\end{align}
where the integral is the derivative of (\ref{eq_4_8}) with respect to $\tau$ at the point $\tau_0$.
Therefore, (\ref{eq_4_8}) holds for any $\tau \in B_{r_0}(\tau_{0})$ but $\tau \neq \tau_{0}$, where $r_0$ is some positive number. Then, by the compactness of the set $\mathcal{T}$, we can select one finite subset $\tilde{\mathcal{T}}$ (must contain $\tau_{0}$) such that
\begin{align*}
	\sum_{\tau_{i}\in \tilde{\mathcal{T}}}\Big \{\int_{0}^{1}{E\Big\{\tilde{w}_{1t}(\tau_i,s) \psi_{\tau}(y_t(s)-Z^{\top}_{t-1}(s){\theta}_0(\tau))  \Big\} ds}\Big\}^2\geq 0,
\end{align*}
for any $\tau\in \mathcal{T}$ and the equality holds if and only if $\tau=\tau_{0}$. Since $\tilde{\mathcal{T}}$ is finite, Assumption \ref{assumption_4_2} holds if we choose the finite elements in the weighting functions
\begin{align}\label{eq_4_9}
	\tilde{w}_{l}(s,\tilde{Y}^{\top}_{t-1}(s))=\tilde{w}_0(s,{Y}^{\top}_{t-1}(s))\times t(y_{t-1}(s))^{d_1}\cdots t(y_{t-\tilde{p}}(s))^{d_{\tilde{p}}}.
\end{align}
The selection of $\tilde{w}_0(\cdot)$ and $t(\cdot)$ ensures that Assumption \ref{assumption_4_1} holds.
		In practice, we recommend selecting a weight function $\tilde{w}_0(\cdot)$ that differs from $w(\cdot)$ in (\ref{eq_3_0}) to improve the efficiency.

%

\section{The Bootstrapped Approximation}\label{section7}
To perform inference on the estimators $\hat{\tau}_n$ and $\hat{\theta}_n$, we must estimate the limiting distributions in Theorems \ref{theorem_4_1}--\ref{theorem_4_2}. However,
  this involves unobserved quantities such as  $f(\cdot)$ and $\sigma_{t}$.  Here we use {the random weighting} approach by \citet{zhu2015} to approximate the distributions.
For any $\tau \in (0,1)$, define
\begin{align}\label{eq_5_1}
	\hat{\theta}^{*}_n(\tau)=\argmin_{\theta \in \mathbb{R}^{p+1}}{ \sum_{t=p+1}^{n}{w^{*}_t w_t \rho_{\tau}(y_t-Z^{\top}_{t-1}\theta)} },
\end{align}
where $\{w_t\}$ is the weight used in (\ref{eq_3_0}) and $\{w^{*}_t\}$ is a sequence of i.i.d. random variables with $P(w^{*}_t=0)=P(w^{*}_t=2)=1/2$.
\begin{assumption}\label{assumption_5_1}
	$\{w^*_t\}$ and $\{y_t\}$ are independent.
\end{assumption}
\begin{theorem}\label{theorem_5_1}
	Suppose that all the conditions in Theorem \ref{theorem_3_3} and Assumption \ref{assumption_5_1} hold. Conditional on $\{y_1,\cdots,y_n\}$, it follows that
	\begin{align*}
		\sqrt{n}(\hat{\theta}^{*}_n(\tau)-\hat{\theta}_n(\tau)) \rightsquigarrow \Sigma^{-1}(\tau) T^{*}_0(\tau)
	\end{align*}
	in probability, where $T^{*}_0(\tau)$ is a centered Gaussian process with  $Cov(T^{*}_0(\tau_{1}),T^{*}_0(\tau_{2}))=\int_{0}^{1}{\mbox{E}[\xi_{0}(\tau_{1},s)\xi^{\top}_{0}(\tau_{2},s)] ds}$, where $\xi_{0}(\tau,s)$ is defined in Lemma \ref{theorem_3_2}.
\end{theorem}
\begin{remark}\label{remark_7_1}
Unlike \citet{zhu2015} and \citet{zhu2019}, who considered the bootstrap distribution only at the point \(\tau_0\), Theorem \ref{theorem_5_1} innovatively extends this to a stochastic process over
 $\tau\in \mathcal{T}$.
  Like Theorem \ref{theorem_3_3}, the key step in Theorem  \ref{theorem_5_1} is to prove the weak convergence of   $T^{*}_n(\tau)=\sum_{t=p+1}^{n}{w^{*}_t\xi_{t}(\tau)}$.
However, given   $\{y_t\}$,  since   $\{{w^{*}_t}\}$ is   random while $\{\xi_{t}(\tau)\}$ is fixed, weak convergence results based on the distribution of $\{y_t\}$, such as Lemma \ref{theorem_3_2} and Theorem \ref{theorem_3_3}, cannot be  applied directly to show the validity of bootstrap methodology. Instead, we require point-wise convergence results, which is a primary challenge.
To address this issue, we utilize Dudley's almost sure representation theorem as in \citet{pollard} and \citet{van1996}. The new method involves creating a new probability space for $\{ y_t\}$ where weak convergence is transformed into almost sure convergence, with distributions identical to the original space.
Thus, our proof procedure is highly innovative and significantly different from that in Lemma \ref{theorem_3_2}.
\end{remark}
As shown in (\ref{eq_a_47_a}) and (\ref{eq_a_49}), the asymptotic distributions in Theorems \ref{theorem_4_1}-\ref{theorem_4_2} only involve  the true    $\tau_{0}$. Therefore, given that $T^{*}_0(\tau_{0})$ and
$T_0(\tau_{0})$ share the same distribution because  $\{\xi_{t}(\tau_{0},s)\}$ is a m.d.s., we define the bootstrapped version of $\hat{\tau}_n$ as
\begin{align}
	\hat{\tau}^{*}_n=\argmin_{\tau \in \mathcal{T}}{\sum_{l=1}^{L}{ \Big\{n^{-1}\sum_{t=p+1}^{n}{r^{*}_t(\tau;\tilde{w}_l)}\Big\}^2}},\label{eq_5_2}
\end{align}
where $r^{*}_t(\tau;\tilde{w}_l)=w^{*}_t\tilde{w}_{lt} \psi_{\tau}(y_t-Z^{\top}_{t-1}\hat{\theta}^{*}_n(\tau))$. Denote $\hat{\theta}^{*}_n=\hat{\theta}^{*}_n(\hat{\tau}^{*}_n)$. The following theorem \ref{theorem_5_2} shows that the limiting distributions  of $\hat{\tau}_n$ and $\hat{\theta}_n$ can be  approximated  effectively by the distributions of their bootstrapped counterparts, $\hat{\tau}_n^{*}$ and $\hat{\theta}^{*}_n$.
\begin{theorem}\label{theorem_5_2}
	Suppose that all the conditions in Theorem \ref{theorem_4_1} and Assumption \ref{assumption_5_1} hold. Conditional on $\{y_1,\cdots,y_n\}$,  we have
	\begin{align*}
		\sqrt{n}(\hat{\tau}^{*}_n-\hat{\tau}_n) \overset{d}{\longrightarrow}N(0,\gamma^2_1)\mbox{ and }\sqrt{n}(\hat{\theta}^{*}_n-\hat{\theta}_n)\overset{d}{\longrightarrow} N(0,\Gamma_1),
	\end{align*}
	in probability.
\end{theorem}
{Thus, the estimation procedure for the covariance matrix can  be summarized as follows:}
\begin{enumerate}
	\item[] \begin{itemize}
		\item[\it Step 1:] Generate $J$ replications of the i.i.d. random weights $\{w^{*}_t\}^{n}_{t=1}$ with the distribution $P(w^{*}_t=0)=P(w^{*}_t=2)=1/2$.
		\item[\it Step 2:] Compute the estimators $\hat{\tau}_n^{*}$ and $\hat{\theta}^{*}_n$ for the $j$-th replication, and denote them as $\hat{\tau}^{*}_{n,j}$ and $\hat{\theta}^{*}_{n,j}$.
		\item[\it Step 3:] Calculate the sample covariance of $\{\hat{\tau}^{*}_{n, j}-\hat{\tau}_n\}^{J}_{j=1}$ and $\{\hat{\theta}^{*}_{n, j}-\hat{\theta}_n\}^{J}_{j=1}$, denoted by $\hat{\gamma}^{2}_{1}$ and $\hat{\Gamma}_1$, which provides a good approximation for the asymptotic covariance matrix.
	\end{itemize}
\end{enumerate}

Let $A$ be an $s\times (p+1)$ constant matrix with rank $s$ and $a$ be a vector of size $s\times 1$. To test the linear hypothesis:
\begin{align*}
	H_0: A \theta_0=a,
\end{align*}
we can establish the Wald test statistic
\begin{align}\label{eq_5_4}
	W_{n}=(A \hat{\theta}_n -a)^{\top}(A \hat{\Gamma}_1 A^{\top})^{-1}(A \hat{\theta}_n -a).
\end{align}
If $W_n$ exceeds the upper critical value of $\chi^{2}_{s}$, the null   $H_0$ is rejected; otherwise, it is not. The test $W_n$ is crucial to  determine whether the true coefficients are zero. Alternatively, to test whether  the true level $\tau_{0}$  equals  a specific level
$\tau_1$, such as $\tau_{1}=0.5$, we can derive  a similar test for $\hat{\tau}_n$:
\begin{align}\label{eq_5_5}
	{w}_n=(\hat{\tau}_n-\tau_{1})^2/\hat{\gamma}^2_1.
\end{align}
In this case, the null hypothesis $\tau_{0}=\tau_{1}$ is rejected when $w_n$ is greater than the critical value of $\chi^2_1$.
In addition to these tests, using $\hat{\gamma}^2_1$ and $\hat{\Gamma}_1$, we can directly construct confidence intervals for the parameters. The simulation results for the size and power of these tests are provided in Section \ref{section8}, while the coverage probabilities of the confidence intervals are detailed in the supplementary material.
\section{Simulation Studies}\label{section8}
In this section, we first examine the finite sample performance of the proposed estimators $\hat{\tau}_n$ and $\hat{\theta}_n$ in the model:
\begin{align}\label{eq_6_1}
	y_t=\mu_0+\phi_{10} y_{t-1}+\varepsilon_{t},
\end{align}
where $(\mu_0,\phi_{10})=(0.1,0.5)$ and the noise is assumed to be the GARCH model as:
\begin{align}\label{eq_6_2}
	\varepsilon_{t}=v(t/n)\times \eta_{t} \sigma_{t}\mbox{ and }\sigma^2_{t}=0.1+(a_1\eta^2_{t-1}+b_1)\sigma^2_{t-1},
\end{align}
where the non-stationary multiplier $v(x)$ satisfies one of the following structures:
\begin{align}
	[\mbox{Increasing change}]\,\,\,\, 	v(x) & =x/4+1/4,\label{eq_6_3}\\
	[\mbox{Symmetric change}]\,\,\,\, 	v(x) & =(x-1/2)^2+1/2.\label{eq_6_4}
\end{align}
Furthermore, we set $\eta_{t}=\eta_{t0}-F_0^{-1}(\tau_{0})$, where $\{\eta_{t0}\}$ is a sequence of i.i.d. random variables with the distribution function $F_0(\cdot)$. Hence, the $\tau_{0}$-th quantile of $\eta_t$ is zero. To generate the heavy-tailed time series, the following settings in model (\ref{eq_6_2}) were used.

1.\quad When $\eta_{t0} \sim N(0,1)$, consider (i) $(a_1,b_1)=(0.7,0.15)$ with $\tau_{0}=0.3$; (ii) $(a_1,b_1)=(0.8,0.2)$ with $\tau_{0}=0.5$; (iii) $(a_1,b_1)=(0.7,0.2)$ with $\tau_{0}=0.7$.

2.\quad When $\eta_{t0} \sim t_3$, consider (i) $(a_1,b_1)=(0.25,0.2)$ with $\tau_{0}=0.3$; (ii) $(a_1,b_1)=(0.4,0.1)$ with $\tau_{0}=0.5$; (iii) $(a_1,b_1)=(0.3,0.05)$ with $\tau_{0}=0.7$.

\noindent
By \citet{zhang2015a}, the above settings  ensure that the tail index of the noise is less than 2.  We choose the estimation weight function in (\ref{eq_3_0}) as
\begin{align}\label{weigth_simulation}
	w_t=1/\prod_{i=1}^{p}{(1+{y _{t-i}^4})}
\end{align}
and for (\ref{weight_practical}), we take
\begin{align}\label{weight2_simulation}
	\Bigg\{\tilde{w}_{lt}=\Big(\prod_{i=1}^{p}{\frac{1}{1+e^{\vert y_{t-i}\vert}}}\Big)\Big(\frac{y_{t-1}}{\sqrt{1+y^2_{t-1}}}\Big)^{d_1} \cdots \Big(\frac{y_{t-p}}{\sqrt{1+y^2_{t-p}}}\Big)^{d_p}\Bigg\},
\end{align}
where $0\leq d_i \leq 3$ for any $i\in \{1,\cdots,p\}$.
For each case, we  perform $1000$ replications, set $\mathcal{T}=[0.1,0.9]$, and use a bootstrap sample size of $J=200$.

Tables \ref{table1}--\ref{table2} report the sample biases (Bias),  sample standard
deviations (SD) and  average bootstrapped sample standard
deviations (BD) of $\hat{\tau}_n$ and $\hat{\theta}_n$ for the fluctuations  (\ref{eq_6_3})--(\ref{eq_6_4}), respectively.
We observe that all estimators  approximate the true parameters closely, with accuracy improving as sample sizes increase. In addition, the differences between SDs and BDs are small across all cases. This indicates that the proposed two-step SQE  approach is both feasible and robust, even in the presence of the heavy tail of  $\varepsilon_t$, the distribution of $\eta_t$, and the non-stationary structure of $v(x)$. For the same model, the additional simulation results regarding the bias of $\hat{\theta}_n(\tau)$ when $\tau\neq \tau_{0}$, as well as the empirical coverage probabilities of the confidence intervals, are presented in the supplementary material.
\begin{table}[h]
	\centering
	\small
	\caption{\footnotesize Biases and SDs of $\hat{\tau}_n$ and $\hat{\theta}_n$ for model (\ref{eq_6_1}) with $v(x)\sim $ (\ref{eq_6_3})}
	\scalebox{0.85}{
		\begingroup
		\setlength{\tabcolsep}{3.5pt} 
		\renewcommand{\arraystretch}{1.1} 
		\begin{tabular}{ccccccccccccccccc}
			\hline
			\quad & \quad & \quad & \quad & \quad & \quad & $\hat{\tau}_n$ & \quad & \quad & \quad & $\hat{\mu}_n$ & \quad & \quad & \quad & $\hat{\phi}_n$ & \quad & \quad \\
			\cline{7-9}\cline{11-13}\cline{15-17}
			$\eta_{t0}\sim $ & \quad & $\tau_{0}$ & \quad & $n$ & \quad & Bias & SD & BD & \quad & Bias & SD & BD & \quad & Bias & SD & BD\\\hline
			$N(0,1)$ & \quad & 0.3 & \quad & 1000 & \quad & -0.0023 & 0.0672 & 0.0674 & \quad & -0.0006 & 0.0284 & 0.0301 & \quad & -0.0083 & 0.0663 & 0.0688\\
			\quad & \quad & \quad & \quad & 2000 & \quad & -0.0003 & 0.0454 & 0.0475 & \quad & -0.0001 & 0.0197 & 0.0204 & \quad & 0.0039 & 0.0452 & 0.0469\\
			\quad & \quad & 0.5 & \quad & 1000 & \quad & 0.0005 & 0.0570 & 0.0587 & \quad & 0.0010 & 0.0315 & 0.0329 & \quad & -0.0021 & 0.0497 & 0.0498\\
			\quad & \quad & \quad & \quad & 2000 & \quad & 0.0007 & 0.0384 & 0.0406 & \quad & 0.0006 & 0.0207 & 0.0224 & \quad & -0.0019 & 0.0323 & 0.0343\\
			\quad & \quad & 0.7 & \quad & 1000 & \quad & 0.0003 & 0.0614 & 0.0642 & \quad & 0.0026 & 0.0420 & 0.0455 & \quad & -0.0083 & 0.0656 & 0.0690\\
			\quad & \quad & \quad & \quad & 2000 & \quad & 0.0001 & 0.0427 & 0.0452 & \quad & 0.0021 & 0.0298 & 0.0312 & \quad & -0.0035 & 0.0443 & 0.0475\\\hline
			$t_3$ & \quad & 0.3 & \quad & 1000 & \quad & -0.0069 & 0.0698 & 0.0725 & \quad & -0.0053 & 0.0395 & 0.0430 & \quad & 0.0091 & 0.0487 & 0.0519\\
			\quad & \quad & \quad & \quad & 2000 & \quad & -0.0043 & 0.0498 & 0.0516 & \quad & -0.0031 & 0.0260 & 0.0283 & \quad & -0.0052 & 0.0329 & 0.0348\\
			\quad & \quad & 0.5 & \quad & 1000 & \quad & -0.0011 & 0.0619 & 0.0656 & \quad & 0.0005 & 0.0307 & 0.0328 & \quad & -0.0026 & 0.0365 & 0.0399\\
			\quad & \quad & \quad & \quad & 2000 & \quad & -0.0002 & 0.0424 & 0.0444 & \quad & -0.0002 & 0.0206 & 0.0210 & \quad & 0.0006 & 0.0265 & 0.0271\\
			\quad & \quad & 0.7 & \quad & 1000 & \quad & 0.0040 & 0.0724 & 0.0762 & \quad & 0.0068 & 0.0443 & 0.0473 & \quad & -0.0071 & 0.0491 & 0.0512\\
			\quad & \quad & \quad & \quad & 2000 & \quad & 0.0025 & 0.0510 & 0.0536 & \quad & 0.0032 & 0.0305 & 0.0323 & \quad & -0.0030 & 0.0319 & 0.0343\\\hline
		\end{tabular}
		\endgroup
	}\label{table1}
\end{table}
\begin{table}[h]
	\centering
	\small
	\caption{\footnotesize Biases and SDs of $\hat{\tau}_n$ and $\hat{\theta}_n$ for model (\ref{eq_6_1}) with $v(x)\sim $ (\ref{eq_6_4})}
	\scalebox{0.85}{
		\begingroup
		\setlength{\tabcolsep}{3.5pt} 
		\renewcommand{\arraystretch}{1.1} 
		\begin{tabular}{ccccccccccccccccc}
			\hline
			\quad & \quad & \quad & \quad & \quad & \quad & $\hat{\tau}_n$ & \quad & \quad & \quad & $\hat{\mu}_n$ & \quad & \quad & \quad & $\hat{\phi}_n$ & \quad & \quad \\
			\cline{7-9}\cline{11-13}\cline{15-17}
			$\eta_{t0}\sim $ & \quad & $\tau_{0}$ & \quad & $n$ & \quad & Bias & SD & BD & \quad & Bias & SD & BD & \quad & Bias & SD & BD\\\hline
			$N(0,1)$ & \quad & 0.3 & \quad & 1000 & \quad & -0.0051 & 0.0654 & 0.0674 & \quad & -0.0035 & 0.0484 & 0.0508 & \quad & -0.0101 & 0.0627 & 0.0660\\
			\quad & \quad & \quad & \quad & 2000 & \quad & -0.0001 & 0.0445 & 0.0474 & \quad & 0.0004 & 0.0329 & 0.0348 & \quad & -0.0034 & 0.0429 & 0.0449\\
			\quad & \quad & 0.5 & \quad & 1000 & \quad & -0.0022 & 0.0593 & 0.0606 & \quad & -0.0009 & 0.0513 & 0.0527 & \quad & 0.0074 & 0.0489 & 0.0500\\
			\quad & \quad & \quad & \quad & 2000 & \quad & 0.0003 & 0.0405 & 0.0415 & \quad & 0.0004 & 0.0337 & 0.0353 & \quad & -0.0013 & 0.0335 & 0.0345\\
			\quad & \quad & 0.7 & \quad & 1000 & \quad & 0.0052 & 0.0631 & 0.0654 & \quad & 0.0076 & 0.0646 & 0.0678 & \quad & -0.0079 & 0.0656 & 0.0686\\
			\quad & \quad & \quad & \quad & 2000 & \quad &0.0038 & 0.0438 & 0.0460 & \quad & 0.0048 & 0.0434 & 0.0465 & \quad & -0.0054 & 0.0447 & 0.0472\\\hline
			$t_3$ & \quad & 0.3 & \quad & 1000 & \quad & -0.0058 & 0.0747 & 0.0759 & \quad & -0.0092 & 0.0687 & 0.0716 & \quad & -0.0089 & 0.0479 & 0.0524\\
			\quad & \quad & \quad & \quad & 2000 & \quad & 0.0003 & 0.0529 & 0.0552 & \quad & -0.0020 & 0.0486 & 0.0505 & \quad & -0.0027 & 0.0325 & 0.0350\\
			\quad & \quad & 0.5 & \quad & 1000 & \quad & -0.0006 & 0.0672 & 0.0733 & \quad & -0.0001 & 0.0512 & 0.0580 & \quad & -0.0037 & 0.0389 & 0.0411\\
			\quad & \quad & \quad & \quad & 2000 & \quad & 0.0007 & 0.0459 & 0.0480 & \quad & 0.0009 & 0.0327 & 0.0356 & \quad & 0.0016 & 0.0263 & 0.0279\\
			\quad & \quad & 0.7 & \quad & 1000 & \quad & 0.0077 & 0.0792 & 0.0819 & \quad & 0.0038 & 0.0736 & 0.0771 & \quad & -0.0069 & 0.0482 & 0.0525\\
			\quad & \quad & \quad & \quad & 2000 & \quad & 0.0043 & 0.0565 & 0.0582 & \quad & 0.0027 & 0.0584 & 0.0535 & \quad & -0.0031 & 0.0336 & 0.0357\\\hline
		\end{tabular}
		\endgroup
	}\label{table2}
\end{table}

Next, we investigate the size and power of the proposed tests of $W_n$ and $w_n$ in Section \ref{section7}. The following AR(2) model is studied:
\begin{align}\label{eq_6_6}
	y_t=0.1+0.5y_{t-1}-\kappa y_{t-2}+\varepsilon_{t},
\end{align}
where the noise $\{\varepsilon_{t}\}$ satisfies model (\ref{eq_6_2}) and $\kappa \in \{0,0.1,0.2\}$. To test the hypothesis $\kappa=0$, we use the Wald statistic $W_n$ in (\ref{eq_5_4}).
Furthermore, for every fixed $\kappa$, we adopt \textcolor{red}{the other} test $w_n$ in (\ref{eq_5_5}) to test the hypothesis $\tau_{0}=\tau_{1}$, where $\tau_{1}=\tau_{0}-\delta$ for $\delta\in \{0,0.1,0.2\}$.
In each case, we set the significance level $\alpha=0.05$. The empirical results are summarized in Tables \ref{table4}--\ref{table6}. The sizes correspond to the case when $\kappa=0$ or $\delta=0$. Notice that the sizes are always close to their nominal values while the powers increase when $\kappa$ or $\delta$ becomes larger as expected. These simulation results indicate that our tests $W_n$ and $w_n$ based on the random weighting approach have a satisfactory performance.
\begin{table}[h]
	\centering
	\small
	\caption{\footnotesize Size and Power of $W_{n}$ for the AR(2) parameter $\kappa$ in model (\ref{eq_6_6})}
	\scalebox{0.85}{
		\begingroup
		\setlength{\tabcolsep}{3pt} 
		\renewcommand{\arraystretch}{1} 
		\begin{tabular}{ccccccccccccc}
			\hline
			\quad & \quad & \quad & \quad & \quad & \quad & \multicolumn{2}{l}{$v(x)\sim (\ref{eq_6_3})$} & \quad & \quad & \multicolumn{2}{l}{$v(x)\sim (\ref{eq_6_4})$} & \quad  \\
			\cline{7-9}\cline{11-13}
			$\eta_{t0}\sim $ & \quad & $\tau_{0}$ & \quad & $n$ & \quad & $\kappa=0$ & $\kappa=0.1$ & $\kappa=0.2$ & \quad & $\kappa=0$ & $\kappa=0.1$ & $\kappa=0.2$   \\\hline
			$N(0,1)$ & \quad & 0.3 & \quad & 1000 & \quad & 0.024 & 0.486 & 0.983 & \quad & 0.040 & 0.436 & 0.967  \\
			\quad & \quad & \quad & \quad & 2000 & \quad & 0.036 & 0.808 & 1.000 & \quad & 0.042 & 0.747 & 1.000  \\
			\quad & \quad & 0.5 & \quad & 1000 & \quad & 0.042 & 0.487 & 0.972 & \quad & 0.038 & 0.493 & 0.987  \\
			\quad & \quad & \quad & \quad & 2000 & \quad & 0.052 & 0.784 & 1.000 & \quad & 0.047 & 0.789 & 1.000 \\
			\quad & \quad & 0.7 & \quad & 1000 & \quad & 0.040 & 0.445 & 0.972 & \quad & 0.026 & 0.437 & 0.951  \\
			\quad & \quad & \quad & \quad & 2000 & \quad & 0.051 & 0.795 & 0.999 & \quad & 0.042 & 0.747 & 1.000  \\\hline
			$t_3$ & \quad & 0.3 & \quad & 1000 & \quad & 0.025 & 0.540 & 0.989 & \quad & 0.017 & 0.454 & 0.977  \\
			\quad & \quad & \quad & \quad & 2000 & \quad & 0.041 & 0.881 & 1.000 & \quad & 0.041 & 0.822 & 1.000  \\
			\quad & \quad & 0.5 & \quad & 1000 & \quad & 0.043 & 0.688 & 1.000 & \quad & 0.039 & 0.664 & 0.995  \\
			\quad & \quad & \quad & \quad & 2000 & \quad & 0.049 & 0.944 & 1.000 & \quad & 0.043 & 0.941 & 1.000  \\
			\quad & \quad & 0.7 & \quad & 1000 & \quad & 0.027 & 0.605 & 0.994 & \quad & 0.026 & 0.470 & 0.982  \\
			\quad & \quad & \quad & \quad & 2000 & \quad & 0.031 & 0.914 & 1.000 & \quad & 0.034 & 0.851 & 1.000  \\\hline
		\end{tabular}
		\endgroup
	}\label{table4}
\end{table}
\begin{table}[h]
	\centering
	\small
	\caption{\footnotesize Size and Power of $w_{n}$ for the true level $\tau_0$ in model (\ref{eq_6_6}) with $v(x)\sim (\ref{eq_6_3})$}
	\scalebox{0.85}{
		\begingroup
		\setlength{\tabcolsep}{3pt} 
		\renewcommand{\arraystretch}{1} 
		\begin{tabular}{ccccccccccccccccc}
			\hline
			\quad & \quad & \quad & \quad & \quad & \quad & \multicolumn{2}{l}{$\kappa=0$} & \quad & \quad & \multicolumn{2}{l}{$\kappa=0.1$} & \quad & \quad & \multicolumn{2}{l}{$\kappa=0.2$} & \quad \\
			\cline{7-9}\cline{11-13} \cline{15-17}
			$\eta_{t0}\sim $ & \quad & $\tau_{0}$ & \quad & $n$ & \quad & $\delta=0$ & $\delta=0.1$ & $\delta=0.2$ & \quad & $\delta=0$ & $\delta=0.1$ & $\delta=0.2$ & \quad & $\delta=0$ & $\delta=0.1$ & $\delta=0.2$  \\\hline
			$N(0,1)$ & \quad & 0.3 & \quad & 1000 & \quad & 0.068 & 0.220 & 0.751 & \quad & 0.050 & 0.266 & 0.851 & \quad & 0.038 & 0.278 & 0.873 \\
			\quad & \quad & \quad & \quad & 2000 & \quad & 0.050 & 0.473 & 0.968 & \quad & 0.042 & 0.545 & 0.990 & \quad & 0.043 & 0.621 & 0.992 \\
			\quad & \quad & 0.5 & \quad & 1000 & \quad & 0.042 & 0.386 & 0.930 & \quad & 0.044 & 0.391 & 0.929 & \quad & 0.047 & 0.422 & 0.944 \\
			\quad & \quad & \quad & \quad & 2000 & \quad & 0.040 & 0.684 & 0.996 & \quad & 0.039 & 0.700 & 0.997 & \quad & 0.045 & 0.718 & 0.997 \\
			\quad & \quad & 0.7 & \quad & 1000 & \quad & 0.042 & 0.357 & 0.843 & \quad & 0.057 & 0.397 & 0.896 & \quad & 0.034 & 0.385 & 0.906 \\
			\quad & \quad & \quad & \quad & 2000 & \quad & 0.037 & 0.600 & 0.979 & \quad & 0.038 & 0.652 & 0.992 & \quad & 0.041 & 0.701 & 0.996 \\\hline
			$t_3$ & \quad & 0.3 & \quad & 1000 & \quad & 0.071 & 0.163 & 0.696 & \quad & 0.068 & 0.176 & 0.734 & \quad & 0.060 & 0.192 & 0.739 \\
			\quad & \quad & \quad & \quad & 2000 & \quad & 0.038 & 0.372 & 0.938 & \quad & 0.040 & 0.403 & 0.957 & \quad & 0.043 & 0.428 & 0.953 \\
			\quad & \quad & 0.5 & \quad & 1000 & \quad & 0.046 & 0.306 & 0.841 & \quad & 0.063 & 0.320 & 0.863 & \quad & 0.055 & 0.335 & 0.868 \\
			\quad & \quad & \quad & \quad & 2000 & \quad & 0.039 & 0.550 & 0.988 & \quad & 0.053 & 0.577 & 0.989 & \quad & 0.042 & 0.596 & 0.989 \\
			\quad & \quad & 0.7 & \quad & 1000 & \quad & 0.080 & 0.329 & 0.707 & \quad & 0.069 & 0.319 & 0.755 & \quad & 0.058 & 0.318 & 0.748 \\
			\quad & \quad & \quad & \quad & 2000 & \quad & 0.052 & 0.437 & 0.933 & \quad & 0.048 & 0.479 & 0.942 & \quad & 0.039 & 0.459 & 0.945 \\\hline
		\end{tabular}
		\endgroup
	}\label{table5}
\end{table}
\begin{table}[h]
	\centering
	\small
	\caption{\footnotesize Size and Power of $w_{n}$ for the true level $\tau_0$ in model (\ref{eq_6_6}) with $v(x)\sim (\ref{eq_6_4})$}
	\scalebox{0.85}{
		\begingroup
		\setlength{\tabcolsep}{3pt} 
		\renewcommand{\arraystretch}{1} 
		\begin{tabular}{ccccccccccccccccc}
			\hline
			\quad & \quad & \quad & \quad & \quad & \quad & \multicolumn{2}{l}{$\kappa=0$} & \quad & \quad & \multicolumn{2}{l}{$\kappa=0.1$} & \quad & \quad & \multicolumn{2}{l}{$\kappa=0.2$} & \quad \\
			\cline{7-9}\cline{11-13} \cline{15-17}
			$\eta_{t0}\sim $ & \quad & $\tau_{0}$ & \quad & $n$ & \quad & $\delta=0$ & $\delta=0.1$ & $\delta=0.2$ & \quad & $\delta=0$ & $\delta=0.1$ & $\delta=0.2$ & \quad & $\delta=0$ & $\delta=0.1$ & $\delta=0.2$  \\\hline
					$N(0,1)$ & \quad & 0.3 & \quad & 1000 & \quad & 0.062 & 0.196 & 0.742 & \quad & 0.057 & 0.206 & 0.800 & \quad & 0.047 & 0.254 & 0.837 \\
			\quad & \quad & \quad & \quad & 2000 & \quad & 0.050 & 0.394 & 0.948 & \quad & 0.054 & 0.489 & 0.972 & \quad & 0.038 & 0.541 & 0.984 \\
			\quad & \quad & 0.5 & \quad & 1000 & \quad & 0.043 & 0.356 & 0.888 & \quad & 0.048 & 0.356 & 0.892 & \quad & 0.039 & 0.376 & 0.894 \\
			\quad & \quad & \quad & \quad & 2000 & \quad & 0.049 & 0.619 & 0.994 & \quad & 0.045 & 0.656 & 0.996 & \quad & 0.039 & 0.678 & 0.995 \\
			\quad & \quad & 0.7 & \quad & 1000 & \quad & 0.063 & 0.370 & 0.817 & \quad & 0.051 & 0.367 & 0.832 & \quad & 0.065 & 0.377 & 0.859 \\
			\quad & \quad & \quad & \quad & 2000 & \quad & 0.037 & 0.556 & 0.966 & \quad & 0.031 & 0.576 & 0.987 & \quad & 0.036 & 0.644 & 0.989 \\\hline
			$t_3$ & \quad & 0.3 & \quad & 1000 & \quad & 0.076 & 0.189 & 0.652 & \quad & 0.074 & 0.141 & 0.589 & \quad & 0.071 & 0.159 & 0.623 \\
			\quad & \quad & \quad & \quad & 2000 & \quad & 0.058 & 0.345 & 0.851 & \quad & 0.045 & 0.334 & 0.883 & \quad & 0.051 & 0.306 & 0.882 \\
			\quad & \quad & 0.5 & \quad & 1000 & \quad & 0.058 & 0.274 & 0.793 & \quad & 0.048 & 0.243 & 0.798 & \quad & 0.059 & 0.277 & 0.794 \\
			\quad & \quad & \quad & \quad & 2000 & \quad & 0.037 & 0.495 & 0.962 & \quad & 0.043 & 0.499 & 0.978 & \quad & 0.038 & 0.499 & 0.978 \\
			\quad & \quad & 0.7 & \quad & 1000 & \quad & 0.084 & 0.315 & 0.683 & \quad & 0.079 & 0.352 & 0.708 & \quad & 0.078 & 0.324 & 0.689 \\
			\quad & \quad & \quad & \quad & 2000 & \quad & 0.048 & 0.385 & 0.856 & \quad & 0.038 & 0.401 & 0.887 & \quad & 0.046 & 0.397 & 0.870 \\\hline
		\end{tabular}
		\endgroup
	}\label{table6}
\end{table}

\section{Real Examples}\label{section9}

 In this section, we analyze three subsets of daily HKD/USD exchange rate data,  denoted as $\{y_{1,t}\}^{400}_{t=1}$, $\{y_{2,t}\}^{400}_{t=1}$ and $\{y_{3,t}\}^{400}_{t=1}$. These represent the log-returns ($\times$100) for samples from three successive periods: February 6, 2003, to March 21, 2005; March 22, 2005, to October 3, 2006; and October 4, 2006, to April 22, 2008, respectively. First, we apply the two-step SQE method to fit an AR(6) model to each time series:
\begin{align}\label{eq_7_1}
	y_{i,t}=\mu_{i}+\sum_{j=1}^{6}{\phi_{ij}y_{i,t-j}}+\varepsilon_{i,t},
\end{align}
 with the weights $w_t$ and $\tilde{w}_{lt}$   in (\ref{weigth_simulation}) and (\ref{weight2_simulation}), respectively. The results are summarized in Table \ref{table8}, with standard deviations shown in parentheses. Some parameters are relatively small compared to their standard deviations. We then use the
test $W_n$ in (\ref{eq_5_4}) to examine whether each parameter is significantly non-zero. Table \ref{table9} reports all the $p$-values of the test.
\begin{table}
	\centering
	\small
	\caption{\footnotesize The estimated AR parameters in model (\ref{eq_7_1})}
	\scalebox{0.85}{
		\begingroup
		\setlength{\tabcolsep}{2pt} 
		\renewcommand{\arraystretch}{1} 
		\begin{tabular}{cccccccc}
			\hline
			\mbox{time series} & $\mu_{i}$ & $\phi_{i1}$ & $\phi_{i2}$ & $\phi_{i3}$ & $\phi_{i4}$ & $\phi_{i5}$ & $\phi_{i6}$\\\hline
			$y_{1,t}$ & -0.019 & -0.366 & -0.155 & -0.143 & -0.099 & -0.114 & -0.013\\
			\quad & (0.019) & (0.113) & (0.089) & (0.080) & (0.097) & (0.068) & (0.070)\\
			$y_{2,t}$ & -0.017 & -0.482 & -0.340 & -0.246 & -0.148 & -0.051 & -0.024\\
			\quad & (0.016) & (0.116) & (0.083) & (0.071) & (0.087) & (0.083) & (0.047)\\
			$y_{3,t}$ & 0.019 & -0.288 & -0.126 & -0.086 & 0.005 & 0.074 & 0.085\\
			\quad & (0.024) & (0.102) & (0.099) & (0.083) & (0.089) & (0.082) & (0.057)\\\hline
		\end{tabular}
		\endgroup
	}\label{table8}
\end{table}
\begin{table}
	\centering
	\small
	\caption{\footnotesize The $p$-values of the test $W_n$ for every parameter of model (\ref{eq_7_1})}
	\scalebox{0.85}{
		\begingroup
		\setlength{\tabcolsep}{2pt} 
		\renewcommand{\arraystretch}{1} 
		\begin{tabular}{cccccccc}
			\hline
			\mbox{time series} & $\mu_{i}$ & $\phi_{i1}$ & $\phi_{i2}$ & $\phi_{i3}$ & $\phi_{i4}$ & $\phi_{i5}$ & $\phi_{i6}$\\\hline
			$y_{1,t}$ & 0.319 & 0.001 & 0.082 & 0.073 & 0.307 & 0.096 & 0.857\\
			$y_{2,t}$ & 0.314 & 0.000 & 0.000 & 0.001 & 0.088 & 0.540 & 0.617\\
			$y_{3,t}$ & 0.429 & 0.005 & 0.202 & 0.303 & 0.955 & 0.367 & 0.137\\\hline
		\end{tabular}
		\endgroup
	}\label{table9}
\end{table}
Thus, at a 5\% significance level, the simplified models for the three datasets are:
\begin{align}\label{eq_7_2}
	\begin{cases}
		y_{1,t} =-0.313y_{1,t-1}+\varepsilon_{1,t},\\
		\qquad \quad (0.074)\\
		y_{2,t}  =-0.430y_{2,t-1}-0.269y_{2,t-2}-0.170y_{2,t-3}+\varepsilon_{2,t},\\
		\qquad \quad (0.126) \qquad \quad (0.066) \qquad \quad (0.055)\\
		y_{3,t} =-0.217y_{3,t-1}+\varepsilon_{3,t}.\\
		\qquad \quad (0.092)
	\end{cases}
\end{align}
\noindent
The $p$-values of the tests $W_n$ are 0.000, (0.001, 0.000, 0.002) and 0.018 for model (\ref{eq_7_2}). So  the models in (\ref{eq_7_2}) are adequate for  three time series. Furthermore, the values of $\hat{\tau}_n$ in (\ref{eq_7_2}) are   $0.558$, $0.564$ and $0.488$. By using the test $w_n$ in (\ref{eq_5_5}), the $p$-values for the null hypothesis that the median is zero (i.e.,$\tau_{0}=0.5$) are $0.033$, $0.019$ and $0.636$, respectively. Thus, the distributions of the first two datasets are asymmetric, whereas the last is nearly symmetric. Finally, we check whether the residuals are heavy-tailed using  Hill's estimator  $\hat{\alpha}_i(k)$:
\begin{align}
	\hat{\alpha}_i(k)=\bigg[\frac{1}{k} \sum_{j=1}^{k}{\log\left(\frac{\hat{\varepsilon}_{(i,n-j)}}{\hat{\varepsilon}_{(i,n-k)}}\right)}\bigg]^{-1}, \ i=1,2,3,
\end{align}
where $\{\hat{\varepsilon}_{(i,t)}\}^{400}_{t=1}$ is the ascending order statistics of  $\{\vert\hat{\varepsilon}_{i,t}
\vert\}^{400}_{t=1}$.
\begin{figure}
	\centering
	\includegraphics[width=0.6\textwidth]{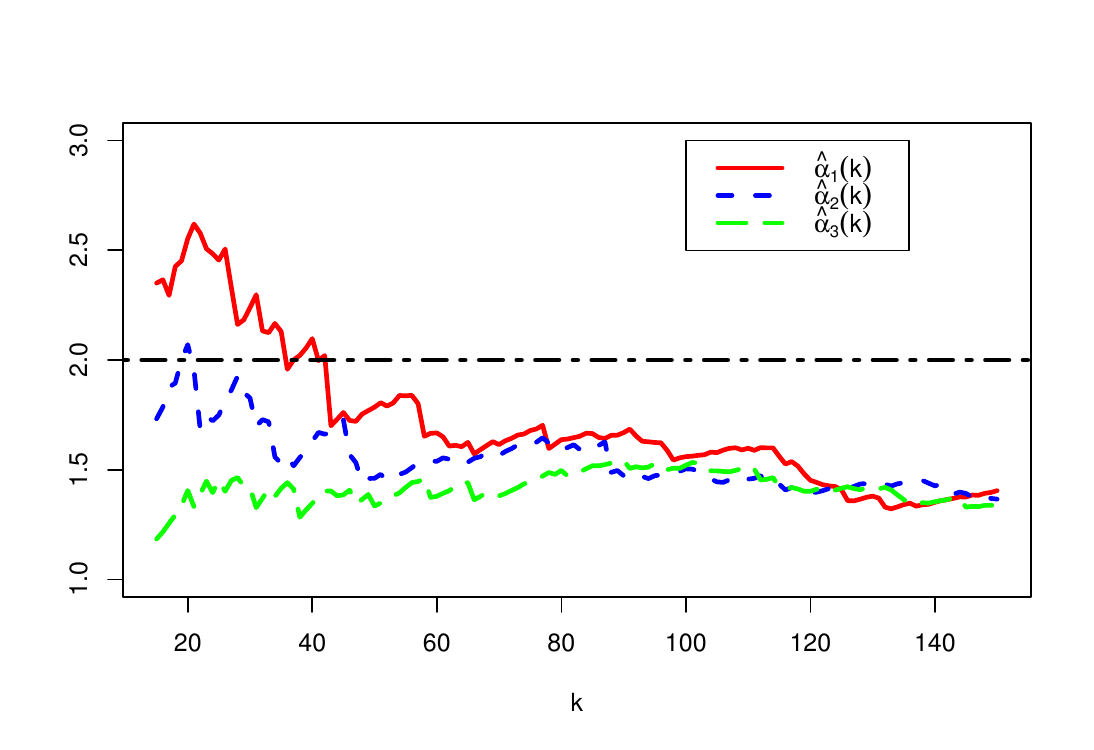}
	\caption{Hill's estimator $\{\hat{\alpha}_i(k)\}$ for the residuals.}
	\label{figure1}
\end{figure}
From Figure \ref{figure1}, we can see that  all the tail indices are obviously less than 2.  Therefore, it is reasonable to use the two-step SQE method to model the HKD/USD exchange rate series, particularly in cases with a non-zero median.

  \begin{appendix}
  	\setcounter{theorem}{0}
  	\renewcommand{\thetheorem}{A.\arabic{theorem}}
  	\setcounter{equation}{0}
  	\renewcommand{\theequation}{A.\arabic{equation}}
  	\setcounter{lemma}{0}
  	\renewcommand{\thelemma}{A.\arabic{lemma}}
  	\section*{Proofs of Lemma \ref{lemma_3_1}, Theorem \ref{theorem_3_1} and Theorems \ref{theorem_4_1}-\ref{theorem_4_2}}
  		\begin{proof}[\textsc Proof of Lemma \ref{lemma_3_1}]
  		(i). Suppose that there exists a $\delta\in\mathbb{R}^{p+1}$ such that   $P(Z^{\top}_{t-1}\delta=\sigma_{t})=1$. Denote $\delta_{i}$ as the $i$-th element in the vector $\delta$. Recall that $\sigma_{t}=\sigma_t(t/n)$, and by the non-degeneracy of $\sigma_{t}(x)$ in Assumption \ref{assumption_2_4} (ii), there exists one smallest $j \in \{1,\cdots,p\}$ such that $\delta_{j+1}\neq 0$.  So, by the fact that $\sigma_{t}\geq \sigma_0>0$ in Assumption \ref{assumption_2_4} (i) and $P(Z^{\top}_{t-1}\delta=\sigma_{t})=1$, we obtain
  		\begin{align}\label{new_a_1}
  			P\bigg(\delta_{j+1}y_{t-j}+\delta_1+\sum_{k=j+1}^{p}{\delta_{k+1}y_{t-k}}\geq \sigma_0\bigg)=1.
  		\end{align}
  		Since $y_{t-j}=Z^{\top}_{t-j-1}\theta_0+\eta_{t-j}\sigma_{t-j}$, (\ref{new_a_1}) implies that
  		\begin{align}\label{new_a_2}
  			\begin{cases}
  				& P(\eta_{t-j}\geq g_1(\mathcal{F}_{t-j-1}) )=1,\,\,\mbox{ if }\delta_{j+1}>0,\\
  				& P(\eta_{t-j}\leq g_1(\mathcal{F}_{t-j-1}) )=1,\,\,\mbox{ if }\delta_{j+1}<0,\\
  			\end{cases}
  		\end{align}
  		where $g_1$ is a deterministic function.
  		Notice that $\eta_{t-j}$ is independent with $\mathcal{F}_{t-j-1}$.  Thus, using the fact that the density of $\eta_{t-j}$ is positive on the whole real line in Assumption \ref{assumption_2_5} (i) and by the law of total expectation, (\ref{new_a_2}) cannot hold and hence (i) in Lemma \ref{lemma_3_1} holds.
  		
  		(ii). We only need to notice that if $P(Z^{\top}_{t-1}\delta=0)=1$ for some vector $\delta\neq 0$, then for some $\delta_{j+1}\neq 0$, we have
  		\begin{align}\label{new_a_3}
  			P\bigg(\delta_{j+1}y_{t-j}+\delta_1+\sum_{k=j+1}^{p}{\delta_{k+1}y_{t-k}}=0\bigg)=1.
  		\end{align}
  		Similarly, this  contradicts  Assumption \ref{assumption_2_5} (i). \\
  	\end{proof}
  	\begin{proof}[\textsc Proof of Theorem \ref{theorem_3_1}]
  		Because $\varepsilon_{t}=\eta_{t} \sigma_{t}$ and $\eta_{t}$ is independent with $\sigma_{t}\in \mathcal{F}_{t-1}$, by the law of total expectation and Assumption \ref{assumption_2_7}, it follows that
  		\begin{align}\label{eq_a_1}
  		\frac{\partial E\tilde{L}_n(\theta;\tau)}{\partial \theta}=-\frac{1}{n} \sum_{t=p+1}^{n}{E\Big(w_t Z_{t-1}\Big[\tau-F\Big( \frac{(\theta-\theta_0)^{\top}Z_{t-1}}{\sigma_{t}} \Big)\Big]     \Big)}.
  		\end{align}
  		Moreover, note that a bounded and Lipschitz continuous function must be $\alpha$-Holder continuous for any $\alpha \in (0,1]$. Therefore, by Assumption \ref{assumption_2_4} (i), Assumption \ref{assumption_2_5} (i) and Assumption \ref{assumption_2_7}, for any $s_1 \in [0,1]$, we can derive that	
  	\begin{eqnarray}\label{eq_a_2}
  && \sup_{\tau \in \mathcal{T}, \theta \in \Theta}\bigg\vert  w_tZ_{t-1} \Big[\tau-F\Big( \frac{ (\theta-\theta_0)^{\top} Z_{t-1}}{\sigma_{t}} \Big)\Big]\nonumber\\
&&\qquad\qquad\qquad\qquad\qquad - w(s_1,Y^{\top}_{t-1}(s_1))Z_{t-1}(s_1)\Big[\tau-F\Big(\frac{(\theta-\theta_0)^{\top} Z_{t-1}(s_1)}{\sigma_{t}(s_1)} \Big) \Big] \bigg\vert\nonumber\\
& &\quad \quad\leq \vert w(s_1,Y^{\top}_{t-1}(s_1))Z_{t-1}(s_1) \vert\times \sup_{\theta\in \Theta}{\bigg\vert F\Big(\frac{(\theta-\theta_0)^{\top} Z_{t-1}}{\sigma_{t}} \Big)-F\Big(\frac{(\theta-\theta_0)^{\top} Z_{t-1}(s_1)}{\sigma_{t}(s_1)} \Big)  \bigg\vert}\nonumber\\
&& \qquad \qquad \qquad+ \vert w_t Z_{t-1}-w(s_1,Y^{\top}_{t-1}(s_1))Z_{t-1}(s_1)\vert\nonumber \\
&& \quad \quad \leq C_1 \times \bigg\{ \vert t/n-s_1\vert^{\alpha_0}+ \sum_{i=1}^{p}{\vert y_{t-i}-y_{t-i}(s_1) \vert^{\alpha_0}}+\vert \sigma_{t}-\sigma_t(s_1) \vert^{\alpha_0}   \bigg\},
\end{eqnarray}
where $C_1$ is a constant only relying on $\sup_{\theta\in \Theta}{\vert \theta-\theta_0\vert}$, $w(\cdot)$, $\sigma_0$, and $\sup_{z\in \mathbb{R}}{f(z)}$.
  		
  		Similarly, we can also prove that, for any $s_1,s_2 \in [0,1]$,
  		\begin{align}\label{eq_a_3}
  		& \sup_{\tau \in \mathcal{T},\theta\in \Theta} \bigg \{ \Big \vert w(s_1,Y^{\top}_{t-1}(s_1))Z_{t-1}(s_1)\Big[\tau-F\Big(\frac{(\theta-\theta_0)^{\top} Z_{t-1}(s_1)}{\sigma_{t}(s_1)} \Big) \Big]\nonumber \\
  		& \qquad \qquad \qquad \qquad -w(s_2,Y^{\top}_{t-1}(s_2))Z_{t-1}(s_2)\Big[\tau-F\Big(\frac{(\theta-\theta_0)^{\top} Z_{t-1}(s_2)}{\sigma_{t}(s_2)} \Big) \Big] \Big \vert\bigg \}\nonumber\\
  		& \qquad \leq C_1 \times \big\{\vert s_1-s_2\vert^{\alpha_0}+ \sum_{i=1}^{p}{\vert y_{t-i}(s_1)-y_{t-i}(s_2) \vert^{\alpha_0}}+\vert \sigma_{t}(s_1)-\sigma_t(s_2) \vert^{\alpha_0}   \big\}.
  		\end{align}
  		On the other hand, by Assumption \ref{assumption_2_1} and using the standard transformation in \citet{hamilton1994}, we can obtain
  		\begin{align}\label{eq_a_4}
  			y_t=\sum_{j=0}^{t-p-1}{\pi_{j0} \varepsilon_{t-j}\Big(\frac{t-j}{n}\Big) }+\sum_{j=0}^{t-p-1}{\pi_{j0} \mu_0}+\Pi^{\top}_{t-p} Y_{p},
  		\end{align}
  		where $t\geq p+1$ and $\Pi_k=O(c^k_0)$. By Assumptions \ref{assumption_2_2}--\ref{assumption_2_3}, it is obvious that
  		\begin{align}\label{eq_a_8_1}
  				\sup\limits_{\substack{x_1 \neq x_2, x_1, x_2 \in [0,1]}}{\{ \Vert \varepsilon_{t}(x_1)-\varepsilon_{t}(x_2) \Vert_{\alpha_0}/\vert x_1-x_2\vert \}}<\infty,\,\,\, \Vert \sup_{x \in [0,1]}{\vert \varepsilon_{t}(x) \vert}\Vert_{\alpha_0}<\infty.
  		\end{align}
  		Consequently, by the definition of \(y_t(x)\) in (\ref{eq_2_1}) and utilizing (\ref{eq_a_2}), (\ref{eq_a_4})--(\ref{eq_a_8_1}), we establish
  		\begin{align}\label{eq_a_5}
  		& \sup_{\tau \in \mathcal{T},\theta\in \Theta}  \bigg \vert   \frac{1}{n}\sum_{t=2p+1}^{n}{E\Big(w_tZ_{t-1} \Big[\tau-F\Big( \frac{(\theta-\theta_0)^{\top} Z_{t-1}}{\sigma_{t}} \Big)\Big]\Big)} \\
  		& \quad -\frac{1}{n} \sum_{t=2p+1}^{n}{E\Big(w(t/n,Y^{\top}_{t-1}(t/n))Z_{t-1}(t/n)\Big[\tau-F\Big(\frac{(\theta-\theta_0)^{\top} Z_{t-1}(t/n)}{\sigma_{t}(t/n)} \Big) \Big]\Big)} \bigg \vert  \nonumber\\
  		& \qquad \leq \frac{C_1}{n}\sum_{i=1}^{p}{\sum_{t=2p+1}^{n}{ \Vert y_{t-i}-y_{t-i}(t/n)  \Vert^{\alpha_0}_{\alpha_0} }}\nonumber\\
  		& \qquad \leq \frac{C_2}{n}\sum_{i=1}^{p}{\sum_{t=2p+1}^{n}{ \bigg\{  \sum_{j=0}^{t-i-p-1}{\tilde{c}^j_0 (\frac{i+j}{n})^{\alpha_0}}+\sum_{j=t-i-p}^{\infty}{\tilde{c}^j_0} \bigg\} }}=O(n^{-\alpha_0}),\nonumber
  		\end{align}
  		where $C_2$ is a constant and $\tilde{c}_0=c^{\alpha_0}_0$. Additionally, by (\ref{eq_a_3}), it is straightforward to show that
  		\begin{eqnarray}\label{eq_a_6}
  		&& \sup_{\tau \in \mathcal{T},\theta\in\Theta} \bigg  \vert   \frac{1}{n}\sum_{t=2p+1}^{n}{E\Big(w(t/n,Y^{\top}_{t-1}(t/n))Z_{t-1}(t/n) \Big[\tau-F\Big( \frac{(\theta-\theta_0)^{\top}Z_{t-1}(t/n)}{\sigma_{t}(t/n)} \Big)\Big]\Big)}\nonumber\\
  		&& \quad\quad\quad - \int_{0}^{1}{E\Big\{w(s,Y^{\top}_{t-1}(s))Z_{t-1}(s)\Big[\tau-F\Big(\frac{(\theta-\theta_0)^{\top}Z_{t-1}(s)}{\sigma_{t}(s)} \Big) \Big]   \Big\} ds} \bigg \vert \nonumber \\
  		&& \quad\quad=O(n^{-\alpha_0}).
  		\end{eqnarray}
  		Hence, Theorem \ref{theorem_3_1} (i) is established by combining (\ref{eq_a_1}) with (\ref{eq_a_5}) and (\ref{eq_a_6}).
  		
  		For (ii), by Assumption \ref{assumption_2_4} (i), Assumption \ref{assumption_2_5} (i) and Assumption \ref{assumption_2_7} and utilizing the dominated convergence theorem, for the limit $g(x,\tau)$ in (i), we have
  		\begin{align}\label{eq_a_7}
  		\frac{\partial g(x,\tau)}{\partial \tau} & =-\int_{0}^{1}{E[w(s,Y^{\top}_{t-1}(s))Z_{t-1}(s)]ds},\nonumber \\
  		\frac{\partial g(x,\tau)}{\partial x}& =\int_{0}^{1}{E\Big[w(s,Y^{\top}_{t-1}(s))Z_{t-1}(s)Z^{\top}_{t-1}(s)f\Big(\frac{x^{\top}Z_{t-1}(s)}{\sigma_{t}(s)}\Big)\sigma^{-1}_{t}(s)  \Big]ds}.
  		\end{align}
  		Furthermore, the continuity of $f(\cdot)$ in Assumption \ref{assumption_2_5} (i) indicates that the vector-value function $g(x,\tau)$ is continuously differentiable. In addition, by Assumption \ref{assumption_2_5} (i) and following the same proof procedures for Lemma \ref{lemma_3_1} (ii), it is easy to show that for any $\theta'\neq 0$ and $s\in [0,1]$, $P(Z^{\top}_{t-1}(s)\theta'= 0)<1$.
        As a result, we must have $\partial g(x,\tau)/ \partial x$ is positive definite for any $x\in \mathbb{R}^{p+1}$. Meanwhile, notice that $g(0,\tau_0)=0$. So, the implicit function theorem derives that there exists a unique and continuously differentiable function $\delta_0(\tau)$ such that
  		\begin{align}\label{eq_a_8}
  			g(\delta_0(\tau),\tau)=0,
  		\end{align}
  		where $\tau \in (\tau_{10},\tau_{20})$ for some $\tau_{10},\,\tau_{20}$ with  $\tau_{10}<\tau_0<\tau_{20}$. Thus, the conclusion (ii) holds for some small open interval containing $\tau_{0}$.
  			
  		Next, we prove that the validity of $\tau$ in (\ref{eq_a_8}) can be extended to the whole interval $(0,1)$. We consider the following two different cases:
  		
  		{\bf Case (i)}: If there exists some $\tau'_0\neq \tau_0$ such that $g(0,\tau'_0)=0$, then it is clear  that $$\int_{0}^{1}{E[w(s,Y^{\top}_{t-1}(s))Z_{t-1}(s)]ds}=0,$$
  		which implies that $g(0,\tau)=0$ for any $\tau \in (0,1)$. Again by the implicit theorem, the uniqueness and the continuity {of $\delta_0(\tau)$ are obvious. Hence,  the conclusion (ii) holds for any $\tau \in (0,1)$.}
  		
  		{\bf Case (ii)}: Assume that $g(0,\tau) \neq 0$ for any $\tau \neq \tau_0$. Define the two boundaries:
  		\begin{align*}
  			& \tilde{\tau}_{l}=\inf\{\tau_1: g(\delta_0(\tau),\tau)=0,\,\forall \tau \in [\tau_1,\tau_0)\},\\
  			& \tilde{\tau}_{u}=\sup\{\tau_2: g(\delta_0(\tau),\tau)=0,\,\forall \tau \in (\tau_0,\tau_2]\}.
  		\end{align*}
  		By (\ref{eq_a_8}), it is obvious that $\tilde{\tau}_{l}<\tau_0<\tilde{\tau}_{u}$ and $g(\delta_0(\tau),\tau)=0$ for any $\tau \in (\tilde{\tau}_{l},\tilde{\tau}_{u})$. In this case, our target is to show that
  		\begin{align}\label{eq_a_9}
  			\tilde{\tau}_{l}=0\mbox{ and }\tilde{\tau}_{u}=1.
  		\end{align}
  		
  		{Due to the similarity in the proof procedure}, we only show that $\tilde{\tau}_{u}=1$.
  		First, we introduce a new function:
  		\begin{align*}
  	\tilde{g}(r,\phi, \tau)=-\int_{0}^{1}{E\bigg\{w(s,Y^{\top}_{t-1}(s))\phi^{\top} Z_{t-1}(s)\bigg[\tau-F\Big(\frac{r\phi^{\top}Z_{t-1}(s)}{\sigma_{t}(s)} \Big) \bigg]   \bigg\} ds},
  	\end{align*}
  		where $r\in \mathbb{R}$, $\phi \in \mathbb{S}^{p}$ and $\tau \in (0,1)$. Notice that $\tilde{g}(r,\phi,\tau)$ is  {a strictly} monotonic function with respect to $r$. By the dominated convergence theorem, and utilizing the fact that for any $\phi\in \mathbb{S}^{p}$ and $s\in [0,1]$, $P(\phi^{\top}Z_{t-1}(s)= 0)<1$, we can easily show that for any fixed $\phi$ and $\tau$,
  		\begin{align*}
  			\lim_{r \to \infty}{\tilde{g}(r,\phi,\tau)}>0 \mbox{ and }\lim_{r \to -\infty}{\tilde{g}(r,\phi,\tau)}<0.
  		\end{align*}
  		Thus, by  {the} mean value theorem, there exists a unique and continuous function $r(\phi,\tau)$ such that
  		\begin{align}\label{eq_a_10}
  			\tilde{g}(r(\phi,\tau),\phi,\tau)=0.
  		\end{align}
  		By the definition of $g(\cdot)$ and $\tilde{g}(\cdot)$ and (\ref{eq_a_10}), for any $\tau \neq \tau_0$ (i.e., $\delta_0(\tau) \neq 0$), we have
  		\begin{align}\label{eq_a_11}
  			g(\delta_0(\tau),\tau)=0  \Rightarrow \tilde{g}\Big(\vert \delta_0(\tau)\vert, \frac{\delta_0(\tau)}{\vert \delta_0(\tau)\vert},\tau\Big)=0 \mbox{ and }\vert \delta_0(\tau)\vert=r\Big(\frac{\delta_0(\tau)}{\vert \delta_0(\tau)\vert},\tau\Big).
  		\end{align}
  		
  		Now, if we assume that $\tilde{\tau}_{u}<1$, then for any sequence $\{\tau_m\}$ with $ \tau_m\in [(\tau_0+\tilde{\tau}_{u})/2,\tilde{\tau}_{u})$ and $\tau_m \to \tilde{\tau}_{u}$, it follows that
  		\begin{align}\label{eq_a_12}
  			\vert \delta_0(\tau_{m_1})-\delta_0(\tau_{m_2})\vert\leq \sup_{\tau \in [(\tau_0+\tilde{\tau}_{u})/2,\tilde{\tau}_{u})}{\Big \{\Big \vert \frac{\partial \delta_0(\tau)}{\partial \tau} \Big \vert\Big \}}\times \vert \tau_{m_1}-\tau_{m_2} \vert.
  		\end{align}
  		{In addition,} by the implicit function theorem, it is obvious that
  		\begin{align}\label{eq_a_13}
  			\frac{\partial \delta_0(\tau)}{\partial \tau}=-\Big \{\frac{\partial g(\delta_0(\tau),\tau)}{\partial x}\Big\}^{-1}\times \Big\{\frac{\partial g(\delta_0(\tau),\tau) }{\partial \tau}\Big\},
  		\end{align}
  		where $\tau \in (\tilde{\tau}_{l},\tilde{\tau}_{u})$. Because $\tilde{\tau}_u<1$, (\ref{eq_a_11}) implies that
  		\begin{align}\label{eq_a_14}
  			\sup_{\tau \in [(\tau_0+\tilde{\tau}_{u})/2,\tilde{\tau}_{u})}{\vert \delta_0(\tau)\vert } \leq \sup_{(\phi,\tau)\in \mathcal{T}_1}{\vert r(\phi, \tau) \vert}<\infty,
  		\end{align}
  		where $\mathcal{T}_1=\mathbb{S}^{p}\times [(\tau_0+\tilde{\tau}_{u})/2,\tilde{\tau}_{u}] $ is a compact subset in $\mathbb{S}^{p}\times (0,1)$.
  		
  		{Hence,} by (\ref{eq_a_7}) and (\ref{eq_a_13}), and using the result that $\partial g(x,\tau)/ \partial x$ is positive definite for any $x\in \mathbb{R}^{p+1}$, (\ref{eq_a_14}) implies that $\sup_{\tau \in [(\tau_0+\tilde{\tau}_{u})/2,\tilde{\tau}_{u})}{\Big \{\Big \vert \frac{\partial \delta_0(\tau)}{\partial \tau} \Big \vert\Big \}}<\infty$ and {thus}
  	   $\delta_0(\tau_{m})$ is a Cauchy sequence. {As a result,} we have shown that for any sequence $\tau_{m} \to \tilde{\tau}_u$, $\delta_0(\tau_{m})$ must converge to a common limit denoted by $\delta_0(\tilde{\tau}_u)$. Obviously, {from the continuity of $g(\cdot)$}, we have
  		\begin{align}\label{eq_a_19}
  			g(\delta_0(\tilde{\tau}_{u}),\tilde{\tau}_u)=0.
  		\end{align}
  		{However, by the implicit function theorem again,} this  contradicts the definition of $\tilde{\tau}_{u}$  as that in (\ref{eq_a_8}). This completes the  proof for (\ref{eq_a_9}).
  		
  		Finally, notice that $\delta_0(\tau_{0})=0$. The form of $\partial \delta_0(\tau_{0})/\partial \tau$ {can be}  derived directly from (\ref{eq_a_13}) and  (\ref{eq_a_7}).\\
  	\end{proof}
	
  \begin{proof}[\textsc Proof of Theorem \ref{theorem_4_1}]
  	By Lemma \ref{lemma_4_1}, it is easy to see that
  	\begin{align}\label{eq_a_38}
  		\sup_{\tau \in \mathcal{T}} \bigg \vert \frac{1}{n}\sum_{t=p+1}^{n}{r_t(\tau;\tilde{w}_l)}-\tilde{g}(\tau;\tilde{w}_l)-\frac{1}{\sqrt{n}}\tilde{S}_n(\tau;\tilde{w}_l)\bigg\vert=o_p(n^{-1/2}),
  	\end{align}
  	where $\tilde{S}_n(\tau;\tilde{w}_l)$ is defined as
  	\begin{align*}
  		\tilde{S}_n(\tau;\tilde{w}_l)=\{ \tilde{T}_n(\tau;\tilde{w}_l)-E\tilde{T}_n(\tau;\tilde{w}_l) \}-b^{\top}(\tau;\tilde{w}_l)\sqrt{n}(\hat{\theta}_n(\tau)-\theta_0(\tau)),
  	\end{align*}
  	where $\tilde{T}_n(\tau;\tilde{w}_l)=n^{-1/2}\sum_{t=p+1}^{n}{r_{t0}(\tau;\tilde{w}_l)}$. Following  proof procedures  similar to  those in Lemma \ref{theorem_3_2}, and by Theorem \ref{theorem_3_3}, one can easily show that
  	\begin{align}\label{eq_a_39}
  		\tilde{S}_n(\tau;\tilde{w}_l) \rightsquigarrow \tilde{S}_{0}(\tau;\tilde{w}_l):=\tilde{T}_0(\tau;\tilde{w}_l)-b^{\top}(\tau;\tilde{w}_l)\Sigma^{-1}(\tau) T_0(\tau),
  	\end{align}
  	where $\tilde{T}_0(\tau;\tilde{w}_l)$ is a Gaussian limit of $\tilde{T}_n(\tau;\tilde{w}_l)-E\tilde{T}_n(\tau;\tilde{w}_l)$.
  	
  	Because $\tilde{g}(\tau;\tilde{w}_{l})$ is continuous on compact set $\mathcal{T}$, by (\ref{eq_a_38}) and (\ref{eq_a_39}), we have
  	\begin{align}\label{eq_a_40}
  		\sup_{\tau \in \mathcal{T}}\bigg\vert \sum_{l=1}^{L}{ \Big\{n^{-1}\sum_{t=p+1}^{n}{r_t(\tau;\tilde{w}_l)}\Big\}^2}-\sum_{l=1}^{L}{\tilde{g}^2(\tau;\tilde{w}_l)} \bigg\vert=O_p(n^{-1/2}).
  	\end{align}
  	By Assumption 6.2 (i), for any $\epsilon>0$, there exists a $\delta>0$ such that
  	\begin{align}\label{eq_a_41}
  		\min_{\vert \tau -\tau_{0}\vert\geq \epsilon ,\tau\in \mathcal{T}}{\sum_{l=1}^{L}{\tilde{g}^2(\tau;\tilde{w}_l)}}>\delta.
  	\end{align}
  	Notice that $\sum_{l=1}^{L}{\tilde{g}^2(\tau_0;\tilde{w}_l)}=0$. So (\ref{eq_a_40})-(\ref{eq_a_41}) further indicate that
  	\begin{align}\label{eq_a_42}
  		\hat{\tau}_n \overset{p}{\longrightarrow} \tau_{0}.
  	\end{align}
  	
  	By Assumption 6.2 (ii), the set $\mathcal{A}=\{l: \partial \tilde{g}(\tau_{0};\tilde{w}_l)/\partial\tau \neq  0\}$ is not empty.
  	Then, by the continuity, there exists a constant $\eta>0$ and $\tilde{c}_1>0$ such that
  	\begin{align}\label{eq_a_43}
  		{\inf_{\vert \tau-\tau_{0}\vert\leq \eta,l\in \mathcal{A}}{\Big\vert \frac{\partial \tilde{g}(\tau;\tilde{w}_l)}{\partial\tau}\Big\vert }}\geq \tilde{c}_1.
  	\end{align}
  	On the other hand, (\ref{eq_a_38})-(\ref{eq_a_39}) imply that for any $\epsilon>0$, there exists {a $M_1>0$} such that for any $l$, it follows that
  	\begin{align}\label{eq_a_44}
  		P\Big (\sup_{\tau \in \mathcal{T}, l\leq L} \Big \vert \frac{1}{\sqrt{n}}\sum_{t=p+1}^{n}{r_t(\tau;\tilde{w}_l)}-\sqrt{n}\tilde{g}(\tau;\tilde{w}_l)\Big\vert>M_1\Big)<\epsilon/2.
  	\end{align}
  	Then by (\ref{eq_a_42})-(\ref{eq_a_44}), for any $M>2M_1/\tilde{c}_1$, using the definition of $\hat{\tau}_n$, we can derive that
  	\begin{align}
  		P(\sqrt{n}\vert\hat{\tau}_n-\tau_{0} \vert>M, \vert\hat{\tau}_n-\tau_{0} \vert\leq \eta )\leq P\Big(\sum_{l\in \mathcal{A}}{[\tilde{c}_1M-M_1]^2}\leq LM^2_1\Big)+\epsilon/2,\label{eq_a_45}
  	\end{align}
  	Therefore, by (\ref{eq_a_42}) and (\ref{eq_a_45}), for any $\epsilon>0$, we can find a large $M$ such that $P(\sqrt{n}\vert\hat{\tau}_n-\tau_{0} \vert>M)<\epsilon$. In other words, we have shown that
  	\begin{align}\label{eq_a_46}
  		\sqrt{n}(\hat{\tau}_n-\tau_{0})=O_p(1).
  	\end{align}
  	Furthermore, let $v=\sqrt{n}(\tau-\tau_{0})$. By (\ref{eq_a_38})-(\ref{eq_a_39}), {for} any compact set $\tilde{\mathcal{T}}_1\subset \mathbb{R}$, {we can easily show that},
  	\begin{align}\label{eq_a_47}
  		\sum_{l=1}^{L}{\Big[\frac{1}{\sqrt{n}}\sum_{t=p+1}^{n}{r_t\Big(\tau_0+\frac{v}{\sqrt{n}};\tilde{w}_l\Big)}\Big]^2}\rightsquigarrow \sum_{l=1}^{L}{\Big[\frac{\partial \tilde{g}(\tau_{0};\tilde{w}_l)}{\partial \tau}v+\tilde{S}_0(\tau_{0};\tilde{w}_l)  \Big]^2}.
  	\end{align}
  	Then from (\ref{eq_a_46})-(\ref{eq_a_47}) and Theorem 3.2.2 in \citet{van1996}, we have shown that
  	\begin{align}\label{eq_a_47_a}
  		\sqrt{n}(\hat{\tau}_n-\tau_{0})=\sum_{l\in \mathcal{A}}{a_l \tilde{S}_n(\tau_{0};\tilde{w}_l)}+o_p(1).
  	\end{align}
  	By the definition of $\tilde{S}_n(\tau_{0};\tilde{w}_l)$, we get the conclusion.\\
  \end{proof}

  \begin{proof}[\textsc Proof of Theorem \ref{theorem_4_2}]
  	For any $\epsilon,\delta>0$, it follows that
  	\begin{align*}
  		& P(\vert \sqrt{n}(\hat{\theta}_n(\hat{\tau}_n)-\theta_0(\hat{\tau}_n))-\sqrt{n}(\hat{\theta}_n({\tau}_0)-\theta_0({\tau}_0))\vert>\epsilon)\\
  		& \quad \leq P(\vert \hat{\tau}_n-\tau_{0} \vert\geq \delta)+P(\sup_{\vert \tau_1-\tau_2 \vert <\delta}\vert \sqrt{n}(\hat{\theta}_n({\tau}_1)-\theta_0({\tau}_1))-\sqrt{n}(\hat{\theta}_n({\tau}_2)-\theta_0({\tau}_2))\vert>\epsilon).\nonumber
  	\end{align*}
  	Furthermore, by the consistency of $\hat{\tau}_n$ in Theorem \ref{theorem_4_1} and using the asymptotic tightness of $\sqrt{n}(\hat{\theta}_n(\tau)-\theta(\tau))$ in Theorem \ref{theorem_3_3}, we can obtain
  	\begin{align}\label{eq_a_48}
  		\sqrt{n}(\hat{\theta}_n(\hat{\tau}_n)-\theta_0(\hat{\tau}_n))=\sqrt{n}(\hat{\theta}_n({\tau}_0)-\theta_0({\tau}_0))+o_p(1).
  	\end{align}
  	Since $\hat{\theta}_n=\hat{\theta}_n(\hat{\tau}_n)$ and {$\theta_0=\theta_0(\tau_{0})$}, (\ref{eq_a_48}) and Theorem \ref{theorem_4_1} further indicate that
  	\begin{align}\label{eq_a_49}
  		\sqrt{n}(\hat{\theta}_n-\theta_0) & =\sqrt{n}(\hat{\theta}_n(\tau_{0})-\theta_0(\tau_{0}))+\sqrt{n}(\theta_{0}(\hat{\tau}_n)-\theta_{0}(\tau_{0}))+o_p(1)\nonumber\\
  		& =\sqrt{n}(\hat{\theta}_n(\tau_{0})-\theta_0(\tau_{0}))+\frac{\partial \theta_0(\tau_{0})}{\partial \tau}\sqrt{n}(\hat{\tau}_n-\tau_{0})+o_p(1)\nonumber\\
  		&
  		=\sqrt{n}(\hat{\theta}_n(\tau_{0})-\theta_0(\tau_{0}))+\frac{\partial \delta_0(\tau_{0})}{\partial \tau}\sqrt{n}(\hat{\tau}_n-\tau_{0})+o_p(1).
  	\end{align}
  	By the martingale central limit theory, the conclusion holds from (\ref{eq_a_49}) and (\ref{eq_a_47_a}) and (B.30) in the supplementary material.\\
  \end{proof}

  \end{appendix}

%
%
%
%
%
	\begin{acks}[Acknowledgments]
		The authors would like to thank two referees, \textcolor{red}{the} Associate
		Editor and the Editor Professor Wang Lan for their constructive comments and Mr
Paul Cooper for his  language editing.
		Rui She's research was supported by  the National Natural Science Foundation of China (Nos. 12201510). Linlin Dai's research was supported by the National Natural Science Foundation of China (Nos. 12001441).
		Shiqing Ling's research was partially supported by   Hong Kong Research Grants Commission Grants (16303118, 16301620, 16300621,  16500522 and SRFS2223-6S02).
	\end{acks}
	
%
	
	\begin{supplement}
		\stitle{Supplementary material to ``A Two-step Estimating Approach for Heavy-tailed AR Models with Non-zero Median GARCH-type Noises"}
		\sdescription{This supplementary material contains   additional simulation results, technical lemmas, the proofs for Sections \ref{section5} and \ref{section7}, and  some further results for the $AR(\infty)$ model.}
	\end{supplement}
	
	
	

\end{document}


\begin{frontmatter}
		\title{Supplementary Material to ``A Two-step Estimating Approach for Heavy-tailed AR Models with Non-zero Median GARCH-type Noises"}
		
		\begin{aug}
			\author[A]{\fnms{Rui}~\snm{She}\ead[label=e1]{rshe@swufe.edu.cn}},
			\author[A]{\fnms{Linlin}~\snm{Dai}\ead[label=e2]{ldaiab@swufe.edu.cn}}
			\and
			\author[B]{\fnms{Shiqing}~\snm{Ling}\ead[label=e3]{maling@ust.hk}}
			\address[A]{School of Statistics,
				The Southwestern University of Finance and Economics\printead[presep={,\ }]{e1,e2}}
			
			\address[B]{Department of Mathematics, The Hong Kong University of Science and Technology\printead[presep={,\ }]{e3}}
		\end{aug}

	\end{frontmatter}

		\setcounter{theorem}{0}
		\renewcommand{\thetheorem}{B.\arabic{theorem}}
		\setcounter{equation}{0}
		\renewcommand{\theequation}{B.\arabic{equation}}
		\setcounter{lemma}{0}
		\renewcommand{\thelemma}{B.\arabic{lemma}}
		\section{Proofs of Lemma 5.1 and Theorem 5.1}
{Recall that   $$T_n(\tau)=\sum_{t=p+1}^{n}{\xi_t(\tau)}, \ \ \, \alpha_n(u,\tau)=\sum_{t=p+1}^{n}{\Big\{ \zeta_t(u,\tau)-E[\zeta_t(u,\tau)\vert \mathcal{F}_{t-1}] \Big\}},$$ where
\begin{align*}
	\xi_t(\tau) & =\frac{1}{\sqrt{n}}w_t Z_{t-1} \psi_{\tau}(\varepsilon_{t}-\delta^{\top}_0(\tau)Z_{t-1}),\\
	\zeta_t(u,\tau) & =w_t \int_{0}^{\frac{u^{\top}}{\sqrt{n}}Z_{t-1}}{[I(\varepsilon_{t}-\delta^{\top}_0(\tau)Z_{t-1}\leq s)-I(\varepsilon_{t}-\delta^{\top}_0(\tau)Z_{t-1}\leq 0)]ds}.
\end{align*}	
}	

To prove Lemma 5.1, we {firstly} need  to show the tightness of $\{T_n(\tau)-ET_n(\tau)\}$, which is given as follows:
		\begin{lemma}\label{lemma_b_3}
			Suppose that Assumptions 2.1--2.4 (i), Assumption 2.5, Assumption 3.1, and Assumption 5.1 hold. Then for any $\epsilon,\eta>0$, we can find some $\delta>0$ such that
			\begin{align*}
				{\limsup_{n \to \infty}{P(\sup_{\vert \tau_{1}-\tau_{2} \vert< \delta}{\vert M_n(\tau_{1})-M_n(\tau_{2}) \vert>\epsilon})} } & <\eta,\\
				{\limsup_{n \to \infty}{P(\sup_{\vert \tau_{1}-\tau_{2} \vert< \delta}{\vert N_n(\tau_{1})-N_n(\tau_{2}) \vert>\epsilon})} } & <\eta,
			\end{align*}
			where $\tau_{1}, \tau_{2} \in \mathcal{T}$ and
		\begin{align}
			M_n(\tau)& =\sum_{t=p+1}^{n}{\{ \xi_t(\tau)-E[\xi_{t}(\tau) \vert \mathcal{F}_{t-1}] \}},\label{eq_b_13} \\
			N_n(\tau) &=\sum_{t=p+1}^{n}{\{ E[\xi_{t}(\tau) \vert \mathcal{F}_{t-1}]-E[\xi_{t}(\tau)]  \}}.\label{eq_b_14}
		\end{align}
		\end{lemma}
		
		\begin{proof}[\textsc Proof of Lemma 5.1]
			{Note that $T_n(\tau)-ET_n(\tau)=M_n(\tau)+N_n(\tau)$, where $M_n(\tau)$ and $N_n(\tau)$ are defined in (\ref{eq_b_13})-(\ref{eq_b_14})}. By Lemma \ref{lemma_b_3}, for any $\epsilon,\eta>0$, there exists some $\delta>0$ such that
			$$\limsup_{n \to \infty}{P(\sup_{\vert \tau_{1}-\tau_{2} \vert<\delta}{\vert T_n(\tau_{1})-ET_n(\tau_{1})-(T_n(\tau_{2})-ET_n(\tau_{2})) \vert>\epsilon})}<\eta.$$
			Therefore, the asymptotic tightness holds.
			
			To establish weak convergence of the centered process, it suffices to show the convergence of finite-dimensional distributions, namely,
			\begin{align}\label{eq_a_16}
				(T_n(\tau_{i})-ET_n(\tau_{i}), 1\leq i \leq k) \overset{d}{\longrightarrow} (T_0(\tau_{i}),1\leq i \leq k),
			\end{align}
			{where $T_0(\tau)$ is a centered Gaussian process defined in Lemma 5.1.}
			Without loss of generality, we only consider the case $k=2$. Let $\beta_i \in \mathbb{R}^{p+1},\,\tau_{i} \in \mathcal{T}$ and define
			\begin{align}\label{eq_a_17}
				S_n=\sum_{t=p+1}^{n}{\{ \beta^{\top}_1 \xi_{t}(\tau_{1})+\beta^{\top}_2 \xi_{t}(\tau_{2})- E[\beta^{\top}_1 \xi_{t}(\tau_{1})+\beta^{\top}_2 \xi_{t}(\tau_{2})] \}}.
			\end{align}
			Notice that for any $X\in \mathcal{F}_t$ with $E\vert X \vert<\infty$, we have $X-EX=\sum_{l=0}^{\infty}{\mathcal{P}_{t-l}(X)}$ with $\mathcal{P}_{t-l}(\cdot)=E(\cdot \vert \mathcal{F}_{t-l})-E(\cdot \vert \mathcal{F}_{t-l-1})$ being the projection on the space $\mathcal{F}_{t-l}$.
			Thus, we further consider the following two truncated version of $S_n$:
			\begin{align}
				S_{n,L} & =\sum_{t=p+1}^{n}{\sum_{l=0}^{L-1}{ \mathcal{P}_{t-l}[\beta^{\top}_1 \xi_{t}(\tau_{1})+\beta^{\top}_2 \xi_{t}(\tau_{2})] }},\label{eq_a_18}\\
				\tilde{S}_{n,L} & =\sum_{t=p+1}^{n}{\sum_{l=0}^{L-1}{ \mathcal{P}_{t-l}[\beta^{\top}_1 \tilde{\xi}_{t}(\tau_{1},t/n)+\beta^{\top}_2 \tilde{\xi}_{t}(\tau_{2},t/n)] }},\label{eq_a_19}
			\end{align}
			where $L\in \mathbb{N}$ and $\tilde{\xi}_{t}(\tau,s)=n^{-1/2} \xi_t(\tau,s)$ with $\xi_{t}(\tau, s)$ given in Lemma 5.1.
			
			{Following} similar proof procedures as those in (\ref{eq_b_22})--(\ref{eq_b_25}), it is easy to derive that
			\begin{align}\label{eq_a_20}
				\sqrt{n}\times \max_{t}\{\Vert \mathcal{P}_{t-l}[\beta^{\top}_i \xi_{t}(\tau_{i})] \Vert+\Vert \mathcal{P}_{t-l}[\beta^{\top}_i \tilde{\xi}_{t}(\tau_{i},t/n)] \Vert\}=O\Big (\sqrt{l\tilde{\kappa}^{l}}\Big),
			\end{align}
			where $\tilde{\kappa}=(c_0\vee \kappa_0)^{\alpha_0}$. {Then, we have}
			\begin{align}\label{eq_a_21}
				\Vert S_n-S_{n,L} \Vert & \leq \sum_{l=L}^{\infty}{\Big \Vert \sum_{t=p+1}^{n}{\mathcal{P}_{t-l}[\beta^{\top}_1 \xi_t(\tau_1)]  } \Big \Vert}+\sum_{l=L}^{\infty}{\Big \Vert \sum_{t=p+1}^{n}{\mathcal{P}_{t-l}[\beta^{\top}_2 \xi_t(\tau_2)]  } \Big \Vert}\nonumber\\
				& \leq \sum_{l=L}^{\infty}{\sqrt{n}  \max_{t}\{\Vert\mathcal{P}_{t-l}[\beta^{\top}_1 \xi_t(\tau_1)] \Vert\}}+\sum_{l=L}^{\infty}{\sqrt{n}  \max_{t}\{\Vert\mathcal{P}_{t-l}[\beta^{\top}_2 \xi_t(\tau_2)] \Vert\}}\nonumber\\
				& \leq C_1\times \sum_{l=L}^{\infty}{ \Big (\sqrt{l\tilde{\kappa}^{l}} \Big ) } \to 0, \mbox{ as }L \to \infty,
			\end{align}
			where $C_1$ is a constant independent of $L$ and $n$.
			
			Now, we proceed to figure out the difference between $S_{n,L}$ and $\tilde{S}_{n,L}$. Clearly, we have
			\begin{align}\label{eq_a_22}
				& \Vert \mathcal{P}_{t-l}[\beta^{\top}_i \xi_{t}(\tau_{i})]-\mathcal{P}_{t-l}[\beta^{\top}_i \tilde{\xi}_{t}(\tau_{i},t/n)]\Vert\\
				& \quad \leq \big( \Vert \mathcal{P}_{t-l}[\beta^{\top}_i \xi_{t}(\tau_{i})] \Vert+\Vert \mathcal{P}_{t-l}[\beta^{\top}_i \tilde{\xi}_{t}(\tau_{i},t/n)] \Vert \big)\wedge \big(\Vert \beta^{\top}_i \xi_{t}(\tau_{i})-\beta^{\top}_i \tilde{\xi}_{t}(\tau_{i},t/n)\Vert \big).\nonumber
			\end{align}
			Notice that $\varepsilon_{t}=\varepsilon_{t}(t/n)$, then similar to (A.5), it is straightforward to show that
			\begin{align}\label{eq_a_23}
				\Vert \beta^{\top}_i \xi_{t}(\tau_{i})-\beta^{\top}_i \tilde{\xi}_{t}(\tau_{i},t/n)\Vert^2 \leq C_2/n \times \sum_{j=1}^{p}{\Vert y_{t-j}-y_{t-j}(t/n) \Vert^{\alpha_0}_{\alpha_0}}.
			\end{align}
			{Thus, combining} (\ref{eq_a_18})-(\ref{eq_a_20}) and (\ref{eq_a_22})-(\ref{eq_a_23}) {yields that}
			\begin{align}\label{eq_a_24}
				& \Vert S_{n,L}-\tilde{S}_{n,L} \Vert\\
				& \qquad \leq \sum_{l=0}^{L-1}{\bigg\{ \Big (\sqrt{l\tilde{\kappa}^{l}} \Big )\wedge\sqrt{n^{-1}\sum_{t=p+1}^{n}{\sum_{j=1}^{p}{ \Vert y_{t-j}-y_{t-j}(t/n) \Vert^{\alpha_0}_{\alpha_0}     }   } } \bigg\}}.\nonumber
			\end{align}
			According to (A.9) and $\alpha_0>0$, (\ref{eq_a_24}) indicates that for any $\epsilon>0$,
			\begin{align}\label{eq_a_25}
				\lim_{L \to \infty}{\limsup_{n \to \infty}{ P( \vert S_{n,L}-\tilde{S}_{n,L} \vert>\epsilon)}}=0.
			\end{align}
			So, (\ref{eq_a_21}) and (\ref{eq_a_25}) indicates that
			\begin{align}\label{eq_a_25_new}
			\lim_{L \to \infty}{\limsup_{n \to \infty}{ P( \vert S_{n}-\tilde{S}_{n,L} \vert>\epsilon)}}=0.
			\end{align}
			\qquad Now we focus on the asymptotic property of $\tilde{S}_{n,L}$ for every fixed $L$. For simplicity, we define {a more homogeneous} version of $\tilde{S}_{n,L}$ as:
			\begin{align}\label{eq_a_26}
				\hat{S}_{n,L} =\sum_{t=p+1}^{n}{\sum_{l=0}^{L-1}{ \mathcal{P}_{t}[\beta^{\top}_1 \tilde{\xi}_{t+l}(\tau_{1},(t+l)/n)+\beta^{\top}_2 \tilde{\xi}_{t+l}(\tau_{2},(t+l)/n)] }},
			\end{align}
			where $\tilde{\xi}_t(\tau,s):=\tilde{\xi}_t(\tau,1)$ for $s>1$. By Assumption 3.1, for a fixed $L$, it is obvious that
			\begin{align}\label{eq_a_27}
				\tilde{S}_{n,L}-\hat{S}_{n,L}=O_p(n^{-1/2}).
			\end{align}
			Notice that the summand $$D_{t,L}=\sum_{l=0}^{L-1}{ \mathcal{P}_{t}[\beta^{\top}_1 \tilde{\xi}_{t+l}(\tau_{1},(t+l)/n)+\beta^{\top}_2 \tilde{\xi}_{t+l}(\tau_{2},(t+l)/n)] }$$ is a m.d.s. and uniformly bounded by $C_3L/\sqrt{n}$ with some constant $C_3$. Then for any $\epsilon>0$,
			\begin{align}\label{eq_a_28}
				\sum_{t=p+1}^{n}{E[D^2_{t,L} I(\vert D_{t,L} \vert>\epsilon)]} \longrightarrow 0, \mbox{ as }n \to \infty.
			\end{align}
			Furthermore, {denote} $$D_{t,L}(s)=\sum_{l=0}^{L-1}{ \mathcal{P}_{t}[\beta^{\top}_1 \tilde{\xi}_{t+l}(\tau_{1},s)+\beta^{\top}_2 \tilde{\xi}_{t+l}(\tau_{2},s)] }.$$  Then, for any $s_1,s_2\in [0,1]$, {it is straightforward to show that}
			\begin{align}\label{eq_b_17_new}
				& E\{\vert E[D^2_{t,L}(s_1)\vert \mathcal{F}_{t-1}]-E[D^2_{t,L}(s_2)\vert \mathcal{F}_{t-1}] \vert\}\\
				& \quad\leq C_4n^{-1}\sum_{l=0}^{L-1}\Big \{ \big \Vert \beta^{\top}_1 [{\xi}_{t+l}(\tau_{1},s_1)-{\xi}_{t+l}(\tau_{1},s_2)] \big \Vert_1\nonumber\\
				& \qquad \qquad+\big \Vert \beta^{\top}_2 [{\xi}_{t+l}(\tau_{2},s_1)-{\xi}_{t+l}(\tau_{2},s_2)] \big\Vert_1\Big\}.\nonumber
			\end{align}
			Obviously, by Assumption 2.4 (i), Assumption 2.5 (i) and Assumption 3.1, it is not hard to obtain that
			\begin{align}\label{eq_b_18_new}
				& \big \Vert \beta^{\top}_i [{\xi}_{t+l}(\tau_{i},s_1)-{\xi}_{t+l}(\tau_{i},s_2)] \big \Vert_1 \leq C_5 \Big \{ \sum_{j=1}^{p}{\Vert y_{t+l-j}(s_1)-y_{t+l-j}(s_2) \Vert^{\alpha_0}_{\alpha_0}}\nonumber\\
				& \qquad \qquad \Vert I(\varepsilon_{t+l}(s_1)\leq \delta^{\top}_0(\tau_{i})Z_{t+l-1}(s_2))-I(\varepsilon_{t+l}(s_2)\leq \delta^{\top}_0(\tau_{i})Z_{t+l-1}(s_2)) \Vert_1\Big\}
			\end{align}
			For any $\delta>0$, we have
			\begin{align}\label{eq_b_19_new}
				& \Vert I(\varepsilon_{t+l}(s_1)\leq \delta^{\top}_0(\tau_{i})Z_{t+l-1}(s_2))-I(\varepsilon_{t+l}(s_2)\leq \delta^{\top}_0(\tau_{i})Z_{t+l-1}(s_2)) \Vert_1 \nonumber\\
				& \quad \leq \delta^{-\alpha_0}\Vert \varepsilon_{t+l}(s_1)-\varepsilon_{t+l}(s_2)  \Vert^{\alpha_0}_{\alpha_0}+P(\vert \varepsilon_{t+l}(s_1)-\delta^{\top}_0(\tau_{i})Z_{t+l-1}(s_2) \vert\leq \delta)\nonumber\\
				& \quad \leq C_6\times [\delta^{-\alpha_0}\vert s_1-s_2\vert^{\alpha_0}+\delta].
			\end{align}
			So, if we set $\delta=\vert s_1-s_2\vert^{\frac{\alpha_0}{\alpha_0+1}}$, then (\ref{eq_b_17_new})--(\ref{eq_b_19_new}) imply that
			\begin{align}\label{eq_b_20_new}
				& E\{\vert E[D^2_{t,L}(s_1)\vert \mathcal{F}_{t-1}]-E[D^2_{t,L}(s_2)\vert \mathcal{F}_{t-1}] \vert\}\nonumber\\
				& \leq  C_7n^{-1} \Big \{ \sum_{l=0}^{L-1}\sum_{j=1}^{p}{\Vert y_{t+l-j}(s_1)-y_{t+l-j}(s_2) \Vert^{\alpha_0}_{\alpha_0}}+L\vert s_1-s_2\vert^{\frac{\alpha_0}{\alpha_0+1}}\Big\}.
			\end{align}
			Similarly, we can also derive that
			\begin{align}\label{eq_b_21_new}
			& E\{\vert E[D^2_{t,L}\vert \mathcal{F}_{t-1}]-E[D^2_{t,L}(s_1)\vert \mathcal{F}_{t-1}] \vert\}\nonumber\\
			& \leq  C_7n^{-1} \Big \{ \sum_{l=0}^{L-1}\sum_{j=1}^{p}{\Vert y_{t+l-j}(s_1)-y_{t+l-j}(\frac{t+l}{n}) \Vert^{\alpha_0}_{\alpha_0}}+L\vert s_1-\frac{t+l}{n}\vert^{\frac{\alpha_0}{\alpha_0+1}}\Big\}.
			\end{align}
			
		{Then, combining (\ref{eq_b_20_new})-(\ref{eq_b_21_new}) and using} a similar proof procedure for Lemma S.2 in \citet{she2022}, it is straightforward to show that
			\begin{align}\label{eq_a_29}
				\sum_{t=p+1}^{n}{E[D^2_{t,L} \vert \mathcal{F}_{t-1}]} \overset{p}{\longrightarrow} \int_{0}^{1}{E\big\{ \sum_{l=0}^{L-1}{ \mathcal{P}_{t}[\beta^{\top}_1 {\xi}_{t+l}(\tau_{1},s)+\beta^{\top}_2 {\xi}_{t+l}(\tau_{2},s)] } \big\}^2 ds}.
			\end{align}
			{Hence,} (\ref{eq_a_26}), (\ref{eq_a_28}) and (\ref{eq_a_29}) imply that
			\begin{align}\label{eq_a_30}
				\hat{S}_{n,L} \overset{d}{\longrightarrow} N\Big(0, \int_{0}^{1}{E\Big\{ \sum_{l=0}^{L-1}{ \mathcal{P}_{t}[\beta^{\top}_1 {\xi}_{t+l}(\tau_{1},s)+\beta^{\top}_2 {\xi}_{t+l}(\tau_{2},s)] } \Big\}^2 ds}\Big).
			\end{align}
			Therefore, by (\ref{eq_a_25_new}), (\ref{eq_a_27}), (\ref{eq_a_30}) and Theorem 3.2 in \citet{billingsley1999convergence}, the conclusion holds. This completes the whole proof.\\
		\end{proof}
To prove  Theorem 5.1,  we	 need  one more lemma as follows.

		\begin{lemma}\label{lemma_b_4}
			Suppose that Assumptions 2.1--2.3, Assumption 2.4 (i), Assumption 2.5 (i), and Assumption 3.1 hold. Then we have $$\sup_{\tau \in \mathcal{T}}{\vert \alpha_n(u,\tau) \vert}=o_p(1),$$
			where $u\in \mathbb{R}^{p+1}$.
		\end{lemma}

		\begin{proof}[\textsc Proof of Theorem 5.1]
			We show that for every fixed $u$,
			\begin{align}\label{eq_a_31}
				\sup_{\tau\in \mathcal{T}}{ \Big\{ \Big \vert \sum_{t=p+1}^{n}{E[\zeta_t(u, \tau) \vert \mathcal{F}_{t-1}]}-u^{\top}\Sigma(\tau)u/2 \Big \vert\Big\}} =o_p(1).
			\end{align}
			By Assumptions 2.5 (i) and 3.1, Fubini's theorem implies that
			\begin{align}\label{eq_a_32}
				& \sum_{t=p+1}^{n}{E[\zeta_t(u, \tau) \vert \mathcal{F}_{t-1}]}=u^{\top}\times \Big[\frac{1}{2n}\sum_{t=p+1}^{n}{\frac{w_t}{\sigma_{t}}Z_{t-1} Z^{\top}_{t-1}f\Big(\frac{\delta^{\top}_0(\tau)Z_{t-1} }{\sigma_{t}} \Big) }   \Big]\times u\\
				& \qquad\qquad\quad + \sum_{t=p+1}^{n}{ \frac{w_t}{\sigma_{t}} \int_{0}^{ \frac{u^{\top}Z_{t-1}}{\sqrt{n} } }{ s \Big[ f\Big(\frac{\delta^{\top}_0(\tau)Z_{t-1} }{\sigma_{t}}+\Delta s \Big)-f\Big(\frac{\delta^{\top}_0(\tau)Z_{t-1} }{\sigma_{t}} \Big) \Big] ds}   },\nonumber
			\end{align}
			where $\vert \Delta s \vert\leq \vert u^{\top} Z_{t-1}\sigma^{-1}_t\vert/\sqrt{n}$.
			
			For the first summation on the right side of (\ref{eq_a_32}), by Assumptions 2.1-2.4 (i), Assumption 2.5, and Assumption 3.1, and following {a similar} proof procedure for Theorem 4.1 (i), it is not hard to show that for every $\tau$,
				\begin{align}\label{eq_a_33}
				\frac{1}{n}\sum_{t=p+1}^{n}{\frac{w_t}{\sigma_{t}}Z_{t-1} Z^{\top}_{t-1}f\Big(\frac{\delta^{\top}_0(\tau)Z_{t-1} }{\sigma_{t}} \Big) }\overset{p}{\longrightarrow} \Sigma(\tau).
			\end{align}
			On the other hand, by Assumptions 2.4 (i), 2.5 and 3.1, and Theorem 4.1 (ii), for any $\tau_{1} \neq \tau_{2}$, we have
			\begin{align}\label{eq_a_34}
				\bigg \vert {\frac{w_t}{\sigma_{t}}Z_{t-1} Z^{\top}_{t-1}f\Big(\frac{\delta^{\top}_0(\tau_1)Z_{t-1} }{\sigma_{t}} \Big) }-{\frac{w_t}{\sigma_{t}}Z_{t-1} Z^{\top}_{t-1}f\Big(\frac{\delta^{\top}_0(\tau_2)Z_{t-1} }{\sigma_{t}} \Big) }\bigg \vert \leq C_1 \vert \tau_{1}-\tau_{2} \vert,
			\end{align}
			where $C_1$ is a constant independent with $t$ and $n$. Thus, (\ref{eq_a_33})-(\ref{eq_a_34}) imply that
			\begin{align}\label{eq_a_35}
				\sup_{\tau\in \mathcal{T}}{\bigg \vert  \frac{1}{n}\sum_{t=p+1}^{n}{\frac{w_t}{\sigma_{t}}Z_{t-1} Z^{\top}_{t-1}f\Big(\frac{\delta^{\top}_0(\tau)Z_{t-1} }{\sigma_{t}} \Big) }-\Sigma(\tau) \bigg \vert} \overset{p}{\longrightarrow} 0.
			\end{align}
			
			For the second item in (\ref{eq_a_32}),  we have
			\begin{align*}
				\bigg \vert \frac{w_t}{\sigma_{t}} \int_{0}^{ \frac{u^{\top}Z_{t-1}}{\sqrt{n} } }{ s \Big[ f\Big(\frac{\delta^{\top}_0(\tau)Z_{t-1} }{\sigma_{t}}+\Delta s \Big)-f\Big(\frac{\delta^{\top}_0(\tau)Z_{t-1} }{\sigma_{t}} \Big) \Big] ds}  \bigg\vert \leq C_2\vert u \vert^3n^{-3/2},
			\end{align*}
			where $C_2$ is independent with $t,\,n$ and $\tau$. Then, it follows that
			\begin{align}\label{eq_a_36}
				\sup_{\tau \in \mathcal{T}}{\bigg\vert \sum_{t=p+1}^{n}{ \frac{w_t}{\sigma_{t}} \int_{0}^{ \frac{u^{\top}Z_{t-1}}{\sqrt{n} } }{ s \Big[ f\Big(\frac{\delta^{\top}_0(\tau)Z_{t-1} }{\sigma_{t}}+\Delta s \Big)-f\Big(\frac{\delta^{\top}_0(\tau)Z_{t-1} }{\sigma_{t}} \Big) \Big] ds}   } \bigg\vert}=o_p(1).
			\end{align}
			{Hence, (\ref{eq_a_31}) holds from (\ref{eq_a_32}) and (\ref{eq_a_35})-(\ref{eq_a_36}).}

			Notice that $L_n(u,\tau)$ is convex in $u$ for each $\tau$, and it is continuous in $\tau$ for each $u$. By Theorem 2 in \citet{kato2009} and Theorem 4.1, Lemma 5.1 and Lemma \ref{lemma_b_4} and (\ref{eq_a_31}), it follows that
			\begin{align}\label{eq_a_36_a}
				\sqrt{n}(\hat{\theta}_n(\tau)-\theta_0(\tau))& =\Sigma^{-1}(\tau)T_n(\tau)+o_p(1).\nonumber\\
				& \rightsquigarrow \Sigma^{-1}(\tau)T_0(\tau).
			\end{align}
			This completes the whole proof.\\
		\end{proof}

		\setcounter{theorem}{0}
		\renewcommand{\thetheorem}{C.\arabic{theorem}}
		\setcounter{equation}{0}
		\renewcommand{\theequation}{C.\arabic{equation}}
		\setcounter{lemma}{0}
		\renewcommand{\thelemma}{C.\arabic{lemma}}
		\setcounter{remark}{0}
		\renewcommand{\theremark}{C.\arabic{remark}}
		\section{Proofs of Lemmas \ref{lemma_b_3}-\ref{lemma_b_4}}
		In this section, we show the details for Lemmas \ref{lemma_b_3}-\ref{lemma_b_4}.
		
 Let $Z_{t-1,i}$ be the $i$-th component in $Z_{t-1}$ and denote
		\begin{align}
			G_{n,i}(x,h) & =\frac{1}{\sqrt{n}}\sum_{t=p+1}^{n}{w_t Z_{t-1,i} \{ \eta_{t}(x,h)-E[\eta_{t}(x,h) \vert \mathcal{F}_{t-1}] \}},\label{eq_b_1}\\
			\eta_{t}(x,h) & =I(\varepsilon_{t} \leq x^{\top}Z_{t-1}+h\vert Z_{t-1}\vert),\label{eq_b_2}
		\end{align}
		where $x\in \mathbb{R}^{p+1}$ and $h\in \mathbb{R}$. {Before presenting the proof of Lemma \ref{lemma_b_3}, we first show the following Lemmas \ref{lemma_b_1}-\ref{lemma_b_2}.}
		\begin{lemma}\label{lemma_b_1}
			Suppose that Assumption 2.4 (i), Assumption 2.5 (i), and Assumption 3.1 hold. Then for any $q>2$, there exists a constant $C_q>0$ such that for any $x,\tilde{x} \in \mathbb{R}^{p+1}$ and $h,\tilde{h}\in \mathbb{R}$, we have
			\begin{align*}
				\Vert G_{n,i}(x,h)-G_{n,i}(\tilde{x},\tilde{h})\Vert^{q}_q \leq C_q n^{\alpha} (\vert x-\tilde{x}\vert+\vert h-\tilde{h}\vert)+C_q (\vert x-\tilde{x}\vert^{q/2}+\vert h-\tilde{h}\vert^{q/2}),
			\end{align*}
			where $i=1,\cdots, p+1$ and $\alpha=[1\vee (q/4)]-q/2$.
		\end{lemma}
		\begin{proof}[\textsc Proof of Lemma \ref{lemma_b_1}]
			Denote $d_t(x,h)=w_t Z_{t-1,i} \{\eta_{t}(x,h)-E(\eta_{t}(x,h)\vert \mathcal{F}_{t-1})\}$ and $d_t=d_t(x,h)-d_t(\tilde{x},\tilde{h})$. Furthermore, let $D_t=d^2_t-E[d^2_t \vert \mathcal{F}_{t-1}]$, $K_n=\sum_{t=p+1}^{n}{D_t}$, and $L_n=\sum_{t=p+1}^{n}{E[d^2_t \vert \mathcal{F}_{t-1}]}$. Since $G_{n,i}(x,h)-G_{n,i}(\tilde{x},\tilde{h})=n^{-1/2}\sum_{t=p+1}^{n}{d_t}$ and $\{d_t\}$ is a martingale difference sequence (m.d.s.), {applying Burkholder's inequality yields that}
			\begin{align}
				\Vert G_{n,i}(x,h)-G_{n,i}(\tilde{x},\tilde{h}) \Vert^{q}_{q} & \leq n^{-q/2}C_q \times E\big[\big(\sum_{t=p+1}^{n}{d^2_t}\big)^{q/2}\big]\nonumber\\
				&  \leq n^{-q/2} C_q 2^{q/2-1}\times (\Vert K_n \Vert^{q/2}_{q/2}+\Vert L_n \Vert^{q/2}_{q/2}),\label{eq_b_3}
			\end{align}
			where the last inequality follows from $\sum_{t=p+1}^{n}{d^2_t}=K_n+L_n$ and the fact that $(a+b)^{q/2}\leq 2^{q/2-1}(a^{q/2}+b^{q/2})$ for any $a,b\geq 0$ and $q>2$.
			
		Given that \(q > 2\) and \(\{D_t\}\) is a m.d.s., we can apply Lemma 3 from \citet{wu2008} and use the inequality \(\| D_t \|^{q/2}_{q/2} \leq 2^{q/2} \| d^2_t \|^{q/2}_{q/2}\) to further deduce that
			\begin{align}
				\Vert K_n \Vert^{q/2}_{q/2}\leq n^{1\vee q/4} C_q \times \max_{t}{\{ \Vert d^2_t \Vert^{q/2}_{q/2} \}}.\label{eq_b_4}
			\end{align}
			Notice that, conditional on $\mathcal{F}_{t-1}$, $\vert \eta_{t}(x,h)-\eta_{t}(\tilde{x},\tilde{h})-E\{[\eta_{t}(x,h)-\eta_{t}(\tilde{x},\tilde{h})]\vert\mathcal{F}_{t-1}  \} \vert\leq 1$. Thus, by Assumption 2.4 (i), Assumption 2.5 (i), and Assumption 3.1, and using the law of total expectation, we have
			\begin{align}\label{eq_b_5}
			\Vert d^2_t \Vert^{q/2}_{q/2} & \leq E\Big \{ \vert w_t Z_{t-1,i}\vert^q E\big ( \vert  \eta_{t}(x,h)-\eta_{t}(\tilde{x},\tilde{h})-E\{[\eta_{t}(x,h)-\eta_{t}(\tilde{x},\tilde{h})]\vert\mathcal{F}_{t-1}  \}  \vert^2   \vert \mathcal{F}_{t-1}\big) \Big \}\nonumber\\
			& \leq E\Big \{ \vert w_t Z_{t-1,i}\vert^q E\big ( \vert  \eta_{t}(x,h)-\eta_{t}(\tilde{x},\tilde{h})\vert^2   \vert \mathcal{F}_{t-1}\big) \Big \}\nonumber\\
			& \leq E\bigg\{ \vert w_t Z_{t-1,i}\vert^{q} \times \Big\vert F\Big(\frac{x^{\top}Z_{t-1}+h\vert Z_{t-1} \vert}{\sigma_{t}} \Big)-F\Big(\frac{\tilde{x}^{\top}Z_{t-1}+\tilde{h}\vert Z_{t-1} \vert}{\sigma_{t}} \Big) \Big\vert  \bigg \}\nonumber\\
				& \leq {C}_q\times (\vert x-\tilde{x}\vert+\vert h-\tilde{h}\vert).
			\end{align}
			So (\ref{eq_b_4}) and (\ref{eq_b_5}) imply that
			\begin{align}\label{eq_b_6}
				\Vert K_n \Vert^{q/2}_{q/2}\leq n^{1\vee q/4} C_q \times (\vert x-\tilde{x}\vert+\vert h-\tilde{h}\vert).
			\end{align}
			In addition, by a similar argument for (\ref{eq_b_5}), it follows that
			\begin{eqnarray}
				&& \Vert L_n \Vert^{q/2}_{q/2} \leq n^{q/2} \times \max_{t}{\{ \Vert E(d^2_t \vert \mathcal{F}_{t-1})\Vert^{q/2}_{q/2} \}}\nonumber\\
				&&\qquad \leq n^{q/2} \max_{t}\Big\{\Big  \Vert w^2_t Z^2_{t-1,i} \Big\vert F\Big(\frac{x^{\top}Z_{t-1}+h\vert Z_{t-1} \vert}{\sigma_{t}} \Big)-F\Big(\frac{\tilde{x}^{\top}Z_{t-1}+\tilde{h}\vert Z_{t-1} \vert}{\sigma_{t}} \Big) \Big\vert \Big \Vert^{q/2}_{q/2} \Big \}\nonumber \\
				&&\qquad \leq n^{q/2} C_q \times (\vert x-\tilde{x}\vert^{q/2}+\vert h-\tilde{h}\vert^{q/2}).\label{eq_b_7}
			\end{eqnarray}
			So the conclusion holds from (\ref{eq_b_3}), (\ref{eq_b_6}) and (\ref{eq_b_7}).
		\end{proof}
		\begin{remark}\label{remark_b_1}
			{Define}
			$$G^{\pm}_{n,i}(x,h) =\frac{1}{\sqrt{n}}\sum_{t=p+1}^{n}{w_t Z^{\pm}_{t-1,i} \{ \eta_{t}(x,h)-E[\eta_{t}(x,h) \vert \mathcal{F}_{t-1}] \}},$$
			where $Z^{+}_{t-1,i}=Z_{t-1,i}\vee 0$ is the positive part of $Z_{t-1,i}$ and $Z^{-}_{t-1,i}=(-Z_{t-1,i})\vee 0$ is its negative part. {Clearly,} under the same conditions, Lemma \ref{lemma_b_1} still holds for $G^{\pm}_{n,i}(x,h)$.
		\end{remark}
		
		\begin{lemma}\label{lemma_b_2}
			Suppose that Assumptions 2.1--2.3, Assumption 2.4 (i), Assumption 2.5 (i), and Assumption 3.1 hold. Then, there exists $C_q>0$ such that for any $b>0$, $\tau_{1} \in (0,1)$, $x_1\in \mathbb{R}^{p+1}$, and $h_1\in \mathbb{R}$, we have
			\begin{align*}
				E&\big\{\sup_{\vert \tau \vert<b}{\vert G_{n,i}(\delta_0(\tau_{1}+\tau)+x_1,h_1)-G_{n,i}(\delta_0(\tau_{1})+x_1,h_1) \vert^q} \big\}\\
				&\qquad \qquad \qquad \qquad\qquad \qquad\qquad \qquad\qquad\leq C_q (n^{\alpha}d^q b+b^{q/2}+n^{q/2}2^{-dq}b^{q}),
			\end{align*}
			where $\alpha$ is given in Lemma \ref{lemma_b_1} and $d$ is any positive integer.
		\end{lemma}
		\begin{proof}[\textsc Proof of Lemma \ref{lemma_b_2}]
		Without loss of generality, we assume that \(\{w_t Z_{t-1,i}\}\) is non-negative. If this is not the case, we can use the decomposition \(w_t Z_{t-1,i} = w_t Z^{+}_{t-1,i} - w_t Z^{-}_{t-1,i}\), and then apply the following same proof procedures, substituting \(w_t Z_{t-1,i}\) with \(w_t Z^{+}_{t-1,i}\) or \(w_t Z^{-}_{t-1,i}\) as appropriate. For theoretical support, see Remark \ref{remark_b_1}.

			By the symmetry, we only focus on the case $\tau>0$. For any positive integer $d$, let $h=b/2^d$ and define $\tau^{(j)}=jh$ for $j=1,\cdots,2^d$. Then for any $\tau\in (0,b)$, we can find some $\tau^{(j)}$ such that $\vert \tau-\tau^{(j)} \vert\leq h$. By the monotone property of the indicator function $\eta_{t}(x,h)$ in (\ref{eq_b_2}) and the non-negative property of $\{w_t Z_{t-1,i}\}$, it is not hard to show that
			\begin{align}
				& G_{n,i}(\delta_0(\tau_{1}+\tau)+x_1,h_1)-G_{n,i}(\delta_0(\tau_{1})+x_1,h_1) \label{eq_b_8} \\
				& \quad \leq   G_{n,i}(\delta_0(\tau_{1}+\tau^{(j)})+x_1,h_1+C_1 h)-G_{n,i}(\delta_0(\tau_{1})+x_1,h_1+C_1h)\nonumber\\
				& \qquad + G_{n,i}(\delta_0(\tau_{1})+x_1,h_1+C_1 h)-G_{n,i}(\delta_0(\tau_{1})+x_1,h_1)\nonumber\\
				& \qquad + \frac{1}{\sqrt{n}}\sum_{t=p+1}^{n}w_t Z_{t-1,i} E\{  [  \eta_{t}(\delta_0(\tau_{1}+\tau^{(j)})+x_1,h_1+C_1h)\nonumber\\
				&\qquad\qquad\qquad\qquad\qquad\qquad\qquad -\eta_{t}(\delta_0(\tau_{1}+\tau^{(j)})+x_1,h_1-C_1h)] \vert \mathcal{F}_{t-1} \},\nonumber\\
				&\nonumber\\
				& G_{n,i}(\delta_0(\tau_{1}+\tau)+x_1,h_1)-G_{n,i}(\delta_0(\tau_{1})+x_1,h_1) \label{eq_b_9}\\
				& \quad \geq   G_{n,i}(\delta_0(\tau_{1}+\tau^{(j)})+x_1,h_1-C_1 h)-G_{n,i}(\delta_0(\tau_{1})+x_1,h_1-C_1h)\nonumber\\
				& \qquad + G_{n,i}(\delta_0(\tau_{1})+x_1,h_1-C_1 h)-G_{n,i}(\delta_0(\tau_{1})+x_1,h_1)\nonumber\\
				& \qquad + \frac{1}{\sqrt{n}}\sum_{t=p+1}^{n}w_t Z_{t-1,i} E\{  [  \eta_{t}(\delta_0(\tau_{1}+\tau^{(j)})+x_1,h_1-C_1h)\nonumber\\
				&\qquad\qquad\qquad\qquad\qquad\qquad\qquad -\eta_{t}(\delta_0(\tau_{1}+\tau^{(j)})+x_1,h_1+C_1h)] \vert \mathcal{F}_{t-1} \},\nonumber
			\end{align}
			where \(C_1 := \sup_{\tau} \left| \frac{\partial \delta_0(\tau)}{\partial \tau} \right| < \infty\), as derived from Theorem 4.1 (ii). By Lemma \ref{lemma_b_1}, we have
			\begin{align}\label{eq_c_10_1}
			   \Vert G_{n,i}(\delta_0(\tau_{1})+x_1,h_1 \pm C_1 h)-G_{n,i}(\delta_0(\tau_{1})+x_1,h_1) \Vert^{q}_q \leq C_q (n^{\alpha}h+h^{q/2}).
			\end{align}
			In addition, Assumptions 2.4 (i), 2.5 (i), and 3.1 imply that
			\begin{align}\label{eq_c_11_1}
			     & \bigg \vert \frac{1}{\sqrt{n}}\sum_{t=p+1}^{n} \Big (w_t Z_{t-1,i} E\{  [  \eta_{t}(\delta_0(\tau_{1}+\tau^{(j)})+x_1,h_1 \pm C_1h)\nonumber\\
			     & \qquad \qquad \qquad \qquad -\eta_{t}(\delta_0(\tau_{1}+\tau^{(j)})+x_1,h_1 \mp C_1h)] \vert \mathcal{F}_{t-1} \} \Big ) \bigg \vert^q\leq C_q n^{q/2}h^q.
			\end{align}
			Therefore, combining (\ref{eq_b_8})--(\ref{eq_c_11_1}), we can derive that
			\begin{eqnarray}
				&& E\Big\{\sup_{0< \tau <b}{\vert G_{n,i}(\delta_0(\tau_{1}+\tau)+x_1,h_1)-G_{n,i}(\delta_0(\tau_{1})+x_1,h_1) \vert^q} \Big\}\nonumber\\
				&& \quad \leq C_q \Big\{ \Vert \max_{j}{\vert \{G_{n,i}(\delta_0(\tau_{1}+\tau^{(j)})+x_1,h_1+C_1 h)-G_{n,i}(\delta_0(\tau_{1})+x_1,h_1+C_1h) \vert} \Vert^q_q\nonumber\\
				&& \qquad   + \Vert \max_{j}{\vert \{G_{n,i}(\delta_0(\tau_{1}+\tau^{(j)})+x_1,h_1-C_1 h)-G_{n,i}(\delta_0(\tau_{1})+x_1,h_1-C_1h) \vert} \Vert^q_q\nonumber\\
				&& \qquad \qquad\qquad  + (n^{\alpha} h+h^{q/2}+n^{q/2} h^{q})\Big\}.\label{eq_b_10}
			\end{eqnarray}
			
			For $j=1,\cdots,2^d$, denote
			\begin{align*}
				S^{(j)}=G_{n,i}(\delta_0(\tau_{1}+\tau^{(j)})+x_1,h_1+C_1 h)-G_{n,i}(\delta_0(\tau_{1})+x_1,h_1+C_1h).
			\end{align*}
			{Then, by} Proposition 1 in \citet{wu2007a} and Lemma \ref{lemma_b_1}, it follows that
			\begin{align*}
				\Vert \max_{j}{\vert S^{(j)} \vert} \Vert_q & \leq \sum_{r=0}^{d}{\big (\sum_{m=1}^{2^{d-r}}{\Vert S^{(2^rm)}-S^{(2^r(m-1))} \Vert^q_q}\big )^{1/q}}\nonumber\\
				& \leq C_q \times \sum_{r=0}^{d}\big(n^{\alpha}2^d h+2^{d-r} (2^r h)^{q/2} \big)^{1/q}\nonumber\\
				& \leq C_q \times (dn^{\alpha/q}b^{1/q}+b^{1/2}).
			\end{align*}
			So, we have shown that
			\begin{align}
				& \Vert \max_{j}{\vert \{G_{n,i}(\delta_0(\tau_{1}+\tau^{(j)})+x_1,h_1+C_1 h)-G_{n,i}(\delta_0(\tau_{1})+x_1,h_1+C_1h) \vert} \Vert^q_q \nonumber
				\\
				& \qquad \qquad \qquad \leq C_q\times (d^q n^{\alpha} b+b^{q/2}).\label{eq_b_11}
			\end{align}
			Similarly,
			\begin{align}
				& \Vert \max_{j}{\vert \{G_{n,i}(\delta_0(\tau_{1}+\tau^{(j)})+x_1,h_1-C_1 h)-G_{n,i}(\delta_0(\tau_{1})+x_1,h_1-C_1h) \vert} \Vert^q_q \nonumber
				\\
				& \qquad \qquad \qquad \leq C_q\times (d^q n^{\alpha} b+b^{q/2}).\label{eq_b_12}
			\end{align}
			Then the conclusion holds from (\ref{eq_b_10})--(\ref{eq_b_12}). \\
		\end{proof}
		
		
		\begin{proof}[\textsc Proof of Lemma \ref{lemma_b_3}]
			For the first statement, {notice that}
			\begin{align*}
				\sup_{\vert \tau_{1}-\tau_{2} \vert<\delta}{\vert M_n(\tau_{1})-M_n(\tau_{2}) \vert} \leq \sum_{i=1}^{p+1}{ \sup_{\vert \tau_{1}-\tau_{2} \vert<\delta}{\vert G_{n,i}(\delta_0(\tau_{1}),0)-G_{n,i}(\delta_0(\tau_{2}),0) \vert} }.
			\end{align*}
			Then, it suffices to show that for any $i=1,\cdots, p+1$ and $\epsilon>0$,
			\begin{align}\label{eq_b_15}
				\lim_{\delta \to 0}{\limsup_{n \to \infty}{ P\Big(\sup_{\vert \tau_{1}-\tau_{2} \vert<\delta}{\vert G_{n,i}(\delta_0(\tau_{1}),0)-G_{n,i}(\delta_0(\tau_{2}),0) \vert}>\epsilon\Big)  }}=0.
			\end{align}
			By Lemma \ref{lemma_b_2} and using the standard chaining method, for any $\delta>0$ and $q>2$, we have
			\begin{align}\label{eq_b_16}
				& P\Big(\sup_{\vert \tau_{1}-\tau_{2} \vert<\delta}{\vert G_{n,i}(\delta_0(\tau_{1}),0)-G_{n,i}(\delta_0(\tau_{2}),0) \vert}>\epsilon\Big)\\
				& \qquad \qquad\qquad \qquad \qquad \qquad  \leq \frac{C_q}{\delta\epsilon^q}\times (n^{\alpha}d^q\delta+\delta^{q/2}+n^{q/2}2^{-dq}\delta^q).\nonumber
			\end{align}
			Let $d=[\log_2(n)]+1$, {and thus} $2^{-d}n\leq 1$. Then (\ref{eq_b_16}) implies that
			\begin{align}\label{eq_b_17}
				\limsup_{n \to \infty}{ P\Big(\sup_{\vert \tau_{1}-\tau_{2} \vert<\delta}{\vert G_{n,i}(\delta_0(\tau_{1}),0)-G_{n,i}(\delta_0(\tau_{2}),0) \vert}>\epsilon\Big)  }\leq C_q \times \frac{\delta^{q/2-1}}{\epsilon^q}.
			\end{align}
			So (\ref{eq_b_15}) follows from (\ref{eq_b_17}) and $q>2$.
			
			For the second statement, we only need to prove that for any $i=1,\cdots, p+1$,
			\begin{align}\label{eq_b_18}
				\lim_{\delta \to 0}{\limsup_{n \to \infty}{P\Big(\sup_{\vert \tau_{1}-\tau_{2} \vert< \delta}{\vert N_{n,i}(\tau_{1})-N_{n,i}(\tau_{2}) \vert>\epsilon}\Big)}}=0,
			\end{align}
			where $N_{n,i}(\tau)$ is the $i$-th component of $N_n(\tau)$. By Assumption 2.4 (i), Assumption 2.5 (i), Assumption 3.1, and the continuity of $\partial \delta_0(\tau)/\partial \tau$ in Theorem 4.1 (ii), it is easy to show that
			\begin{align}\label{eq_b_19}
				\sup_{\vert \tau_{1}-\tau_{2} \vert<\delta}{\vert  N_{n,i}(\tau_{1})-N_{n,i}(\tau_{2}) \vert}\leq \delta \times  \sup_{\tau \in \mathcal{T}}{\vert H_n(\tau) \vert},
			\end{align}
			where $H_n(\tau)=\sum_{t=p+1}^{n}{\{  Q'_{t}(\tau)-E[Q'_t(\tau)] \}}$ with $Q'_t(\tau)$ being the first derivative of $Q_t(\tau):=n^{-1/2}w_tZ_{t-1,i}[\tau-F(\delta^{\top}_0(\tau)Z_{t-1}/\sigma_{t})]$. Thus, $Q'_t(\tau)$ can be expressed as
			\begin{align}\label{eq_c_18_new}
				Q'_t(\tau)=\frac{1}{\sqrt{n}}w_t Z_{t-1,i}\Big[1-f(\delta^{\top}_0(\tau)Z_{t-1}/\sigma_{t})Z^{\top}_{t-1} \frac{\partial \delta_0(\tau)}{\partial \tau} /\sigma_{t}\Big].
			\end{align}
			In addition, by leveraging the boundedness and continuity of $f^{(1)}(\cdot)$ in Assumption 2.5 (ii) and using the result in equation (A.11), we can easily show that the second-order partial derivatives of $g(x,\tau)$ in Theorem 4.1 are continuous. Then, the implicit function theorem derives that the second-order derivative $\partial^2 \delta_0(\tau)/\partial \tau^2$ is continuous. As a result, $Q'_t(\tau)$ in (\ref{eq_c_18_new}) has the following continuous derivative
			\begin{align}\label{eq_c_19_new}
				Q''_t(\tau)& =-\frac{1}{\sqrt{n}}w_t Z_{t-1,i}\Big \{ f(\delta^{\top}_0(\tau)Z_{t-1}/\sigma_{t})Z^{\top}_{t-1} \frac{\partial^2 \delta_0(\tau)}{\partial \tau^2}/\sigma_{t}  \nonumber\\
				 & \qquad +f^{(1)}(\delta^{\top}_0(\tau)Z_{t-1}/\sigma_{t})\big[Z^{\top}_{t-1} \frac{\partial \delta_0(\tau)}{\partial \tau}/\sigma_{t}\big]^2 \Big\}.
			\end{align}
			This implies that $H'_n(\tau)=\sum_{t=p+1}^{n}{\{ Q''_t(\tau)-E[Q''_t(\tau)] \}}$ is continuous.
			
			Furthermore, by the continuity of $H_n(\tau)$ and $H'_n(\tau)$, and using the equation (27) from \citet{wu2008}, it follows that
			\begin{align}\label{eq_b_20}
				E\Big[\sup_{\tau \in \mathcal{T}}{H^2_n(\tau)}\Big]\leq \frac{2}{\tau_{u}-\tau_{l}}\int_{\tau_{l}}^{\tau_{u}}{\Vert H_n(\tau) \Vert^2 d\tau}+2\int_{\tau_{l}}^{\tau_{u}}{\Vert H'_n(\tau) \Vert^2 d\tau},
			\end{align}
			where $\tau_{l}$ and $\tau_{u}$ are the lower bound and the upper bound of $\mathcal{T}$, respectively.
			
			Let $\mathcal{P}_{t-m}(\cdot)=E(\cdot \vert \mathcal{F}_{t-m})-E(\cdot \vert \mathcal{F}_{t-m-1})$ be the projection on the space $\mathcal{F}_{t-m}$. Then,
			\begin{align}\label{eq_b_21}
				H_n(\tau)=\sum_{m=1}^{\infty}{\Big\{  \sum_{t=p+1}^{n}{\mathcal{P}_{t-m}[Q'_t(\tau)]} \Big \}}.
			\end{align}
			Notice that $\mathcal{P}_{i}(\cdot)$ must be uncorrelated with $\mathcal{P}_{j}(\cdot)$ for any $i \neq j$, so we have
			\begin{align}\label{eq_b_22}
				\Big\Vert  \sum_{t=p+1}^{n}{\mathcal{P}_{t-m}[Q'_t(\tau)]} \Big\Vert^2=\sum_{t=p+1}^{n}{\Vert \mathcal{P}_{t-m}[Q'_t(\tau)] \Vert^2}\leq n \times \max_{t}{ \Vert \mathcal{P}_{t-m}[Q'_t(\tau)] \Vert^2  }.
			\end{align}
			As shown in \citet{wu2007a}, the inequality $\Vert \mathcal{P}_{t-m}[X]\Vert^2\leq \Vert X-X^{*(t-m)}\Vert^2$ holds for any random varibale $X$ with $\Vert X \Vert<\infty$.
			Then, by utilizing the boundedness and the Lipschitz continuity of $f(\cdot)$ in Assumption 2.5 and Assumption 3.1, it is not hard to prove that there exists a constant $C$ independent of $t, n, m,\tau$, such that
			\begin{align}\label{eq_b_23}
					\Vert \mathcal{P}_{t-m}[Q'_t(\tau)] \Vert^2 \leq C/n \times \{  \Vert Y^{*(t-m)}_{t-1}-Y_{t-1} \Vert^{\alpha_0}_{\alpha_0}+\Vert \sigma_{t}-\sigma^{*(t-m)}_{t}\Vert^{\alpha_0}_{\alpha_0} \}.
			\end{align}
			By Assumption 5.1 and (A.7), it is straightforward to verify that
			\begin{align}\label{eq_b_24}
				\max_{t}{\{ \Vert Y^{*(t-m)}_{t-1}-Y_{t-1} \Vert^{\alpha_0}_{\alpha_0}+\Vert \sigma_{t}-\sigma^{*(t-m)}_{t}\Vert^{\alpha_0}_{\alpha_0} \}}=O(m\tilde{\kappa}^{m}),
			\end{align}
			where $\tilde{\kappa}=(c_0\vee \kappa_0)^{\alpha_0}$. {Therefore,} (\ref{eq_b_21})-(\ref{eq_b_24}) imply that
			\begin{align}\label{eq_b_25}
				\sup_{\tau \in \mathcal{T}}{\Vert H_n(\tau) \Vert^2} \leq  C_1\times\Big(\sum_{m=1}^{\infty}{\sqrt{m\tilde{\kappa}^{m}}}\Big)^2<\infty,
			\end{align}
			where $C_1$ is a constant independent of $n$.
			
			Recall that $H'_n(\tau)=\sum_{t=p+1}^{n}{\{ Q''_t(\tau)-E[Q''_t(\tau)] \}}$. By (\ref{eq_c_19_new}), the boundedness and Lipschitz continuity of $f(\cdot)$ and $f^{(1)}(\cdot)$ in Assumption 2.5, Assumption 3.1, and Assumption 5.1, we can similarly derive that
			\begin{align}\label{eq_b_26}
				\sup_{\tau \in \mathcal{T}}{\Vert H'_n(\tau) \Vert^2}\leq C_2<\infty,
			\end{align}
			where $C_2$ is a constant independent of $n$.
			
			Thus, (\ref{eq_b_18}) holds from (\ref{eq_b_19}), (\ref{eq_b_20}), and (\ref{eq_b_25})--(\ref{eq_b_26}). This completes the proof.
		\end{proof}
		\begin{remark}\label{remark_b_2}
		{If we generalize} the item $Q_t(\tau)$ in (\ref{eq_b_19}) as follows:
			\begin{align*}
				Q_t(\tau)=n^{-1/2}w_tZ_{t-1,i}[\tau-F(\{\delta^{\top}_0(\tau)Z_{t-1}+x_1Z_{t-1}+h_1\vert Z_{t-1}\vert\}/\sigma_{t})],
			\end{align*}
			where $(x_1,h_1)\in \mathbb{R}^{p+2}$, {then} we can also show that $E[\sup_{\tau \in \mathcal{T}}{H^2_n(\tau)}]<C_4$, where $C_4$ is a constant relying on $(x_1,h_1)$ and uniformly bounded on any compact subset.
			
		\end{remark}
%
		\begin{proof}[\textsc Proof of Lemma \ref{lemma_b_4}]
			By the definition of $\alpha_n(u,\tau)$, we can rewrite $\alpha_n(u,\tau)$ as
			\begin{align}
				\alpha_n(u,\tau) & =\sum_{i=1}^{p+1}{u_i \int_{0}^{1}{\bigg[G_{n,i}\Big(\delta_0(\tau)+\frac{us}{\sqrt{n}},0\Big)-G_{n,i}\big(\delta_0(\tau),0\big)\bigg]ds}}\nonumber\\
				& \equiv \sum_{i=1}^{p+1}{u_i \tilde{G}_{n,i}(\tau)},\label{eq_b_25_1}
			\end{align}
			where $u_i$ is the $i$-th component of $u$ and $G_{n,i}(\cdot)$ is defined in (\ref{eq_b_1}). (\ref{eq_b_25_1}) indicates that it {suffices} to show that, for any $i\leq p+1$,
			\begin{align}\label{eq_b_26_1}
				\sup_{\tau\in \mathcal{T}}{\vert \tilde{G}_{n,i}(\tau)\vert }=o_p(1).
			\end{align}
			For every fixed $\tau$, Lemma \ref{lemma_b_1} derives that
			\begin{align}\label{eq_b_27}
				\Vert \tilde{G}_{n,i}(\tau) \Vert^q_q & \leq \int_{0}^{1}{   \Big\Vert G_{n,i}\Big(\delta_0(\tau)+\frac{us}{\sqrt{n}},0\Big)-G_{n,i}\big(\delta_0(\tau),0\big) \Big\Vert^q_q ds}\nonumber\\
				& \leq C_q \times (\vert u \vert n^{\alpha-1/2}+ \vert u \vert^{q/2} n^{-q/4}) \to 0, \mbox{ as }n \to \infty.
			\end{align}
			Furthermore, for any $\tau_{1}, \tau_{2} \in \mathcal{T}$, we have
			\begin{align}\label{eq_b_28}
				\tilde{G}_{n,i}(\tau_{1})-\tilde{G}_{n,i}(\tau_{2}) & =\int_{0}^{1}{\Big[G_{n,i}\Big(\delta_0(\tau_1)+\frac{us}{\sqrt{n}},0\Big)-G_{n,i}\Big(\delta_0(\tau_2)+\frac{us}{\sqrt{n}},0\Big)\Big]ds}\nonumber\\
				& \quad -  [G_{n,i}\big(\delta_0(\tau_1),0\big)-G_{n,i}\big(\delta_0(\tau_2),0\big)]\nonumber\\
				& \equiv \tilde{G}^{(1)}_{n,i}(\tau_1)-\tilde{G}^{(1)}_{n,i}(\tau_2)-[G_{n,i}\big(\delta_0(\tau_1),0\big)-G_{n,i}\big(\delta_0(\tau_2),0\big)].
			\end{align}
			{By Lemma \ref{lemma_b_1} and Theorem 4.1 (ii), it is clear that}
			\begin{align*}
				\Vert \tilde{G}^{(1)}_{n,i}(\tau_1)-\tilde{G}^{(1)}_{n,i}(\tau_2) \Vert^q_q \leq \tilde{C}_q \times (n^{\alpha} \vert \tau_{1}-\tau_{2} \vert+\vert \tau_{1}-\tau_{2} \vert^{q/2}).
			\end{align*}
			{By following a similar} proof procedure in Lemma \ref{lemma_b_2} and {applying} the first inequality in (\ref{eq_b_27}), we can easily show that for any $\epsilon>0$, there exists a  $\delta>0$ such that
			\begin{align}\label{eq_b_29}
				\lim_{\delta \to 0}{\limsup_{n \to \infty}{ P(\sup_{\vert \tau_{1}-\tau_{2} \vert<\delta}{\vert \tilde{G}^{(1)}_{n,i}(\tau_{1})-\tilde{G}^{(1)}_{n,i}(\tau_{2}) \vert}>\epsilon)  }}=0.
			\end{align}
			Therefore, by (\ref{eq_b_15}) and (\ref{eq_b_28})-(\ref{eq_b_29}), the asymptotic tightness of $\tilde{G}_{n,i}(\tau)$ holds. By combining (\ref{eq_b_27}), (\ref{eq_b_26_1}) is proved. This completes the proof.
		\end{proof}
		\begin{remark}\label{remark_b_3}
			On the other hand, {one may be} interested in the following summation:
			\begin{align}\label{eq_b_32}
				\tilde{\alpha}_n(u,\tau)=\sum_{t=p+1}^{n}{\{ E[\zeta_t(u,\tau)\vert \mathcal{F}_{t-1}]-E[\zeta_t(u,\tau)] \}}.
			\end{align}
			{By} repeating the proof procedure for  (\ref{eq_b_18}) and using the fact described in Remark \ref{remark_b_2}, it is straightforward  to get that $\sup_{\tau \in \mathcal{T}}{\vert \tilde{\alpha}_n(u,\tau)\vert }=o_p(1)$.
		\end{remark}
		
		\setcounter{theorem}{0}
		\renewcommand{\thetheorem}{D.\arabic{theorem}}
		\setcounter{equation}{0}
		\renewcommand{\theequation}{D.\arabic{equation}}
		\setcounter{lemma}{0}
		\renewcommand{\thelemma}{D.\arabic{lemma}}
		\section{Proofs of Lemma 6.1 }
		
 Denote
		\begin{align}\label{eq_c_1}
			\Delta_t(\theta, \tau; \tilde{w}_l)&=\tilde{w}_{lt} [I(\varepsilon_{t}-\delta^{\top}_0(\tau)Z_{t-1}\leq \theta^{\top}Z_{t-1} )-I(\varepsilon_{t}-\delta^{\top}_0(\tau)Z_{t-1}\leq 0 )]\nonumber\\
			&=\tilde{w}_{lt} [\eta_{t}(\delta_0(\tau)+\theta,0)-\eta_{t}(\delta_0(\tau),0)],
		\end{align}
		where $\eta_{t}(\cdot)$ is defined in (\ref{eq_b_2}). {We first prove the following lemma.}
		\begin{lemma}\label{lemma_c_1}
			Suppose that all the conditions in Lemma 6.1 hold. Then we have
			\begin{align}
				& \sup_{\theta \in \mathcal{C}_n,\,\tau \in \mathcal{T}}\Big \vert \frac{1}{\sqrt{n}}\sum_{t=p+1}^{n}{\big\{\Delta_t(\theta,\tau;\tilde{w}_l)-E[\Delta_t(\theta,\tau;\tilde{w}_l) \vert \mathcal{F}_{t-1}]  \big\}}  \Big \vert=o_p(1),\label{eq_c_2}\\
				& \sup_{\theta \in \mathcal{C}_n,\,\tau \in \mathcal{T}}\Big \vert \frac{1}{\sqrt{n}}\sum_{t=p+1}^{n}{E[\Delta_t(\theta,\tau;\tilde{w}_l) \vert \mathcal{F}_{t-1}]}-b^{\top}(\tau;\tilde{w}_l)\sqrt{n}\theta  \Big \vert=o_p(1),\label{eq_c_3}
			\end{align}
			where $\mathcal{C}_n=\{\theta\in \mathbb{R}^{p+1}: \vert \theta \vert<M/\sqrt{n}\}$ {with} $M$ being any positive number.
		\end{lemma}
		\begin{proof}[\textsc Proof of Lemma \ref{lemma_c_1}]
			To prove the equation (\ref{eq_c_2}), we use the standard argument as that for Lemma S.4 in \citet{she2022}. For any $\delta \in (0,1)$, by the covering number theory in \citet{van1996}, there exists  a sequence of open balls $\{B_{M\delta}(u_i)\}^{K_\delta}_{i=1}$ with $\vert u_i \vert<M$ and $K_{\delta}\leq (3/\delta)^{p+1}$. Denote $\theta_i=u_i/\sqrt{n}$ and then we have $\mathcal{C}_n \subset \cup_{i=1}^{K_{\delta}}{\mathcal{C}_{in}}$ and $\mathcal{C}_{in}=B_{M\delta/\sqrt{n}}(\theta_i)$.
			
			Similar to the proof procedure for Lemma \ref{lemma_b_2}, we assume that $\tilde{w}_{lt}$ is always non-negative. For any $\theta\in \mathcal{C}_{in} $, by the monotone property of indicator function, it follows that
			\begin{align}\label{eq_c_4}
				\Delta^{-}_t(\theta_i,\delta,\tau;\tilde{w}_l)\leq \Delta_t(\theta,\tau;\tilde{w}_l) \leq \Delta^{+}_t(\theta_i,\delta,\tau;\tilde{w}_l),
			\end{align}
			where $\Delta^{\pm}_t(\theta_i,\delta,\tau;\tilde{w}_l)=\tilde{w}_{lt} [\eta_t(\delta_0(\tau)+\theta_i,\pm M\delta/\sqrt{n})-\eta_{t}(\delta_0(\tau),0)]$. Notice that we can find a constant $C_1$ only relying on $f(\cdot)$, $\sigma_0$ and $\tilde{w}_l(\cdot)$, such that
			\begin{align}\label{eq_c_5}
				E\{[\Delta^{+}_t(\theta_i,\delta,\tau;\tilde{w}_l)-\Delta^{-}_t(\theta_i,\delta,\tau;\tilde{w}_l)]\vert \mathcal{F}_{t-1}\}\leq C_1M\delta/\sqrt{n}.
			\end{align}
			So (\ref{eq_c_4})-(\ref{eq_c_5}) imply that
			\begin{eqnarray}
				&& \sup_{\theta \in \mathcal{C}_n,\,\tau \in \mathcal{T}} \frac{1}{\sqrt{n}}\sum_{t=p+1}^{n}{\big\{\Delta_t(\theta,\tau;\tilde{w}_l)-E[\Delta_t(\theta,\tau;\tilde{w}_l) \vert \mathcal{F}_{t-1}]  \big\}} \nonumber \\
				&& \qquad\quad \leq \max_{i}\Big\{ \sup_{\tau \in \mathcal{T}} \frac{1}{\sqrt{n}}\sum_{t=p+1}^{n}{\big\{\Delta^{+}_t(\theta_i,\delta,\tau;\tilde{w}_l)-E[\Delta^{+}_t(\theta_i,\delta,\tau;\tilde{w}_l) \vert \mathcal{F}_{t-1}]  \big\}}\Big\}+C_1M\delta.\label{eq_c_6}
			\end{eqnarray}
			For every fixed $i$ and $\tau$, we have
			\begin{align}\label{eq_c_7}
				& E\Big[	\frac{1}{\sqrt{n}}\sum_{t=p+1}^{n}{\big\{\Delta^{+}_t(\theta_i,\delta,\tau;\tilde{w}_l)-E[\Delta^{+}_t(\theta_i,\delta,\tau;\tilde{w}_l) \vert \mathcal{F}_{t-1}]  \big\}}\Big]^2\nonumber\\
				& \qquad \qquad\qquad\leq \frac{1}{n} \sum_{t=p+1}^{n}{E[\Delta^{+}_t(\theta_i,\delta,\tau;\tilde{w}_l)]^2}=O(1/\sqrt{n}).
			\end{align}
			On the other hand, the following fact holds:
			\begin{align}\label{eq_c_8}
				& \frac{1}{\sqrt{n}}\sum_{t=p+1}^{n}{\big\{\Delta^{+}_t(\theta_i,\delta,\tau;\tilde{w}_l)-E[\Delta^{+}_t(\theta_i,\delta,\tau;\tilde{w}_l) \vert \mathcal{F}_{t-1}]  \big\}}\\
				& \quad =\frac{1}{\sqrt{n}}\sum_{t=p+1}^{n}{\tilde{w}_{lt}\big\{\eta_t(\delta_0(\tau)+\theta_i,M\delta/\sqrt{n})-E[\eta_t(\delta_0(\tau)+\theta_i, M\delta/\sqrt{n})\vert \mathcal{F}_{t-1}]  \big\} }\nonumber \\
				& \qquad\qquad\qquad - \frac{1}{\sqrt{n}}\sum_{t=p+1}^{n}{\tilde{w}_{lt}\big\{\eta_t(\delta_0(\tau),0)-E[\eta_t(\delta_0(\tau),0)\vert \mathcal{F}_{t-1}]  \big\}. }\nonumber
			\end{align}
			By (\ref{eq_c_7})-(\ref{eq_c_8}) and using a similar procedure for (\ref{eq_b_26_1}), it is easy to prove that
			\begin{align}\label{eq_c_9}
				\sup_{\tau \in \mathcal{T}} \Big \vert \frac{1}{\sqrt{n}}\sum_{t=p+1}^{n}{\big\{\Delta^{+}_t(\theta_i,\delta,\tau;\tilde{w}_l)-E[\Delta^{+}_t(\theta_i,\delta,\tau;\tilde{w}_l) \vert \mathcal{F}_{t-1}]  \big\}}\Big \vert=o_p(1).
			\end{align}
			Therefore, {combining} (\ref{eq_c_6}) and (\ref{eq_c_9}) yields that
			\begin{align}\label{eq_c_10}
				P\Big( \sup_{\theta \in \mathcal{C}_n,\,\tau \in \mathcal{T}} \frac{1}{\sqrt{n}}\sum_{t=p+1}^{n}{\big\{\Delta_t(\theta,\tau;\tilde{w}_l)-E[\Delta_t(\theta,\tau;\tilde{w}_l) \vert \mathcal{F}_{t-1}]  \big\}}>2C_1M\delta \Big) \to 0.
			\end{align}
			Similarly, {we have}
			\begin{align}\label{eq_c_11}
				P\Big( \inf_{\theta \in \mathcal{C}_n,\,\tau \in \mathcal{T}} \frac{1}{\sqrt{n}}\sum_{t=p+1}^{n}{\big\{\Delta_t(\theta,\tau;\tilde{w}_l)-E[\Delta_t(\theta,\tau;\tilde{w}_l) \vert \mathcal{F}_{t-1}]  \big\}}<-2C_1M\delta \Big) \to 0.
			\end{align}
		{Hence}, (\ref{eq_c_2}) follows from (\ref{eq_c_10})-(\ref{eq_c_11}) and the arbitrary of $\delta$.
			
			For (\ref{eq_c_3}), similar to (\ref{eq_a_32})-(\ref{eq_a_35}), it is easy to derive that
			\begin{eqnarray}
				&& \sup_{\theta \in \mathcal{C}_n,\,\tau \in \mathcal{T}}\bigg \vert \frac{1}{\sqrt{n}}\sum_{t=p+1}^{n}{E[\Delta_t(\theta,\tau;\tilde{w}_l) \vert \mathcal{F}_{t-1}]}\nonumber\\
				&& \qquad \qquad \qquad \qquad -\frac{1}{n}\sum_{t=p+1}^{n}{{\tilde{w}_{lt}Z^{\top}_{t-1}} f(\delta^{\top}_0(\tau)Z_{t-1}/\sigma_{t})\sigma^{-1}_{t} } \sqrt{n}\theta  \bigg \vert=O_p(n^{-1/2}),\nonumber\\
				&& \quad \sup_{\tau \in \mathcal{T}}{\bigg\vert \frac{1}{n}\sum_{t=p+1}^{n}{{\tilde{w}_{lt} Z_{t-1}} f(\delta^{\top}_0(\tau)Z_{t-1}/\sigma_{t})\sigma^{-1}_{t} } -b(\tau;\tilde{w}_l)\bigg\vert}=o_p(1).\nonumber
			\end{eqnarray}
			This completes the proof by the fact that $\sqrt{n}\vert\theta\vert<M$.\\
		\end{proof}
		
		\begin{proof}[\textsc Proof of Lemma 6.1]
			By the definition of $\Delta_t(\theta,\tau;\tilde{w}_l)$ in (\ref{eq_c_1}), it is obvious that
			\begin{align}\label{eq_a_37}
				R_t(\tau;\tilde{w}_l)=-\Delta_t(\hat{\theta}_n(\tau)-\theta_0(\tau),\tau;\tilde{w}_l).
			\end{align}
			Then the conclusion holds from Lemma \ref{lemma_c_1} and Theorem 5.1. On the other hand, the second statement can be easily shown by similar proof procedures in Theorem 4.1 and then the details are omitted.\\
		\end{proof}

		\setcounter{theorem}{0}
		\renewcommand{\thetheorem}{E.\arabic{theorem}}
		\setcounter{equation}{0}
		\renewcommand{\theequation}{E.\arabic{equation}}
		\setcounter{lemma}{0}
		\renewcommand{\thelemma}{E.\arabic{lemma}}
		\section{Proofs of Theorems 7.1-7.2}
 Denote
		\begin{align}\label{eq_d_1}
			L^{*}_n(u,\tau)=-u^{\top}T^{*}_n(\tau)+\sum_{t=p+1}^{n}{w^{*}_t\zeta_t(u,\tau)},
		\end{align}
		where $T^{*}_n(\tau)=\sum_{t=p+1}^{n}{w^{*}_t\xi_t(\tau)}$, and $\xi_t(\tau)$ and $\zeta_t(u,\tau)$ are defined in (5.1)-(5.2). Then $\hat{u}^{*}_n(\tau)=\sqrt{n}(\hat{\theta}^{*}_n(\tau)-\theta_0(\tau))$ is the minimizer of $L^{*}_n(u,\tau)$.
		
		We focus on the weak convergence of $T^{*}_n(\tau)$ under the condition that the samples $\{y_t\}$ are given. Furthermore, we rewrite $T^{*}_n(\tau)$ as $T^{*}_n(\tau)=T^{*}_{n1}(\tau)+T^{*}_{n2}(\tau)-T^{*}_{n3}(\tau)$, where
		\begin{align*}
			T^{*}_{n1}(\tau)& =\frac{1}{\sqrt{n}}\sum_{t=p+1}^{n}{w^{*}_tE[w_t Z_{t-1} \psi_{\tau}(\varepsilon_{t}-\delta^{\top}_0(\tau)Z_{t-1})]},\\
			T^{*}_{n2}(\tau)& =\frac{\tau}{\sqrt{n}} \sum_{t=p+1}^{n}{w^{*}_t[w_tZ_{t-1}-E(w_tZ_{t-1})] },\\
			T^{*}_{n3}(\tau)& =\frac{1}{\sqrt{n}}\sum_{t=p+1}^{n}{w^{*}_t\{w_tZ_{t-1}\eta_{t}(\delta_0(\tau),0)-E[w_tZ_{t-1}\eta_{t}(\delta_0(\tau),0)] \}},
		\end{align*}
		where $\eta_{t}(\cdot)$ is defined in (\ref{eq_b_2}). To make the proof process easy to understand, we introduce one well-known a.s. representation theorem presented on page $59$ in \citet{van1996}.
		\begin{theorem}\label{theorem_d_1}
			Suppose that $(\Omega,\mathcal{F},P)$ is a probability space. Let $X_n (n\geq 0)$ be a mapping from sample space $\Omega$ to a general metric space $\mathbb{D}$. Then, if $X_n \rightsquigarrow X_0$ on space $\mathbb{D}$ and $X_0$ is Borel and separable, then there exist a richer probability space $(\tilde{\Omega},\tilde{\mathcal{F}},\tilde{P})$ and a sequence of measurable and perfect mapping $\phi_n$ such that
			\begin{align*}
				(i).\,\tilde{X}_n({\tilde{\omega}}) \mbox{ converges to }\tilde{X}_0(\tilde{\omega}) \mbox{ in space }\mathbb{D}, \mbox{a.s.; }\quad  (ii).\, P(\cdot)=\tilde{P}(\phi^{-1}_n(\cdot)),
			\end{align*}
			where $\tilde{X}_n({\tilde{\omega}})=X_n(\phi_n(\tilde{\omega}))$.
		\end{theorem}
		In the rest of this section, denote $(\Omega,\mathcal{F},P)$ as the original probability space for $\{y_t\}$ and $\{\eta_{t}\}$, and  $(\tilde{\Omega},\tilde{\mathcal{F}},\tilde{P})$ as a richer probability where the corresponding weak convergence on $(\Omega,\mathcal{F},P)$ is transformed to a.s. convergence on $(\tilde{\Omega},\tilde{\mathcal{F}},\tilde{P})$. Meanwhile, let $E^{*}$, $P^{*}$ be the conditional expectation and conditional probability with respect to $\{w^{*}_t\}$. Now, we describe the asymptotic tightness of $T^{*}_{n}(\tau)$ as follows.
		\begin{lemma}\label{lemma_d_1}
			Suppose that all the conditions in Theorem 7.1 hold, then on a richer probability space, for any $\epsilon>0$, we have
			\begin{align*}
				\lim_{\delta\to 0}\limsup_{n \to \infty}{P^{*}(\sup_{\vert \tau_{1}-\tau_{2}\vert<\delta }{\vert T^{*}_{ni}(\tau_{1})-T^{*}_{ni}(\tau_{2}) \vert>\epsilon})}=0,
			\end{align*}
			almost surely, where $i\in \{1,2,3\}$.
		\end{lemma}
		\begin{proof}[\textsc Proof of Lemma \ref{lemma_d_1}]
			Because the proof procedures are close, we mainly show the details for the process $T^{*}_{n3}(\tau)$. Without loss of generality, we still assume that $\{w_t Z_{t-1}\}$ is non-negative as that for Lemma \ref{lemma_b_2}.
			
			For any $\tau_{1}\in \mathcal{T}$ and $\delta \in (0,1)$, let $h=\delta/2^d$ (i.e., $d=d_n$ is a sequence only relies on the sample size) and $\tau^{(j)}=jh$. Then, for any $\tau \in (0,\delta)$, there exists some $\tau^{(j)}$ such that $\vert \tau-\tau^{(j)}\vert\leq h$. So we can imply that
			\begin{align}\label{eq_d_5}
				T^{*}_{n3}(\tau_{1}+\tau)-T^{*}_{n3}(\tau_{1})\leq {G}^{*}_{n3}(\tau_{1}+\tau^{(j)},C_1h;\tau_{1},C_1 h)+{G}^{*}_{n3}(\tau_{1},C_1 h;\tau_{1},0)+C_2h\sqrt{n},
			\end{align}
			where  $C_1$ and $C_2$ are two deterministic positive constants and
			\begin{align}
				G^{*}_{n3}(\tau',r';\tau'',r'')& =\frac{1}{\sqrt{n}}\sum_{t=p+1}^{n}{w^{*}_t \{\Delta_{t3}(\tau',r';\tau'',r'')-E[\Delta_{t3}(\tau',r';\tau'',r'')]\}},\\
				\Delta_{t3}(\tau',r';\tau'',r'') & = w_t Z_{t-1} [\eta_{t}(\delta_0(\tau'),r')-\eta_{t}(\delta_0(\tau''),r'')].
			\end{align}
			Similarly, we have
			\begin{align}\label{eq_d_8}
				T^{*}_{n3}(\tau_{1}+\tau)-T^{*}_{n3}(\tau_{1})\geq {G}^{*}_{n3}(\tau_{1}+\tau^{(j)},-C_1h;\tau_{1},-C_1 h)+{G}^{*}_{n3}(\tau_{1},-C_1 h;\tau_{1},0)-C_2h\sqrt{n}.
			\end{align}
			On the other hand, notice that we have the following decomposition:
			\begin{align}\label{eq_d_9}
				G^{*}_{n3}(\tau',r';\tau'',r'') & =\frac{1}{\sqrt{n}}\sum_{t=p+1}^{n}{(w^{*}_t-1) \{\Delta_{t3}(\tau',r';\tau'',r'')-E[\Delta_{t3}(\tau',r';\tau'',r'')]\}}\nonumber\\
				& \quad + \frac{1}{\sqrt{n}}\sum_{t=p+1}^{n}{\{\Delta_{t3}(\tau',r';\tau'',r'')-E[\Delta_{t3}(\tau',r';\tau'',r'')]\}}\nonumber\\
				& \equiv G^{*}_{n3,0}(\tau',r';\tau'',r'')+G_{n3,1}(\tau',r';\tau'',r'').
			\end{align}
			Thus, by (\ref{eq_d_5}), (\ref{eq_d_8}) and (\ref{eq_d_9}), when $\delta<C_1/2$, we can further derive that
			\begin{align}\label{eq_d_10}
				& \sup_{\tau \in (0,\delta)}{\vert T^{*}_{n3}(\tau_{1}+\tau)-T^{*}_{n3}(\tau_{1}) \vert}\leq \sup_{\vert \tau'-\tau'' \vert<\delta,\vert r'\vert+\vert r''\vert<1/2^{d}}{\{2\vert G_{n3,1}(\tau',r';\tau'',r'') \vert\}}\\
				& \quad +\max_{j}{\{\vert {G}^{*}_{n3,0}(\tau_{1}+\tau^{(j)},C_1h;\tau_{1},C_1 h) \vert+ \vert {G}^{*}_{n3,0}(\tau_{1}+\tau^{(j)},-C_1h;\tau_{1},-C_1 h) \vert\}}\nonumber\\
				& \quad + \{\vert {G}^{*}_{n3,0}(\tau_{1},C_1h;\tau_{1},0) \vert+ \vert {G}^{*}_{n3,0}(\tau_{1},-C_1h;\tau_{1},0) \vert\}+C_2h\sqrt{n}.\nonumber
			\end{align}
			
			For the first term on the right side of (\ref{eq_d_10}), under the condition that $2^d/\sqrt{n}\to \infty$, following the proof procedures for Lemma \ref{lemma_c_1} and Theorem 5.1, it is easy to prove
			\begin{align}\label{eq_d_11}
				\chi_{n3,1}& =\sup_{\tau \in \mathcal{T}, \vert r \vert<1/2^d}{\Big\vert \frac{1}{\sqrt{n}}\sum_{t=p+1}^{n}{\{w_tZ_{t-1}\eta_{t}(\delta_0(\tau),r)-E[w_tZ_{t-1}\eta_{t}(\delta_0(\tau),r)]\}}-T_{n3,1}(\tau) \Big\vert}=o_p(1),\nonumber\\
				T_{n3,1}(\tau)& =\frac{1}{\sqrt{n}}\sum_{t=p+1}^{n}{\{w_tZ_{t-1}\eta_{t}(\delta_0(\tau),0)-E[w_tZ_{t-1}\eta_{t}(\delta_0(\tau),0)]\}}\rightsquigarrow T_{03,1}(\tau),
			\end{align}
			where $T_{03,1}(\tau)$ is some Gaussian limit. Therefore, it follows that
			\begin{align}\label{eq_d_12}
				\sup_{\vert \tau'-\tau'' \vert<\delta,\vert r'\vert +\vert r''\vert<1/2^{-d}}{\vert G_{n3,1}(\tau',r';\tau'',r'') \vert}\leq 2\chi_{n3,1}+\sup_{\vert \tau'-\tau'' \vert<\delta}{\vert T_{n3,1}(\tau')-T_{n3,1}(\tau'') \vert}.
			\end{align}
			
			For the second term on the right side of (\ref{eq_d_10}), notice that $\{w^{*}_t-1\}$ is a sequence of i.i.d. random variables satisfying the radematcher distribution. Then, for any $q>2$, it follows that
			\begin{align}\label{eq_d_13}
				E^{*}\vert G^{*}_{n3,0}(\tau',r';\tau'',r'') \vert^q\leq C_q \Big\{\frac{1}{n} \sum_{t=p+1}^{n}{\{\Delta^2_{t3}(\tau',r';\tau'',r'')+E[\Delta^2_{t3}(\tau',r';\tau'',r'')]\}}  \Big\}^{q/2}.
			\end{align}
			Furthermore, note that
			\begin{align}\label{eq_d_14}
				& \Delta^2_{t3}(\tau',r';\tau'',r'')=w^2_t Z^2_{t-1}[I(\varepsilon_{t} \leq \delta^{\top}_0(\tau')Z_{t-1}+r'\vert Z_{t-1}\vert)+I(\varepsilon_{t} \leq \delta^{\top}_0(\tau'')Z_{t-1}+r''\vert Z_{t-1}\vert )\nonumber\\
				& \qquad \qquad -2 I(\varepsilon_{t}\leq (\delta^{\top}_0(\tau')Z_{t-1}+r'\vert Z_{t-1}\vert )\wedge (\delta^{\top}_0(\tau'')Z_{t-1}+r''\vert Z_{t-1}\vert))].
			\end{align}
			Then, using the expression (\ref{eq_d_14}) and by the standard argument as that in Lemma \ref{lemma_c_1} (In this case, we should set the ball radius in the covering number theory as $\delta_n$ instead of a constant in Lemma \ref{lemma_c_1}), it is not hard to show that, there exists a positive number $\xi_0\in (0,1/2)$ such that
			\begin{align}\label{eq_d_15}
				\chi_{n3,0}& =\sup_{\tau',\tau''\in \mathcal{T};r'r''\in [-M,M]} \Big \vert \frac{1}{n^{1-\xi_{0}}} \sum_{t=p+1}^{n}{\{\Delta^2_{t3}(\tau',r';\tau'',r'')-E[\Delta^2_{t3}(\tau',r';\tau'',r'')] \}}\Big\vert\nonumber\\
				& =o_p(1),
			\end{align}
			where $M$ is any positive real number. Meanwhile, it is obvious to check that
			\begin{align}\label{eq_d_16}
				E[\Delta^2_{t3}(\tau',r';\tau'',r'')]\leq C_3 [\vert \tau'-\tau'' \vert+\vert r'-r'' \vert].
			\end{align}
			So (\ref{eq_d_13}), (\ref{eq_d_15}) and (\ref{eq_d_16}) imply that
			\begin{align}\label{eq_d_17}
				E^{*}\vert G^{*}_{n3,0}(\tau',r';\tau'',r'') \vert^q\leq  C_q \{\vert \tau'-\tau''\vert^{q/2}+\vert r'-r''\vert^{q/2}+[\chi_{n3,0}n^{-\xi_{0}}]^{q/2}  \}.
			\end{align}
			Therefore, by using the proof procedure in Lemma \ref{lemma_b_2}, (\ref{eq_d_17}) indicates that
			\begin{align}\label{eq_d_18}
				E^{*}[\max_{j}{\vert {G}^{*}_{n3,0}(\tau_{1}+\tau^{(j)},\pm C_1h;\tau_{1},\pm C_1 h) \vert^q}]\leq C_q [\delta^{q/2}+(\chi_{n3,0} n^{-\xi_{0}})^{q/2}2^d].
			\end{align}
			
			For the third one on the right side of (\ref{eq_d_10}), (\ref{eq_d_17}) derives that
			\begin{align}\label{eq_d_19}
				E^{*}[{\vert {G}^{*}_{n3,0}(\tau_{1},\pm C_1h;\tau_{1},0) \vert^q}]\leq C_q [2^{-dq/2}+(\chi_{n3,0} n^{-\xi_{0}})^{q/2}].
			\end{align}
			
			Notice that the upper bounds in (\ref{eq_d_12}), (\ref{eq_d_18}) and (\ref{eq_d_19}) are independent with $\tau_{1}$ and only rely on $\delta$ and $n$. Then, by the symmetry of the interval $(0,\delta)$ and $(-\delta,0)$, we can get the same conclusion as that in (\ref{eq_d_10})-(\ref{eq_d_19}). Thus, we actually have shown that
			\begin{align}\label{eq_d_20}
				& P^{*}(\sup_{\vert \tau_1-\tau_{2}\vert <\delta}{\vert T^{*}_{n3}(\tau_{1})-T^{*}_{n3}(\tau_{2}) \vert>\epsilon})\leq \frac{C_q}{\delta\epsilon^q}[\delta^{q/2}+(\chi_{n3,0}n^{-\xi_{0}})^{q/2}2^d+2^{-qd/2}]\nonumber\\
				& \qquad + \frac{2}{\delta}P^{*}(6\chi_{n3,1}+2\sup_{\vert \tau'-\tau'' \vert<\delta}{\vert T_{n3,1}(\tau')-T_{n3,1}(\tau'') \vert}+2C_2\sqrt{n}2^{-d}>\epsilon/4).
			\end{align}
			
			Let $\mathbb{D}=l^{\infty}(\mathcal{T})\times \mathbb{R}^2$ be the product space, then by Theorem \ref{theorem_d_1} and (\ref{eq_d_11}) and (\ref{eq_d_15}),  on a richer probability space $(\tilde{\Omega},\tilde{\mathcal{F}},\tilde{P})$, for almost every $\tilde{\omega} \in \tilde{\Omega}$, we have
			\begin{align}\label{eq_d_21}
				(T_{n3,1}(\tau;\tilde{\omega}),\chi_{n3,0}(\tilde{\omega}),\chi_{n3,1}(\tilde{\omega})) \mbox{ converges to } (T_{03,1}(\tau),0,0).
			\end{align}
			So, by the uniform metric definition in $l^{\infty}(\mathcal{T})$, we must can find a large $n(\tilde{\omega})$ and $\delta(\tilde{\omega})$ such that, for any $n >n(\tilde{\omega})$,
			\begin{align}\label{eq_d_22}
				6\chi_{n3,1}+2\sup_{\vert \tau'-\tau'' \vert<\delta(\tilde{\omega})}{\vert T_{n3,1}(\tau')-T_{n3,1}(\tau'') \vert}+2C_2\sqrt{n}2^{-d}<\epsilon/4,
			\end{align}
			where we choose $d=[\log_2(n)]+1$.
			
			On the other hand, we only need choose $q=4/\xi_{0}+2$, then by (\ref{eq_d_20})-(\ref{eq_d_22}), for every $\tilde{\omega}\in \tilde{\Omega}$, when $\delta<\delta(\tilde{\omega})$, we have
			\begin{align}
				\limsup_{n \to \infty}{P^{*}(\sup_{\vert \tau_1-\tau_{2}\vert <\delta}{\vert T^{*}_{n3}(\tau_{1})-T^{*}_{n3}(\tau_{2}) \vert>\epsilon})}\leq C_q\delta^{q/2-1}\epsilon^{-q}.
			\end{align}
			This implies that Lemma \ref{lemma_d_1} holds for $T^{*}_{n3}(\tau)$.
			
			For $T^{*}_{n1}(\tau)$ and $T^{*}_{n2}(\tau)$, we only need to notice that
			\begin{align}
				T_{n1,1}(\tau)&=\frac{1}{\sqrt{n}}\sum_{t=p+1}^{n}{E[w_t Z_{t-1} \psi_{\tau}(\varepsilon_{t}-\delta^{\top}_0(\tau)Z_{t-1})]}\rightsquigarrow 0,\label{eq_d_24}\\
				T_{n2,1}(\tau)&=\frac{1}{\sqrt{n}} \sum_{t=p+1}^{n}{[w_tZ_{t-1}-E(w_tZ_{t-1})] }\rightsquigarrow T_{02,1},\label{eq_d_25}
			\end{align}
			where $T_{02,1}$ is a Gaussian random variable. Then following the same proof process as that for $T^{*}_{n3}(\tau)$, the conclusion holds. This completes the whole proof.\\
		\end{proof}
		\begin{lemma}\label{lemma_d_2}
			Suppose that all the conditions in Theorem 7.1 hold, then
			\begin{align*}
				\sup_{\tau\in \mathcal{T}}{\Big\vert \sum_{t=p+1}^{n}{w^{*}_t\zeta_t(u,\tau)}-u^{\top}\Sigma(\tau)u/2 \Big\vert }=o^{*}_p(1)
			\end{align*}
			holds for every $u \in \mathbb{Q}^{p+1}$ at the same time, almost surely on a richer probability space.
		\end{lemma}
		\begin{proof}[\textsc Proof of Lemma \ref{lemma_d_2}]
		By (\ref{eq_a_32}), it is clear that
		\begin{align}\label{eq_d_26}
			&\sum_{t=p+1}^{n}{w^{*}_tE[\zeta_t(u,\tau)\vert \mathcal{F}_{t-1}]}=u^{\top}\times \Big[\frac{1}{2n}\sum_{t=p+1}^{n}{w^{*}_t\frac{w_t}{\sigma_{t}}Z_{t-1} Z^{\top}_{t-1}f\Big(\frac{\delta^{\top}_0(\tau)Z_{t-1} }{\sigma_{t}} \Big) }   \Big]\times u\nonumber\\
			& \quad + \sum_{t=p+1}^{n}{ w^{*}_t \frac{w_t}{\sigma_{t}} \int_{0}^{ \frac{u^{\top}Z_{t-1}}{\sqrt{n} } }{ s \Big[ f\Big(\frac{\delta^{\top}_0(\tau)Z_{t-1} }{\sigma_{t}}+\Delta s \Big)-f\Big(\frac{\delta^{\top}_0(\tau)Z_{t-1} }{\sigma_{t}} \Big) \Big] ds}   }.
		\end{align}
		By the boundedness of $w^{*}_t$, we have
		\begin{align}\label{eq_d_27}
			\sum_{t=p+1}^{n}{\Big \vert  \frac{w^{*}_tw_t}{\sigma_{t}} \int_{0}^{ \frac{u^{\top}Z_{t-1}}{\sqrt{n} } }{ s \Big[ f\Big(\frac{\delta^{\top}_0(\tau)Z_{t-1} }{\sigma_{t}}+\Delta s \Big)-f\Big(\frac{\delta^{\top}_0(\tau)Z_{t-1} }{\sigma_{t}} \Big) \Big] ds}   \Big \vert}\leq \frac{C_1\vert u\vert^3}{\sqrt{n}}.
		\end{align}
		On the other hand, it is easy to see that
		\begin{align}\label{eq_d_28}
			E^{*}\Big\vert \frac{1}{2n}\sum_{t=p+1}^{n}{(w^{*}_t-1)\frac{w_t}{\sigma_{t}}Z_{t-1} Z^{\top}_{t-1}f\Big(\frac{\delta^{\top}_0(\tau)Z_{t-1} }{\sigma_{t}} \Big) }   \Big\vert^2\leq C_2 n^{-1}.
		\end{align}
		By Theorem \ref{theorem_d_1}, on a richer probability space, (\ref{eq_a_35}) holds almost surely and thus by (\ref{eq_d_27})-(\ref{eq_d_28}), we have
		\begin{align}\label{eq_d_29}
			\sup_{\tau\in \mathcal{T}}{\Big\vert \sum_{t=p+1}^{n}{w^{*}_tE[\zeta_t(u,\tau)\vert \mathcal{F}_{t-1}]}-u^{\top}\Sigma(\tau)u/2\Big\vert }=o^{*}_p(1).
		\end{align}
		
		On the other hand, on a richer probability space, Lemma \ref{lemma_b_4} holds almost surely for all $u\in \mathbb{Q}^{p+1}$ (i.e., For every sample point $\tilde{w}$ in a measurable set with probability one, we have for any $u\in \mathbb{Q}^{p+1}$, $\sup_{\tau \in \mathcal{T}}{\vert \alpha_n(u,\tau) \vert}\to 0$). Then using the procedure as that in Lemma \ref{lemma_d_1}, it is not hard to show that
		\begin{align}\label{eq_d_30}
			\sup_{\tau\in \mathcal{T}}{\vert  \sum_{t=p+1}^{n}{w^{*}_t\{\zeta_t(u,\tau)-E[\zeta_t(u,\tau)\vert \mathcal{F}_{t-1}]\}} \vert}=o^{*}_p(1)
		\end{align}
		almost surely. So the conclusion follows from  (\ref{eq_d_29}) and (\ref{eq_d_30}).
		\end{proof}
		\begin{proof}[\textsc Proof of Theorem 7.1]
			By a similar process for (\ref{eq_d_15}), we can easily show that
			\begin{align}\label{eq_d_32}
				& \sup_{\tau_1,\tau_{2} \in \mathcal{T}}\bigg\{\Big\vert \frac{1}{n}\sum_{t=p+1}^{n}{\{w^2_tZ_{t-1} Z^{\top}_{t-1}\psi_{\tau_1}(\varepsilon_{t}-\delta^{\top}_0(\tau_1)Z_{t-1})\psi_{\tau_2}(\varepsilon_{t}-\delta^{\top}_0(\tau_2)Z_{t-1})\}}\nonumber \\
				& \qquad \qquad \qquad \qquad \qquad \quad-\int_{0}^{1}{\mbox{E}(\xi_{0}(\tau_{1},s)\xi^{\top}_{0}(\tau_{2},s)) ds} \Big\vert\bigg\}=o_p(1).
			\end{align}
			Review that the proofs in Theorem 5.1 actually imply that
			\begin{align}\label{eq_d_33}
				\sqrt{n}(\hat{\theta}_n(\tau)-\theta_0(\tau))=\Sigma^{-1}(\tau)T_n(\tau)+o_p(1).
			\end{align}
			Therefore, by carefully checking the stochastic processes involved in the original samples $\{y_t\}$, (i.e.(\ref{eq_d_21}), (\ref{eq_d_24}), (\ref{eq_d_25}), (\ref{eq_d_32}) and (\ref{eq_a_35}) and Lemma \ref{lemma_b_4}), we can find one enough rich probability space, such that in this new probability space, Lemmas \ref{lemma_d_1}-\ref{lemma_d_2}, (\ref{eq_d_32})-(\ref{eq_d_33}), and the result that $T_n(\tau)$ converges to $T_0(\tau)$ hold simultaneously, almost surely. That is, for almost every fixed sample point $\tilde{\omega}$ in the new probability space, the deterministic elements only relying on $\{y_t\}$ have the following convergence:
			\begin{align}
			& \sup_{\tau \in \mathcal{T}}{\vert T_n(\tau)-T_0(\tau)\vert} \longrightarrow 0,\label{eq_d_31_new_new}\\
			& \sup_{\tau \in \mathcal{T}}{\vert \sqrt{n}(\hat{\theta}_n(\tau)-\theta_0(\tau))-\Sigma^{-1}(\tau)T_n(\tau)\vert} \longrightarrow 0,\label{eq_d_31_new}\\
		    & \sup_{\tau_1,\tau_{2} \in \mathcal{T}}\bigg\{\Big\vert \frac{1}{n}\sum_{t=p+1}^{n}{\{w^2_tZ_{t-1} Z^{\top}_{t-1}\psi_{\tau_1}(\varepsilon_{t}-\delta^{\top}_0(\tau_1)Z_{t-1})\psi_{\tau_2}(\varepsilon_{t}-\delta^{\top}_0(\tau_2)Z_{t-1})\}}\nonumber \\
			& \qquad \qquad \qquad \qquad \qquad -\int_{0}^{1}{\mbox{E}(\xi_{0}(\tau_{1},s)\xi^{\top}_{0}(\tau_{2},s)) ds} \Big\vert\bigg\}\longrightarrow 0,\label{eq_d_33_new}
			\end{align}
			and  the random elements generated by $\{w^{*}_t\}$ have the followng weak convergences:
			\begin{align}
				& T^{*}_n(\tau) \mbox{ is asymptotically tightness, }\label{eq_d_34_new}\\
				& 	\sup_{\tau\in \mathcal{T}}{\Big\vert \sum_{t=p+1}^{n}{w^{*}_t\zeta_t(u,\tau)}-u^{\top}\Sigma(\tau)u/2 \Big\vert }=o^{*}_p(1).\label{eq_d_35_new}
			\end{align}
			So, by the central limit theorem, (\ref{eq_d_31_new_new}), (\ref{eq_d_33_new}) and (\ref{eq_d_34_new}) further indicate that for almost every fixed sample point $\tilde{\omega}$, we have
			\begin{align}
				T^{*}_n(\tau)-T_n(\tau)=\frac{1}{\sqrt{n}}\sum_{t=p+1}^{n}{(w^{*}_t-1)w_tZ_{t-1}\psi_{\tau}(\varepsilon_{t}-\delta^{\top}_0(\tau)Z_{t-1})} \rightsquigarrow T^{*}_0(\tau).\label{eq_d_36_new}
			\end{align}
			Furthermore, by (\ref{eq_d_31_new_new})--(\ref{eq_d_31_new}), (\ref{eq_d_35_new}) and (\ref{eq_d_36_new}) and using the convex argument in \citet{kato2009}, it follows that
			\begin{align}\label{eq_d_34}
			\sqrt{n}(\hat{\theta}^{*}_n(\tau)-{\theta}_0(\tau)) & = \Sigma^{-1}(\tau)\times T^{*}_n(\tau)+o^{*}_p(1),\nonumber \\
				\sqrt{n}(\hat{\theta}^{*}_n(\tau)-\hat{\theta}_n(\tau)) & =\Sigma^{-1}(\tau)\times \frac{1}{\sqrt{n}}\sum_{t=p+1}^{n}{(w^{*}_t-1)w_tZ_{t-1}\psi_{\tau}(\varepsilon_{t}-\delta^{\top}_0(\tau)Z_{t-1})}+o^{*}_p(1)\nonumber \\
				& {\rightsquigarrow}\Sigma^{-1}(\tau)T^{*}_0(\tau).
			\end{align}
		Finally, notice that the limiting distribution $T^{*}_0(\tau)$ does not rely on the sample point $\tilde{\omega}$, then by the statement (ii) in Theorem \ref{theorem_d_1} and the independence of $\{w^{*}_t\}$ and $\{y_t\}$, the result (\ref{eq_d_34}) can then be transformed to hold in probability for the original sample space. This completes the whole proof.
		\end{proof}
		\begin{proof}[\textsc Proof of Theorem 7.2]
			Similar to that in Section 6, denote
			\begin{align}\label{eq_d_35}
				r^{*}_{t0}(\tau;\tilde{w}_l)=w^{*}_t\tilde{w}_{lt} \psi_{\tau}(y_t-Z^{\top}_{t-1}{\theta}_0(\tau)).
			\end{align}
			Then, by the proof processes as those in Lemma \ref{lemma_d_1} and Lemma \ref{lemma_c_1} and using the representations in (\ref{eq_d_31_new_new})-(\ref{eq_d_34}), we can easily show that on a richer probability space,
			\begin{align*}
				& \sup_{\tau \in \mathcal{T}}{\Big\vert \frac{1}{\sqrt{n}}\sum_{t=p+1}^{n}{[r^{*}_t(\tau;\tilde{w}_l)-r^{*}_{t0}(\tau;\tilde{w}_l)]} + b^{\top}(\tau;\tilde{w}_l)\sqrt{n}(\hat{\theta}^{*}_n(\tau)-\theta_0(\tau))\Big \vert}=o^{*}_p(1),\nonumber\\
				& \sup_{\tau\in \mathcal{T}}{\Big\vert \frac{1}{\sqrt{n}}\sum_{t=p+1}^{n}{r^{*}_{t0}(\tau;\tilde{w}_l)}-\sqrt{n}\tilde{g}(\tau;\tilde{w}_l)-[\tilde{T}_n(\tau;\tilde{w}_l)-E\tilde{T}_n(\tau;\tilde{w}_l)]-\tilde{T}^{*}_n(\tau;\tilde{w}_l)\Big\vert}=o^{*}_p(1),
			\end{align*}
			where $\tilde{T}^{*}_n(\tau;\tilde{w}_l)=n^{-1/2}\sum_{t=p+1}^{n}{[(w^{*}_t-1)\tilde{w}_{lt}\psi_{\tau}(y_t-Z^{\top}_{t-1}\theta_0(\tau))]}$ converges weakly to a Gaussian process. By combining the proof procedures for Theorem 6.1, we can further get that
			\begin{align}
				\sqrt{n}(\hat{\tau}^{*}_n-\tau_{0})& =\sum_{l\in \mathcal{A}}{a_l \{ \tilde{S}_n(\tau_{0};\tilde{w}_l)+\tilde{S}^{*}_n(\tau_{0};\tilde{w}_l)\}}+o^{*}_p(1),\nonumber\\
				\tilde{S}^{*}_n(\tau;\tilde{w}_l)&=\tilde{T}^{*}_n(\tau;\tilde{w}_l)-b^{\top}(\tau;\tilde{w}_l)\times \sqrt{n}(\hat{\theta}^{*}_n(\tau)-\hat{\theta}_n(\tau)).
			\end{align}
			Thus, by (A.30), (\ref{eq_d_34}) and the central limit theorem, we have
			\begin{align}\label{eq_d_37}
				\sqrt{n}(\hat{\tau}^{*}_n-\hat{\tau}_n) & =\sum_{l\in \mathcal{A}}{a_l \tilde{S}^{*}_n(\tau_{0};\tilde{w}_l)}+o^{*}_p(1),\nonumber\\
				& =\frac{1}{\sqrt{n}}\sum_{t=p+1}^{n}{\Big[(w^{*}_t-1)\Big\{\sum_{l\in \mathcal{A}}{a_l[ \tilde{w}_{lt}-b^{\top}(\tau_{0};\tilde{w}_l)\Sigma^{-1}(\tau_{0})w_tZ_{t-1}] }\Big\}\psi_{\tau_{0}}(\varepsilon_{t})\Big]}+o^{*}_p(1),\nonumber\\
				& \overset{d}{\longrightarrow} N(0,\gamma^2_1),
			\end{align}
			where we use the fact that $\{\psi_{\tau_{0}}(\varepsilon_{t})\}$ is a  martingale difference sequence. Furthermore, we can easily show that
			\begin{align}\label{eq_d_38}
				\sqrt{n}(\hat{\theta}^{*}_n-\theta_0)=\sqrt{n}(\hat{\theta}^{*}_n(\tau_{0})-\theta_0 )+\frac{\partial \delta_0(\tau_{0})}{\partial\tau}\sqrt{n}(\hat{\tau}^{*}_n-\tau_{0})+o^{*}_p(1).
			\end{align}
			So (A.32) and (\ref{eq_d_38}) imply that
			\begin{align}\label{eq_d_39}
				\sqrt{n}(\hat{\theta}^{*}_n-\hat{\theta}_n)&=\sqrt{n}(\hat{\theta}^{*}_n(\tau_{0})-\hat{\theta}_n(\tau_{0}))+\frac{\partial \delta_0(\tau_{0})}{\partial\tau}\sqrt{n}(\hat{\tau}^{*}_n-\hat{\tau}_{n})+o^{*}_p(1),\nonumber\\
				& \overset{d}{\longrightarrow} N(0,\Gamma_1).
			\end{align}
			Finally, the conclusion holds from (\ref{eq_d_37}) and (\ref{eq_d_39}). 		
		\end{proof}
		
		\setcounter{theorem}{0}
		\renewcommand{\thetheorem}{F.\arabic{theorem}}
		\setcounter{equation}{0}
		\renewcommand{\theequation}{F.\arabic{equation}}
		\setcounter{lemma}{0}
		\renewcommand{\thelemma}{F.\arabic{lemma}}
		\setcounter{table}{0}
		\renewcommand{\thetable}{F.\arabic{table}}

		\section{Additional Simulation Results}
		In this section, we still focus on the model (8.1) with the same parameteric settings as those in Section 8. Initially, we investigate the finite sample performance of the SQE $\hat{\theta}_n(\tau_1)$ when $\tau_{1}\neq \tau_{0}$. For $\delta \in \{0.1,0.2,0.3\}$, we choose $\tau_{1}=\tau_{0}+\delta$ when $\tau_{0}=0.3,0.5$ and $\tau_{1}=\tau_{0}-\delta$ when $\tau_{0}=0.7$. The sample biases (i.e., $\hat{\theta}_n(\tau_1)-\theta_0$) are reported in Tables \ref{table_e_1}--\ref{table_e_2}. It is clear to see that all the biases are much larger than those of the proposed estimator $\hat{\theta}_n$ as shown in Tables 1--2. Meanwhile, the biases are very stable as the sample size is increasing. The phenomenon shows that the traditional SQE $\hat{\theta}_n(\tau)$ in (3.5) is inconsistent to the true parameter $\theta_0$ when $\tau\neq \tau_{0}$.
		Subsequently, we report the empirical coverage probabilities of the confidence intervals for the parameters $\tau_{0}$, $\mu_0$, and $\phi_{10}$
		in Tables \ref{table_e_3}--\ref{table_e_4}. We observe that the empirical coverage probabilities consistently approximate the nominal values, regardless of the specific nominal level chosen. This result demonstrates that the proposed method provides reliable inference for the parameters of interest.
		
		\begin{table}[H]
			\centering
			\small
			\caption{\footnotesize Biases of $\hat{\theta}_n(\tau_1)$ for model (8.1) with $v(x)\sim $ (8.3)}
			\scalebox{0.85}{
				\begingroup
				\setlength{\tabcolsep}{3.5pt} 
				\renewcommand{\arraystretch}{1} 
				\begin{tabular}{ccccccccccccc}
					\hline
					\quad & \quad & \quad & \quad & \quad & \quad & $\hat{\mu}_n(\tau_{1})$ & \quad & \quad & \quad & $\hat{\phi}_n(\tau_{1})$ & \quad & \quad\\
					\cline{7-9}\cline{11-13}
					$\eta_{t0}\sim $ & \quad & $\tau_{0}$ & \quad & $n$ & \quad & $\delta=0.1$ & $\delta=0.2$ & $\delta=0.3$ & \quad & $\delta=0.1$ & $\delta=0.2$ & $\delta=0.3$\\\hline
					$N(0,1)$ & \quad & 0.3 & \quad & 1000 & \quad & 0.0383 & 0.0722 & 0.1093 & \quad & 0.0443 & 0.0917 & 0.1358\\
					\quad & \quad & \quad & \quad & 2000 & \quad & 0.0377 & 0.0724 & 0.1096 & \quad & 0.0457 & 0.0906 & 0.1339 \\
					\quad & \quad & 0.5 & \quad & 1000 & \quad & 0.0571 & 0.1188 & 0.1984 & \quad & -0.0172 & -0.0294 & -0.0447\\
					\quad & \quad & \quad & \quad & 2000 & \quad & 0.0561 & 0.1188 & 0.1995 & \quad & -0.0139 & -0.0276 & -0.0454 \\
					\quad & \quad & 0.7 & \quad & 1000 & \quad & -0.0619 & -0.1213 & -0.1821 & \quad & 0.0674 & 0.1312 & 0.1960\\
					\quad & \quad & \quad & \quad & 2000 & \quad & -0.0618 & -0.1210 & -0.1811 & \quad & 0.0687 & 0.1352 & 0.1953 \\\hline
					$t_3$ & \quad & 0.3 & \quad & 1000 & \quad & 0.0471 & 0.0895 & 0.1332 & \quad & 0.0133 & 0.0278 & 0.0424 \\
					\quad & \quad & \quad & \quad & 2000 & \quad & 0.0468 & 0.0885 & 0.1331 & \quad & 0.0148 & 0.0301 & 0.0429 \\
					\quad & \quad & 0.5 & \quad & 1000 & \quad & 0.0484 & 0.1031 & 0.1793 & \quad & -0.0135 & -0.0232 & -0.0348 \\
					\quad & \quad & \quad & \quad & 2000 & \quad & 0.0479 & 0.1029 & 0.1792 & \quad & -0.0101 & -0.0209 & -0.0368 \\
					\quad & \quad & 0.7 & \quad & 1000 & \quad & -0.0480 & -0.0917 & -0.1370 & \quad & 0.0297 & 0.0568 & 0.0849\\
					\quad & \quad & \quad & \quad & 2000 & \quad & -0.0479 & -0.0917 & -0.1368 & \quad & 0.0306 & 0.0590 & 0.0872 \\\hline
				\end{tabular}
				\endgroup
			}\label{table_e_1}
		\end{table}
		
		\begin{table}[H]
			\centering
			\small
			\caption{\footnotesize Biases of $\hat{\theta}_n(\tau_1)$ for model (8.1) with $v(x)\sim $ (8.4)}
			\scalebox{0.85}{
				\begingroup
				\setlength{\tabcolsep}{3.5pt} 
				\renewcommand{\arraystretch}{1} 
				\begin{tabular}{ccccccccccccc}
					\hline
					\quad & \quad & \quad & \quad & \quad & \quad & $\hat{\mu}_n(\tau_{1})$ & \quad & \quad & \quad & $\hat{\theta}_n(\tau_{1})$ & \quad & \quad\\
					\cline{7-9}\cline{11-13}
					$\eta_{t0}\sim $ & \quad & $\tau_{0}$ & \quad & $n$ & \quad & $\delta=0.1$ & $\delta=0.2$ & $\delta=0.3$ & \quad & $\delta=0.1$ & $\delta=0.2$ & $\delta=0.3$\\\hline
					$N(0,1)$ & \quad & 0.3 & \quad & 1000 & \quad & 0.0699 & 0.1362 & 0.2048 & \quad & 0.0283 & 0.0569 & 0.0814\\
					\quad & \quad & \quad & \quad & 2000 & \quad & 0.0711 & 0.1353 & 0.2038 & \quad & 0.0275 & 0.0570 & 0.0836\\
					\quad & \quad & 0.5 & \quad & 1000 & \quad & 0.0858 & 0.1796 & 0.2979 & \quad & -0.0178 & -0.0321 & -0.0520\\
					\quad & \quad & \quad & \quad & 2000 & \quad & 0.0858 & 0.1796 & 0.2968 & \quad & -0.0163 & -0.0335 & -0.0527\\
					\quad & \quad & 0.7 & \quad & 1000 & \quad & -0.0894 & -0.1734 & -0.2604 & \quad & 0.0516 & 0.1038 & 0.1536\\
					\quad & \quad & \quad & \quad & 2000 & \quad & -0.0892 & -0.1728 & -0.2588 &\quad & 0.0534 & 0.1048 & 0.1549\\\hline
					$t_3$ & \quad & 0.3 & \quad & 1000 & \quad & 0.0777 & 0.1506 & 0.2222 & \quad & 0.0054 & 0.0073 & 0.0131\\
					\quad & \quad & \quad & \quad & 2000 & \quad & 0.0782 & 0.1487 & 0.2209 & \quad & 0.0049 & 0.0102 & 0.0149\\
					\quad & \quad & 0.5 & \quad & 1000 & \quad & 0.0731 & 0.1569 & 0.2693 & \quad & -0.0139 & -0.0262 & -0.0417\\
					\quad & \quad & \quad & \quad & 2000 & \quad & 0.0734 & 0.1560 & 0.2673 & \quad & -0.0135 & -0.0258 & -0.0401\\
					\quad & \quad & 0.7 & \quad & 1000 & \quad & -0.0717 & -0.1368 & -0.2040 & \quad & 0.0202 & 0.0435 & 0.0667\\
					\quad & \quad & \quad & \quad & 2000 & \quad & -0.0719 & -0.1367 & -0.2039 & \quad & 0.0231 & 0.0472 & 0.0668\\\hline
				\end{tabular}
				\endgroup
			}\label{table_e_2}
		\end{table}

	\begin{table}[H]
		\centering
		\small
		\caption{\footnotesize Empirical coverage probabilities of $\hat{\tau}_n$ and $\hat{\theta}_n$ for model (8.1) with $v(x)\sim $ (8.3)}
		\scalebox{0.85}{
			\begingroup
			\setlength{\tabcolsep}{3.5pt} 
			\renewcommand{\arraystretch}{1} 
			\begin{tabular}{ccccccccccccc}
				\hline
				\quad & \quad & \quad & \quad & \quad & \quad & $0.90$ & \quad & \quad & \quad & $0.95$ & \quad & \quad\\
				\cline{7-9}\cline{11-13}
				$\eta_{t0}\sim $ & \quad & $\tau_{0}$ & \quad & $n$ & \quad & $\hat{\tau}_n$ & $\hat{\mu}_n$ & $\hat{\phi}_n$ & \quad & $\hat{\tau}_n$ & $\hat{\mu}_n$ & $\hat{\phi}_n$\\\hline
				$N(0,1)$ & \quad & 0.3 & \quad & 1000 & \quad & 0.901 & 0.927 & 0.895 & \quad & 0.940 & 0.968 & 0.941\\
				\quad & \quad & \quad & \quad & 2000 & \quad & 0.897 & 0.919 & 0.901 & \quad & 0.944 & 0.957 & 0.949\\
				\quad & \quad & 0.5 & \quad & 1000 & \quad & 0.922 & 0.931 & 0.902 & \quad & 0.965 & 0.967 & 0.954\\
				\quad & \quad & \quad & \quad & 2000 & \quad & 0.911 & 0.923 & 0.898 & \quad & 0.951 & 0.956 & 0.954\\
				\quad & \quad & 0.7 & \quad & 1000 & \quad & 0.909 & 0.923 & 0.929 & \quad & 0.951 & 0.961 & 0.964\\
				\quad & \quad & \quad & \quad & 2000 & \quad & 0.917 & 0.913 & 0.916 & \quad & 0.962 & 0.969 & 0.957\\\hline
				$t_3$ & \quad & 0.3 & \quad & 1000 & \quad & 0.896 & 0.922 & 0.926 & \quad & 0.930 & 0.967 & 0.966\\
				\quad & \quad & \quad & \quad & 2000 & \quad & 0.917 & 0.902 & 0.916 & \quad & 0.958 & 0.963 & 0.957\\
				\quad & \quad & 0.5 & \quad & 1000 & \quad & 0.913 & 0.930 & 0.904 & \quad & 0.962 & 0.970 & 0.952\\
				\quad & \quad & \quad & \quad & 2000 & \quad & 0.911 & 0.902 & 0.911 & \quad & 0.951 & 0.950 & 0.961\\
				\quad & \quad & 0.7 & \quad & 1000 & \quad & 0.890 & 0.916 & 0.927 & \quad & 0.930 & 0.949 & 0.966\\
				\quad & \quad & \quad & \quad & 2000 & \quad & 0.913 & 0.903 & 0.917 & \quad & 0.944 & 0.954 & 0.951\\\hline
			\end{tabular}
			\endgroup
		}\label{table_e_3}
	\end{table}

		\begin{table}[H]
			\centering
			\small
			\caption{\footnotesize Empirical coverage probabilities of $\hat{\tau}_n$ and $\hat{\theta}_n$ for model (8.1) with $v(x)\sim $ (8.4)}
			\scalebox{0.85}{
				\begingroup
				\setlength{\tabcolsep}{3.5pt} 
				\renewcommand{\arraystretch}{1} 
				\begin{tabular}{ccccccccccccc}
					\hline
					\quad & \quad & \quad & \quad & \quad & \quad & $0.90$ & \quad & \quad & \quad & $0.95$ & \quad & \quad\\
					\cline{7-9}\cline{11-13}
					$\eta_{t0}\sim $ & \quad & $\tau_{0}$ & \quad & $n$ & \quad & $\hat{\tau}_n$ & $\hat{\mu}_n$ & $\hat{\phi}_n$ & \quad & $\hat{\tau}_n$ & $\hat{\mu}_n$ & $\hat{\phi}_n$\\\hline
					$N(0,1)$ & \quad & 0.3 & \quad & 1000 & \quad & 0.912 & 0.916 & 0.909 & \quad & 0.946 & 0.960 & 0.959\\
					\quad & \quad & \quad & \quad & 2000 & \quad & 0.910 & 0.910 & 0.911 & \quad & 0.953 & 0.958 & 0.963\\
					\quad & \quad & 0.5 & \quad & 1000 & \quad & 0.923 & 0.929 & 0.920 & \quad & 0.969 & 0.966 & 0.961\\
					\quad & \quad & \quad & \quad & 2000 & \quad & 0.909 & 0.917 & 0.906 & \quad & 0.954 & 0.958 & 0.960\\
					\quad & \quad & 0.7 & \quad & 1000 & \quad & 0.909 & 0.913 & 0.916 & \quad & 0.951 & 0.956 & 0.965\\
					\quad & \quad & \quad & \quad & 2000 & \quad & 0.898 & 0.911 & 0.912 & \quad & 0.936 & 0.947 & 0.956\\\hline
					$t_3$ & \quad & 0.3 & \quad & 1000 & \quad & 0.901 & 0.923 & 0.933 & \quad & 0.943 & 0.955 & 0.965\\
					\quad & \quad & \quad & \quad & 2000 & \quad & 0.907 & 0.917 & 0.913 & \quad & 0.951 & 0.960 & 0.963\\
					\quad & \quad & 0.5 & \quad & 1000 & \quad & 0.929 & 0.938 & 0.930 & \quad & 0.968 & 0.971 & 0.968\\
					\quad & \quad & \quad & \quad & 2000 & \quad & 0.910 & 0.911 & 0.912 & \quad & 0.962 & 0.966 & 0.954 \\
					\quad & \quad & 0.7 & \quad & 1000 & \quad & 0.915 & 0.926 & 0.928 & \quad & 0.934 & 0.946 & 0.964\\
					\quad & \quad & \quad & \quad & 2000 & \quad & 0.916 & 0.903 & 0.917 & \quad & 0.956 & 0.967 & 0.959\\\hline
				\end{tabular}
				\endgroup
			}\label{table_e_4}
		\end{table}

		\setcounter{theorem}{0}
		\renewcommand{\thetheorem}{G.\arabic{theorem}}
		\setcounter{equation}{0}
		\renewcommand{\theequation}{G.\arabic{equation}}
		\setcounter{lemma}{0}
		\renewcommand{\thelemma}{G.\arabic{lemma}}
		\setcounter{table}{0}
		\renewcommand{\thetable}{G.\arabic{table}}
		\setcounter{assumption}{0}
		\renewcommand{\theassumption}{G.\arabic{assumption}}
		\setcounter{remark}{0}
		\renewcommand{\theremark}{G.\arabic{remark}}
		\setcounter{corollary}{0}
		\renewcommand{\thecorollary}{G.\arabic{corollary}}

		\section{Further Results for  the AR($\infty$) Model}
		In this section, we mainly focus on the fundamental asymptotic properties of self-weighted quantile regression estimator (SQE) when dealing with high-dimensional heavy-tailed time series.
		We attempt to extend the AR(p) model, as presented in (1.1), to the following AR($\infty$) model:
		\begin{align}\label{high_model}
			y_t = \mu_0 + \sum_{j=1}^{\infty} \phi_{j0} y_{t-j} + \varepsilon_t \mbox{ and }\varepsilon_t=\eta_{t} \sigma_{t},
		\end{align}
		where the volatility $\{\sigma_{t}\}$ is strictly stationary. Here, we impose no restrictions on the form of \( \sigma_{t} \). In addition, $\phi_{j0}=O(c^{j}_0)$ for some $c_0 \in (0,1)$.

		Let $p:=p_n$ be the truncated order with $p=o(n)$ and denote $\theta_0=(\mu_0,\phi_{10},\cdots,\phi_{p0})^{\top}$. Then the SQE $\hat{\theta}_n(\tau)$ for $\theta_0$ in model (\ref{high_model}) is defined as
		\begin{align}\label{g_2}
			\hat{\theta}_n(\tau)=\argmin_{\theta \in \mathbb{R}^{p+1}}{\tilde{L}_n(\theta,\tau)} \mbox{ with }\tilde{L}_n(\theta,\tau)=n^{-1} \sum_{t=p+1}^{n}{[w_t \rho_{\tau}(y_t-Z^{\top}_{t-1}\theta)]},
		\end{align}
		where $Z_{t-1}=(1,Y^{\top}_{t-1})^{\top}$ with $Y_{t-1}=(y_{t-1},\cdots,y_{t-p})^{\top}$ and $w_t=w(y_{t-1},\cdots,y_{t-p})$.
		
		\begin{assumption}\label{assumption_g_1}
			$\phi_0(z)=1-\sum_{j=1}^{\infty}{\phi_{j0}z^j}$ has all roots outside the unit circle.
		\end{assumption}
			
	\begin{assumption}\label{assumption_g_2}
			There exists $\alpha_0 \in (0,1]$ such that $\Vert \sigma_{t} \Vert_{\alpha_0}<\infty$ and $\Vert \eta_{t} \Vert_{\alpha_0}<\infty$.
	\end{assumption}

	\begin{assumption}\label{assumption_g_3}
	There exists a constant $\sigma_0>0$ such that $\sigma_{t}\geq \sigma_0$, and $\sigma_{t}$ is non-degenerate.
	\end{assumption}
   \begin{assumption}\label{assumption_g_4}
   	(i) The continuous density $f(z)$ is positive for any $z\in \mathbb{R}$ and uniformly bounded; (ii) $f^{(1)}(z)$ is Lipschitz continuous and uniformly bounded.
   \end{assumption}
	
\begin{assumption}\label{assumption_g_1_1}
	For $k,l\geq 0$, $\sup_{j\geq p+1}{E[w^l_t \vert Z_{t-1} \vert^{k} (1+\vert y_{t-j} \vert^{\alpha_0})]}=O(p^{{\beta}_{l,k}})$ for some ${\beta}_{l,k}\in \mathbb{R}$.
\end{assumption}
Assumptions \ref{assumption_g_1}--\ref{assumption_g_4} can be viewed as the stationary counterparts of Assumptions 2.1--2.5 in the manuscript. Recall that $w_t=w(y_{t-1},\cdots,y_{t-p})$ and $\vert Z_{t-1} \vert=\sqrt{1+\sum_{i=1}^{p}{y^2_{t-i}}}$, Assumption \ref{assumption_g_1_1} is introduced to characterize the magnitude of the corresponding moment conditions associated with the truncated order $p$.

\subsection{\bf The existence of the bias when $p \to \infty$}
Denote \( R_{t-1} = \sum_{j=p+1}^{\infty} \phi_{j0} y_{t-j} \), then
\begin{align*}
	& \frac{\partial E\tilde{L}_n(\theta,\tau)}{\partial \theta}  =-(1-p/n) {E[w_tZ_{t-1}\psi_{\tau}(y_{t}-Z^{\top}_{t-1}\theta) ]} \\
	& \qquad \qquad \quad= -(1-p/n) \Big \{E\Big (w_tZ_{t-1}[\tau- F( \frac{Z^{\top}_{t-1}(\theta-\theta_0)}{\sigma_{t}} )]\Big) \nonumber \\
	& \qquad \qquad\qquad+E\Big(w_tZ_{t-1}[F( \frac{Z^{\top}_{t-1}(\theta-\theta_0)}{\sigma_{t}} )-F( \frac{Z^{\top}_{t-1}(\theta-\theta_0)-R_{t-1}}{\sigma_{t}} )]\Big)\Big\}.\nonumber
\end{align*}
Furthermore, we can derive the counterpart as that in Theorem 4.1 in the manuscript.

\begin{theorem}\label{lemma_g_1}
	Suppose that Assumptions \ref{assumption_g_1}--\ref{assumption_g_1_1} hold. Then for any \(\tau \in \mathcal{T}\), there exists a unique \(\delta_0(\tau) \in \mathbb{R}^{p+1}\) such that
	\[
	E\Big [w_tZ_{t-1}\big (\tau-F(\frac{\delta^{\top}_0(\tau)Z_{t-1}}{\sigma_{t}})\big )\Big ]=0\mbox{ and }\,\,\sup_{\tau \in \mathcal{T}}{\Big \vert \frac{\partial E\tilde{L}_n(\theta_0(\tau),\tau)}{\partial \theta} \Big \vert} = O(c^{p}_1p^{\beta_{1,1}}),
	\]
	where \(\theta_0(\tau) = \theta_0 + \delta_0(\tau)\) and $c_1=c^{\alpha_0}_0 \in (0,1)$.
\end{theorem}
For the proof of the above theorem, we only need to repeat the process for Theorem 4.1 (ii) and use the Lipschitz continuity of \( F(\cdot) \) derived from Assumption \ref{assumption_g_4}. Thus, the details are omitted.
Based on Theorem \ref{lemma_g_1}, denote $\hat{u}_n(\tau)=\sqrt{n/p}(\hat{\theta}_n(\tau)-\theta_0(\tau))$ and then it is obvious that $\hat{u}_n(\tau)$ is the minimizer of the following loss function:
\begin{align}\label{g_3}
	L_n(u,\tau)=-u^{\top} T_n(\tau)+\sum_{t=p+1}^{n}{E[\zeta_t(u,\tau)\vert \mathcal{F}_{t-1}]}+\alpha_n(u,\tau),
\end{align}
where $T_n(\tau)=\sum_{t=p+1}^{n}{\xi_t(\tau)}$, $\alpha_n(u,\tau)=\sum_{t=p+1}^{n}{\Big\{ \zeta_t(u,\tau)-E[\zeta_t(u,\tau)\vert \mathcal{F}_{t-1}] \Big\}}$ and
\begin{align*}
	& \xi_t(\tau) =\frac{1}{\sqrt{np}}w_t Z_{t-1} \psi_{\tau}(\varepsilon_{t}+R_{t-1}-\delta^{\top}_0(\tau)Z_{t-1}),\\
	& \zeta_t(u,\tau) =\frac{1}{p} w_t\int_{0}^{\frac{\sqrt{p}u^{\top}}{\sqrt{n}}Z_{t-1}}{[I(\varepsilon_{t}+R_{t-1}-\delta^{\top}_0(\tau)Z_{t-1}\leq s)-I(\varepsilon_{t}+R_{t-1}-\delta^{\top}_0(\tau)Z_{t-1}\leq 0)]ds}.
\end{align*}


\subsection{\bf The asymptotic property of $T_n(\tau)-ET_n(\tau)$ when $p \to \infty$}
Now we study the property of $T_n(\tau)-ET_n(\tau)$. The centered process can be expressed as  $T_n(\tau)-ET_n(\tau)=M_n(\tau)+N_n(\tau)$, where $M_n(\tau)$ and $N_n(\tau)$ are defined as
\begin{align*}
M_n(\tau)=\sum_{t=p+1}^{n}{\{ \xi_t(\tau)-E[\xi_{t}(\tau) \vert \mathcal{F}_{t-1}] \}},\,\,N_n(\tau)=\sum_{t=p+1}^{n}{\{ E[\xi_{t}(\tau) \vert \mathcal{F}_{t-1}]-E[\xi_{t}(\tau)]  \}}.
\end{align*}

Let $C_{p,1}=\sup_{\tau \in \mathcal{T}}{\{\vert \partial \delta_0(\tau)/\partial \tau \vert\}}$, we make the next assumption as Assumption \ref{assumption_g_1_1}.
\begin{assumption}\label{assumption_g_5}
	There exists some $\beta_{\delta_0}\in \mathbb{R}$ such that $C_{p,1}=O(p^{\beta_{\delta_0}})$.
\end{assumption}
\begin{theorem}\label{corollary_g_1}
	Suppose that Assumptions \ref{assumption_g_1}-\ref{assumption_g_5} hold. Then it follows that: when $p=M_0\log(n)$ for some $M_0>0$, we have $$\sup_{\tau}{\vert M_n(\tau)\vert}=O_p({\log(n)}^{\beta_M}),$$
	where $\beta_{M}$ is a constant.
\end{theorem}

For $N_n(\tau)$, it is easy to obtain that
\begin{align}\label{g_4_1}
	N_n(\tau)& =\frac{1}{\sqrt{np}}\sum_{t=p+1}^{n}\Big\{ w_tZ_{t-1}[\tau-F(\delta^{\top}_0(\tau)Z_{t-1}/\sigma_{t})]\nonumber\\
	& \qquad \qquad -E \big(w_tZ_{t-1}[\tau-F(\delta^{\top}_0(\tau)Z_{t-1}/\sigma_{t})] \big)   \Big\}+O_p(\sqrt{n/p}\times c^p_1 p^{\beta_{1,1}})\nonumber\\
	& := \tilde{N}_n(\tau)+O_p(\sqrt{n/p}\times c^p_1 p^{\beta_{1,1}}).
\end{align}
Then, following the proof procedures from (\ref{eq_b_19}) to (\ref{eq_b_26}), we can derive that
\begin{align*}
\big \Vert \sup_{\vert \tau_{1}-\tau_2\vert <\delta}{\vert \tilde{N}_n(\tau_{1})-\tilde{N}_n(\tau_{2}) \vert}\big \Vert \leq \delta \sqrt{\sum_{i=1}^{p+1}{E[\sup_{\tau\in \mathcal{T}}{H^2_{n,i}(\tau)}]}}.
\end{align*}
where $H_{n,i}(\tau)$ is defined as that in (\ref{eq_b_19}) and relies on $w_t$, $\partial \delta_0(\tau)/\partial \tau$ and the dependence of $Z_{t-1}$ and $\sigma_{t}$ (i.e.,$\Vert Z_{t-1}-Z^{*(t-m)}_{t-1} \Vert^{\alpha_0}_{\alpha_0}+\Vert \sigma_{t}-\sigma^{*(t-m)}_{t}\Vert^{\alpha_0}_{\alpha_0}$). For the sake of simplicity, the following condition is assumed to hold throughout this section.
\begin{assumption}\label{assumption_g_6_1}
	For some $\beta_N\in \mathbb{R}$, $\sqrt{\sum_{i=1}^{p+1}{E[\sup_{\tau\in \mathcal{T}}{H^2_{n,i}(\tau)}]}}=O(p^{\beta_N})$.
\end{assumption}
Review that $\tilde{N}_n(\tau_{0})=0$, then when $p=M_0\log(n)$ for some $M_0>0$, by Assumption \ref{assumption_g_6_1} and (\ref{g_4_1}), we have
\begin{align}\label{g_12_1}
\sup_{\tau \in \mathcal{T}}{\vert N_n(\tau) \vert}=O_p({\log(n)}^{\beta_N}).
\end{align}

\subsection{\bf The asymptotic property of $\hat{\theta}_n(\tau)$ when $p \to \infty$}

For $\sum_{t=p+1}^{n}{E[\zeta_t(u,\tau)\vert \mathcal{F}_{t-1}]}$, we can obtain Theorem \ref{theorem_g_3} as follows.
\begin{theorem}\label{theorem_g_3}
Suppose that Assumptions \ref{assumption_g_1}--\ref{assumption_g_1_1} hold. Then for any compact set $U_n\subset\mathbb{R}^{p+1}$, it follows that $$\sup_{\tau\in \mathcal{T},u\in U_n}{\vert \sum_{t=p+1}^{n}{E[\zeta_t(u,\tau)\vert \mathcal{F}_{t-1}]}-u^{\top}\Sigma_n(\tau)u/2  \vert }=O_p( p^{{\beta}_{1,3}+1/2}n^{-1/2}+c^{p}_1 p^{\beta_{1,2}}),$$ where $\Sigma_n(\tau)=n^{-1}\sum_{t=p+1}^{n} w_tZ_{t-1}Z^{\top}_{t-1} f(\delta^{\top}_0(\tau)Z_{t-1}/\sigma_{t})\sigma^{-1}_{t}$.
\end{theorem}

Then, for the matrix $\Sigma_n(\tau)$ and the remainder term $\alpha_n(u,\tau)$, we make the regular assumptions in high dimensional time series.
\begin{assumption}\label{assumption_g_8}
	(i). There exist $\beta_{\zeta} \in \mathbb{R}$ and a positive constant $\lambda_{1}$ such that $\lambda_{1} p^{\beta_{\zeta}} \vert u \vert^2 \leq u^{\top}\Sigma_n(\tau) u$ holds with probability to one; (ii). For any compact set $U_n \in \mathbb{R}^{p+1}$, $\sup_{\tau \in \mathcal{T}, u\in U_n}{\vert \alpha_n(u,\tau) \vert}$ $=o_p( p^{\beta_{\zeta}} \sup_{u \in U_n}{\vert u \vert^2})$.
\end{assumption}

Based on these fundamental results, we now derive the result for $\hat{\theta}_n(\tau)$.
\begin{theorem}\label{theorem_g_4}
	Suppose that Assumptions \ref{assumption_g_1}--\ref{assumption_g_8} hold. Then when $p=M_0\log(n)$ for some $M_0>0$, we have $$\sup_{\tau \in \mathcal{T}}{\vert \hat{\theta}_n(\tau)-\theta_{0}(\tau) \vert}=O_p({\log(n)}^{\beta_0+1/2}n^{-1/2}),$$ where $\beta_0=(\beta_{M}\vee\beta_{N})-\beta_{\zeta}$.
\end{theorem}

From Theorem \ref{theorem_g_4},  we can see that the convergence rate of the SQE is slightly slower than that when $p$ is fixed. Similarly, we can further demonstrate that \(\hat{\tau}_n\) will converge to \(\tau_0\) at a similar rate. Thus, in the AR($\infty)$ setting, we cannot obtain the limiting distributions as in Theorems 5.1 and 6.1--6.2 and hence, the two-step statistical inference method is not available to this scenario.
  We may need to employ methodologies specific to high-dimensional settings to address the issues; for example, see \citet{belloni2019conditional}. Given that our current paper is the first to address the open problem of how to conduct inference for the AR($p$) model with unspecified volatility \(\sigma_{t}\) and unknown \(\tau_0\),  we will further explore alternative approaches for the AR(\(\infty\)) model in the future.

\subsection{\bf Proofs of the Results of AR($\infty$) Model}
Let $Z_{t-1,i}$ be the $i$-th component in $Z_{t-1}$ and $\delta_{0,i}(\tau)$ be the $i$-th element in $\delta_0(\tau)$. Meanwhile, denote $J_{p}=\{i: \delta_{0,i}(\tau) \neq 0 \mbox{ for some }\tau\in \mathcal{T}\}$ and $C_{p,1}=\sup_{\tau \in \mathcal{T}}{\{\vert \partial \delta_0(\tau)/\partial \tau \vert\}}$.
\begin{lemma}\label{theorem_g_1}
	Suppose that Assumptions \ref{assumption_g_1}--\ref{assumption_g_1_1} hold. Then for any $q>2$, there exists a constant $C_q>0$ such that for any $\tau_1 \in (0,1)$, $\delta>0$ and $d\in \mathbb{N}$, we have
	\begin{align*}
	E &\big\{\sup_{\vert \tau \vert<\delta}{\vert [M_{n}(\tau_{1}+\tau)-M_{n}(\tau_1)] \vert^q} \big\}\leq p^{-1}C_q  \Big \{ \big[\sum_{i=1}^{p+1}{D_{p,2,i}(J_p)}\big] C_{p,1}\times n^{\alpha}d^q \delta\\
	& \qquad \qquad + \big[\sum_{i=1}^{p+1}{D_{p,3,i}(J_p)}\big ]C^{q/2}_{p,1}\times {\delta}^{q/2} +\big[\sum_{i=1}^{p+1}{D_{p,4,i}(J_p)}\big] C^{q}_{p,1}\times n^{q/2}2^{-dq}{\delta}^{q}\Big \},
	\end{align*}
	where $\alpha=[1\vee (q/4)]-q/2$, and $D_{p,2,i}(J_p)$$=E[\vert w_t Z_{t-1,i} \vert^{q} \vert  Z_{t-1,J_p}\vert]$ and $D_{p,3,i}(J_p)=E[\vert w_t Z_{t-1,i} \vert^{q} \vert  Z_{t-1,J_p}\vert^{q/2}]$ and $D_{p,4,i}(J_p)=E[\vert w_t Z_{t-1,i} \vert^{q} \vert  Z_{t-1,J_{p}}\vert^{q}]$, where $Z_{t-1,J_p}$ is the subvector of $Z_{t-1}$ by picking out the elements with the index set $J_p$.	
\end{lemma}
\begin{proof}[Proof of Lemma \ref{theorem_g_1}]
	Define
	\begin{align}
		G_{n,i}(x,h,J) & =\frac{1}{\sqrt{np}}\sum_{t=p+1}^{n}{w_t Z_{t-1,i} \{ \eta_{t}(x,h,J)-E[\eta_{t}(x,h,J) \vert \mathcal{F}_{t-1}] \}},\label{g_5}\\
		\eta_{t}(x,h,J) & =I(\varepsilon_{t} \leq x^{\top}Z_{t-1,J}+h\vert Z_{t-1,J}\vert-R_{t-1}),\label{g_6}
	\end{align}
	where $J$ is a subset of $\{1,\cdots,p+1\}$.
	Using the proof procedures in Lemma \ref{lemma_b_1} and carefully checking the equations (\ref{eq_b_5}) and (\ref{eq_b_7}), we can directly obtain one more accurate result as
	\begin{align}\label{g_7}
		& \Vert \sqrt{p}[G_{n,i}(x,h,J)-G_{n,i}(\tilde{x},\tilde{h},J)]\Vert^{q}_q\nonumber\\
		& \qquad \leq C_q \big \{D_{p,2,i}(J)n^{\alpha} (\vert x-\tilde{x}\vert+\vert h-\tilde{h}\vert)+D_{p,3,i}(J)(\vert x-\tilde{x}\vert^{q/2}+\vert h-\tilde{h}\vert^{q/2})\big\}.
	\end{align}
	Furthermore, note that $\vert \delta_0(\tau_1)-\delta_0(\tau_{2})\vert \leq C_{p,1} \vert \tau_{1}-\tau_{2}\vert$. By (\ref{g_7}) and following the proof procedures for Lemma \ref{lemma_b_2}, we can derive that
	\begin{align}\label{g_8}
		E &\big\{\sup_{\vert \tau \vert<\delta}{\vert \sqrt{p} [M_{n,i}(\tau_{1}+\tau)-M_{n,i}(\tau_1)] \vert^q} \big\}\\
		&\qquad \leq C_q \big \{ D_{p,2,i}(J_p)C_{p,1}n^{\alpha}d^q \delta+D_{p,3,i}(J_p)C^{q/2}_{p,1}{\delta}^{q/2}+D_{p,4,i}(J_p) C^{q}_{p,1}n^{q/2}2^{-dq}{\delta}^{q}\big \},\nonumber
	\end{align}
	where $M_{n,i}(\tau)$ is the $i$-th component of $M_n(\tau)$. Then, by Jensen's inequality  that $\vert M_n(\tau_1+\tau)-M_n(\tau_{1})\vert^{q} \leq C_q  p^{q/2-1} [\sum_{i=1}^{p+1}{ \vert M_{n,i}(\tau_1+\tau)-M_{n,i}(\tau_{1})\vert^{q}  }]$, the conclusion holds from (\ref{g_8}).
\end{proof}

\begin{proof}[Proof of Theorem \ref{corollary_g_1}]
	By Lemma \ref{theorem_g_1}, for $q>2$, we can define
	$$D_{p,2}=p^{-1}\sum_{i=1}^{p+1}{E[\vert w_t Z_{t-1,i}\vert^q\vert Z_{t-1,J_p}\vert]}\times C_{p,1}$$
	$$D_{p,3}=p^{-1}\sum_{i=1}^{p+1}{E[\vert w_t Z_{t-1,i}\vert^q\vert Z_{t-1,J_p}\vert^{q/2}]}\times C^{q/2}_{p,1}$$
	$$D_{p,4}=p^{-1}\sum_{i=1}^{p+1}{E[\vert w_t Z_{t-1,i}\vert^q\vert Z_{t-1,J_p}\vert^q]}\times C^q_{p,1}.$$
	By Assumptions \ref{assumption_g_1_1}--\ref{assumption_g_5}, there exists $\beta_i \in \mathbb{R}$ such that $D_{p,i}=O(p^{\beta_{i}})$ for any $i=2,3,4$.
	
	Choose $\delta=(1-\tau_{0})\vee \tau_{0}$, then for any $\epsilon>0$, by Lemma \ref{theorem_g_1}, it follows that
	\begin{align*}
		& P(\sup_{\tau \in \mathcal{T}}{\vert M_n(\tau)\vert}>\epsilon) \leq P(\vert M_n(\tau_{0})\vert>\epsilon/2)+ P(\sup_{\vert \tau\vert<\delta}{\vert M_n(\tau_{0})-M_n(\tau_{0}+\tau)\vert}>\epsilon/2)\\
		& \qquad \leq \frac{C_q\times p^{-1}E[w^2_t \vert Z_{t-1}\vert^2 ]}{\epsilon^2}+\frac{C_q}{\epsilon^q}\Big \{ D_{p,2}n^{\alpha}d^q \delta+ D_{p,3}{\delta}^{q/2} +D_{p,4} n^{q/2}2^{-dq}{\delta}^{q}\Big \}.
	\end{align*}
	Let $d=[\log_2(n)]+1$. Using the fact that $D_{p,i}=O(p^{\beta_{i}})$ and $p=M_0\log(n)$ and by Assumption \ref{assumption_g_1_1}, we can obtain $$\limsup_{n \to \infty}{P(\sup_{\tau \in \mathcal{T}}{\vert M_n(\tau)\vert}>\epsilon)}\leq C_q [p^{\beta_{2,2}-1}/\epsilon^2+p^{\beta_{3}}/\epsilon^q].$$
	This implies the conclusion.
\end{proof}

\begin{proof}[Proof of Theorem \ref{theorem_g_3}]
	By the definition of $\zeta_t(u,\tau)$, it is straightforward to show that
	\begin{align}
		& E \bigg \{\sup_{\tau,u}\Big \vert \sum_{t=p+1}^{n}{E[\zeta_t(u,\tau)\vert \mathcal{F}_{t-1}]}-u^{\top}\times \frac{1}{2n} \sum_{t=p+1}^{n} w_tZ_{t-1}Z^{\top}_{t-1} f(\delta^{\top}_0(\tau)Z_{t-1}/\sigma_{t})\sigma^{-1}_{t}\times u \Big \vert \bigg\}\nonumber\\
		& \qquad \leq \frac{\sqrt{p}\sup _{u}\vert u \vert^3}{2n^{3/2}} \sum_{t=p+1}^{n}{E[w_t \vert Z_{t-1}\vert^3]}+\frac{C_1\sup_{u}\vert u\vert^2}{2n}\sum_{t=p+1}^{n}{E[w_t\vert R_{t-1}\vert^{\alpha_0}\vert Z_{t-1} \vert^2]},\nonumber\\
		& \qquad\leq \frac{\sqrt{p}\sup_{u}\vert u \vert^3}{2n^{1/2}} {E[w_t \vert Z_{t-1}\vert^3]}+{C_1\sup_{u}\vert u\vert^2}\sum_{j=p+1}^{\infty}{[\vert \phi_{j0}\vert^{\alpha_0} E(w_t\vert Z_{t-1} \vert^2 \vert y_{t-j} \vert^{\alpha_0})]},\nonumber\\
		& \qquad=O(\sup_{u}\vert u \vert^{3} p^{{\beta}_{1,3}+1/2}n^{-1/2}+\sup_{u}\vert u \vert^2 c^{p}_1 p^{\beta_{1,2}}).\nonumber
	\end{align}
	where $C_1$ is a constant.
\end{proof}

\begin{proof}[Proof of Theorem \ref{theorem_g_4}]
	Notice that $L_n(0,\tau)=0$ for any $\tau \in \mathcal{T}$, then by the convex property of $L_n(u,\tau)$, it is enough to show that for any $\epsilon>0$, there exists a $C>0$ such that
	\begin{align*}
		P(\inf_{\tau \in \mathcal{T}, u\in U_n}{L(u,\tau)}\leq 0 )<\epsilon,
	\end{align*}
	where $U_n=\{u: \vert u \vert=Cp^{\beta_0}\}$.
	
	Note that $ET_n(\tau)=-\sqrt{n/p}{\partial {E\tilde{L}_n(\theta_{0}(\tau),\tau)}}/{\partial \theta}$, Theorem \ref{lemma_g_1} indicates that
	\begin{align}\label{g_4}
		\sup_{\tau \in \mathcal{T}}{\Big \vert   ET_n(\tau)\Big \vert}=O(\sqrt{n/p}\times c^p_1 p^{\beta_{1,1}}).
	\end{align}
	Thus, by Theorem \ref{corollary_g_1}, (\ref{g_12_1}) and (\ref{g_4}), we can choose a large $M_0>0$ such that $c^{p}_1n \to 0$ and then
	\begin{align}\label{g_14}
		\sup_{\tau \in \mathcal{T}}{\vert T_n(\tau)\vert}=O_p(p^{\beta_M\vee \beta_N}).
	\end{align}
	So we can further find a constant $M_1$ such that
	\begin{align}\label{g_15}
		P(\sup_{\tau \in \mathcal{T}}{\vert T_n(\tau)\vert}>M_1 p^{\beta_M \vee \beta_N})<\epsilon/2.
	\end{align}
	In addition, by $p=M_0\log(n)$ and $c^{p}_1n \to 0$ and Assumption \ref{assumption_g_8}, Theorem \ref{theorem_g_3} indicates that
	\begin{align}\label{g_16}
		P(\inf_{\tau \in \mathcal{T},u\in U_n}{\Big \vert \sum_{t=p+1}^{n}{\zeta_t(u,\tau)} \Big \vert}>\lambda_{1}C^2p^{2\beta_0+\beta_{\zeta}}/3) \to 1.
	\end{align}
	By (\ref{g_15})--(\ref{g_16}), we can choose $C>3M_1/\lambda_{1}$, then
	\begin{align}
		& P(\inf_{\tau \in \mathcal{T}, u\in U_n}{L(u,\tau)}\leq 0 ) \leq P( \inf_{\tau \in \mathcal{T},u\in U_n}{\Big \vert \sum_{t=p+1}^{n}{\zeta_t(u,\tau)} \Big \vert} \leq \sup_{\tau \in \mathcal{T}, u\in U_n}{\vert u^{\top} T_n(\tau)\vert})\nonumber\\
		& \leq P( \inf_{\tau \in \mathcal{T},u\in U_n}{\Big \vert \sum_{t=p+1}^{n}{\zeta_t(u,\tau)} \Big \vert}\leq CM_1 p^{\beta_0+\beta_M\vee \beta_N} )+\epsilon/2\nonumber\\
		& \leq P(\lambda_{1}C^2p^{2\beta_0+\beta_\zeta}/3<CM_1 p^{\beta_0+\beta_M \vee \beta_N})\nonumber\\
		& \qquad \quad+P(\inf_{\tau \in \mathcal{T},u\in U_n}{\Big \vert \sum_{t=p+1}^{n}{\zeta_t(u,\tau)} \Big \vert}\leq \lambda_{1}C^2p^{2\beta_0+\beta_{\zeta}}/3)+\epsilon/2\nonumber\\
		& = P(\inf_{\tau \in \mathcal{T},u\in U_n}{\Big \vert \sum_{t=p+1}^{n}{\zeta_t(u,\tau)} \Big \vert}\leq \lambda_{1}C^2p^{2\beta_0+\beta_\zeta}/3)+\epsilon/2 \longrightarrow \epsilon/2, \mbox{ as }n\to \infty.
	\end{align}
	This completes the whole proof.
\end{proof}

	\setcounter{theorem}{0}
\renewcommand{\thetheorem}{H.\arabic{theorem}}
\setcounter{equation}{0}
\renewcommand{\theequation}{H.\arabic{equation}}
\setcounter{lemma}{0}
\renewcommand{\thelemma}{H.\arabic{lemma}}
\setcounter{table}{0}
\renewcommand{\thetable}{H.\arabic{table}}
\setcounter{assumption}{0}
\renewcommand{\theassumption}{H.\arabic{assumption}}
\setcounter{remark}{0}
\renewcommand{\theremark}{H.\arabic{remark}}
\setcounter{corollary}{0}
\renewcommand{\thecorollary}{H.\arabic{corollary}}

\section{Further Extension to General Noise Structure}
In this section, we present theoretical details of the proposed two-step estimation procedure when the noise is extended to the general noise structure as discussed in Remark 3.2. Consider the AR($p$) model with non-stationary noises as follows:
\begin{align}\label{eq_h_1}
	y_t=\mu_0+\sum_{j=1}^{p}{\phi_{j0} y_{t-j}}+\varepsilon_{t} \mbox{ and }\varepsilon_{t}=G_0(t/n,\mathcal{F}_t),
\end{align}
where $\mathcal{F}_t$ denotes the filtration $(\eta_{t},\eta_{t-1},\cdots)$ with $\{\eta_{t}\}$ being a sequence of independent and identically distributed (i.i.d.) random variables, and $G_0: [0,1]\times \mathbb{R}^{\infty} \to \mathbb{R}$ is a measurable function. Similar to the main manuscript, we make some assumptions.
\begin{assumption}\label{assumption_h_1}
	$\phi_0(z)=1-\sum_{j=1}^{p}{\phi_{j0}z^j}$ has all roots outside the unit circle.
\end{assumption}
\begin{assumption}\label{assumption_h_2}
	There exists $\alpha_0 \in (0,1]$ such that $\Vert \sup_{x\in [0,1]}{\vert \varepsilon_{t}(x) \vert}\Vert_{\alpha_0}<\infty$ and $\Vert Y_p \Vert_{\alpha_0}<\infty$, where $\varepsilon_{t}(x)=G_0(x,\mathcal{F}_{t})$ and $Y_p=(y_p,\cdots,y_1)^{\top}$.
\end{assumption}
\begin{assumption}\label{assumption_h_3}
	$\sup\limits_{\substack{x_1 \neq x_2, x_1, x_2 \in [0,1]}}{\{ \Vert \varepsilon_{t}(x_1)-\varepsilon_{t}(x_2) \Vert_{\alpha_0}/\vert x_1-x_2\vert \}}<\infty$.
\end{assumption}
Under the above assumptions, the following stationary time series is well defined:
\begin{align}\label{eq_h_2}
	y_t(x)=\pi_0(1)\mu_0+\sum_{j=0}^{\infty}{\pi_{j0} \varepsilon_{t-j}(x)}, \mbox{ for } x \in [0,1],
\end{align}
where $\pi_0(z)=\phi^{-1}_0(z)=\sum_{j=0}^{\infty}{\pi_{j0}z^j}$ with $\pi_{j0}=O(c^j_0)$ for some $c_0 \in (0,1)$.

We further give three assumptions on the conditional distribution of $\varepsilon_t(x)$.
Define $F(y,x;\mathcal{F}_{t-1})=P(\varepsilon_t(x)\leq y \vert \mathcal{F}_{t-1})$ with its inverse  regarding $y$ denoted by $F^{-1}(y,x;\mathcal{F}_{t-1})$ and let $F_k(y,x;\mathcal{F}_{t-1})=\partial^kF(y,x;\mathcal{F}_{t-1})/\partial y^k $ for $k\in \{0,1,2\}$, where $F_0(y,x;\mathcal{F}_{t-1})=F(y,x;\mathcal{F}_{t-1})$.
\begin{assumption}\label{assumption_h_4}
	There exists a constant $C_0>0$ such that for $k=0,1$ and $x_1,x_2\in [0,1]$, we have $\big \Vert \sup_{y}{\{{\vert F_k(y,x_1;\mathcal{F}_{t-1})-F_k(y,x_2;\mathcal{F}_{t-1})\vert }/{(\vert y \vert+1)}\}} \big \Vert_1\leq C_0 \vert x_1-x_2\vert^{\alpha_0}$.
\end{assumption}
\begin{assumption}\label{assumption_h_5}
	(i) For any $x\in [0,1]$, $F_1(y,x;\mathcal{F}_{t-1})$ is positive on the whole real line almost surely; (ii) for $k=1,2$, $\sup_{y_1\neq y_2, x\in [0,1]} \big \{ \vert F_2(y_1,x;\mathcal{F}_{t-1})-F_2(y_2,x;\mathcal{F}_{t-1})\vert/\vert y_1-y_2 \vert \big \}\leq C_0$ almost surely, and $\sup_{y\in \mathbb{R},x\in[0,1]}{\vert F_k(y,x;\mathcal{F}_{t-1})\vert}\leq C_0$ almost surely.
\end{assumption}
\begin{assumption}\label{assumption_h_6}
	(i) There exists an unknown $\tau_{0} \in (0,1)$ such that $F^{-1}(\tau_{0},x;\mathcal{F}_{t-1})=0$; (ii) for any $\tau \neq \tau_{0}$ and $x\in [0,1]$, $F^{-1}(\tau,x;\mathcal{F}_{t-1})$ is non-degenerate. 
\end{assumption}
Denote the unknown value of $\tau_{0}$ to be $\tau\in (0,1)$.  Then the form of the $\tau$-th conditional quantile of $y_t$ is changed to be
\begin{align}\label{eq_h_3}
	Q_{y_t}(\tau\vert\mathcal{F}_{t-1})=Z^{\top}_{t-1}\theta_0+F_t^{-1}(\tau),
\end{align}
where   $F^{-1}_t(\cdot)$ is the inverse of $F(\cdot,t/n;\mathcal{F}_{t-1})$. Now, we can derive a significant result as follows.
\begin{lemma}\label{lemma_h_1}
	Given Assumption \ref{assumption_h_5} (i) and Assumption \ref{assumption_h_6}, it follows that
	(i) for any $\delta\in \mathbb{R}^{p+1}$ and $\tau \neq \tau_{0}$, $P(Z^{\top}_{t-1}\delta=F_t^{-1}(\tau))<1;$ (ii) $P(Z^{\top}_{t-1}\delta=0)=1$ if and only if $\delta=0$.
\end{lemma}
Given Lemma \ref{lemma_h_1}, Equation (3.3) in Section 3 still holds. Therefore, we proceed with the proposed two-step estimation approach to estimate the parameters of model (\ref{eq_h_1}). For convenience and clarity, notations used in the rest of this section maintain the same definitions as in the main manuscript, with exceptions explicitly noted.

\subsection{\bf The existence of the bias}
\begin{assumption}\label{assumption_h_7}
	$\{w(x_0,x_1,\cdots,x_p)x^{l_i}_{i}x^{l_j}_j x^{l_k}_k x^{l_m}_m \}$ are bounded and Lipschitz continuous on $[0,1]\times \mathbb{R}^p$, where $i,j,k,m\in \{1,\cdots,p\}$ and $l_i, l_j, l_k,l_m \in \{0,1\}$.
\end{assumption}
\begin{theorem}\label{theorem_h_1}
	Suppose that Assumptions \ref{assumption_h_1}--\ref{assumption_h_5}, Assumption \ref{assumption_h_6} (i) and Assumption \ref{assumption_h_7} hold. Then it follows that
	
	(i) for any bounded set $\Theta \subset \mathbb{R}^{p+1}$, we have
	\begin{align*}
		\sup_{\tau \in \mathcal{T},\theta \in \Theta}{\Big \vert   \frac{\partial E\tilde{L}_n(\theta,\tau)}{\partial \theta}-g(\theta-\theta_{0},\tau)  \Big \vert}=O(n^{-\alpha_0}),
	\end{align*}
	where $g(x,\tau)=-\int_{0}^{1}{E\Big\{w(s,Y^{\top}_{t-1}(s))Z_{t-1}(s)\big[\tau-F\big({x^{\top}Z_{t-1}(s)},s;\mathcal{F}_{t-1}\big) \big]   \Big\} ds}$.
	
	(ii) for any $\tau \in \mathcal{T}$, in the whole space $\mathbb{R}^{p+1}$, there exists a unique $\delta_0(\tau)$ such that $g(\delta_0(\tau),\tau)=0$ and  $\delta_0(\tau)$ has the continuous derivative with respect to $\tau$ with
	\begin{align*}
		\frac{\partial\delta_0(\tau_{0})}{\partial \tau} & =\Big\{\int_{0}^{1}{E\big[w(s,Y^{\top}_{t-1}(s))Z_{t-1}(s)Z^{\top}_{t-1}(s)F_1(0,s;\mathcal{F}_{t-1})  \big]ds}\Big\}^{-1}\\
		& \qquad \quad \times \int_{0}^{1}{E[w(s,Y^{\top}_{t-1}(s))Z_{t-1}(s)]ds}.
	\end{align*}
\end{theorem}

\subsection{\bf The asymptotic property of $\hat{\theta}_n(\tau)$}
\begin{assumption}\label{assumption_h_8}
	There exists a $\kappa_0 \in (0,1)$ such that, for any $m \geq 0$ and $k=0,1,2$,
	\begin{align*}
		& (i). \sup_{x \in [0,1]}{\Vert \varepsilon_t(x)-\varepsilon^{*(t-m)}_t(x) \Vert_{\alpha_0}}+\Vert Y_p-Y^{*(p-m)}_p\Vert_{\alpha_0}=O(\kappa_0^m),\\
		& (ii). \sup_{x\in [0,1]}{\big\Vert \sup_{y \in \mathbb{R}}{\big[\vert F_k(y,x;\mathcal{F}_{t})-F_k(y,x;\mathcal{F}^{*(t-m)}_{t})\vert/(\vert y \vert+1)\big]} \big \Vert}=O(\kappa^{m}_0).
	\end{align*}
\end{assumption}

\begin{lemma}\label{lemma_h_2}
	Suppose that Assumptions \ref{assumption_h_1}-\ref{assumption_h_8} hold. Then it follows that
	\begin{align*}
		\{T_n(\tau)-ET_n(\tau)\}_{\tau \in \mathcal{T}} \rightsquigarrow T_0(\tau),
	\end{align*}
	as $n \to \infty$, where $\rightsquigarrow$ {denotes} weak convergence on the space $l^{\infty}(\mathcal{T})$, and $T_0(\tau)$ is a centered Gaussian process with $Cov(T_0(\tau_{1}),T_0(\tau_{2}))=\sum_{k\in \mathbb{Z}}{\int_{0}^{1}{\mbox{Cov}(\xi_{0}(\tau_{1},s),\xi_{k}(\tau_{2},s)) ds}}$,  where $\tau_{1},\tau_{2} \in \mathcal{T}$
	and $\xi_{t}(\tau,s)=w(s,Y_{t-1}(s)) Z_{t-1}(s)$ $\psi_{\tau}(\varepsilon_{t}(s)-\delta^{\top}_0(\tau)Z_{t-1}(s))$.
\end{lemma}

\begin{theorem}\label{theorem_h_2}
	Suppose that all the conditions in Lemma \ref{lemma_h_2} hold. If $\alpha_0>1/2$, then
	\begin{align*}
		\sqrt{n}(\hat{\theta}_n(\tau)-\theta_0(\tau)) \rightsquigarrow \Sigma^{-1}(\tau) T_0(\tau),
	\end{align*}
	as $n \to \infty$, where  $\Sigma(\tau)=\int_{0}^{1}{E\big[{w(s,Y^{\top}_{t-1}(s))}Z_{t-1}(s)Z^{\top}_{t-1}(s)F_1({\delta^{\top}_0(\tau)Z_{t-1}(s)},s;\mathcal{F}_{t-1})  \big]ds}$.
\end{theorem}

\subsection{\bf The asymptotic property of $\hat{\theta}_n$ and $\hat{\tau}_n$}
\begin{assumption}\label{assumption_h_9}
	$\{\tilde{w}_l(x_0,x_1,\cdots,x_{\tilde{p}})x^{l_i}_{i}x^{l_j}_j x^{l_k}_k x^{l_m}_m \}$ are bounded and Lipschitz continuous on $[0,1]\times \mathbb{R}^{\tilde{p}\vee p}$, where $i,j,k,m\in \{1,\cdots,p\}$ and $l_i, l_j, l_k,l_m \in \{0,1\}$.
\end{assumption}

\begin{lemma}\label{lemma_h_3}
	Suppose that all the conditions in Theorem \ref{theorem_h_2} and Assumption \ref{assumption_h_9} hold. Then for any $l\in \{1,\cdots,L\}$, we have
	\begin{align*}
		& \sup_{\tau \in \mathcal{T}} \Big \vert  \frac{1}{\sqrt{n}}\sum_{t=p+1}^{n}{R_t(\tau;\tilde{w}_l)}+b^{\top}(\tau;\tilde{w}_l)\sqrt{n}(\hat{\theta}_n(\tau)-\theta_0(\tau)) \Big \vert=o_p(1),\\
		& \sup_{\tau\in \mathcal{T}} \Big \vert \frac{1}{\sqrt{n}}\sum_{t=p+1}^{n}{E[r_{t0}(\tau;\tilde{w}_l)]}-\sqrt{n}\tilde{g}(\tau;\tilde{w}_l)\Big \vert=o(1),
	\end{align*}
	as $n \to \infty$, where $b(\tau;\tilde{w}_l)=\int_{0}^{1}$ $ E\big[ {\tilde{w}_l(s,\tilde{Y}^{\top}_{t-1}(s))}Z_{t-1}(s)F_1( {\delta^{\top}_0(\tau)Z_{t-1}(s)},s;\mathcal{F}_{t-1})   \big] ds$ and $\tilde{g}(\tau;\tilde{w}_l)=\int_{0}^{1}{E\big\{\tilde{w}_l(s,\tilde{Y}^{\top}_{t-1}(s))[\tau-F\big({\delta_0^{\top}(\tau)Z_{t-1}(s)},s;\mathcal{F}_{t-1} \big) ]   \big\} ds}$ .
\end{lemma}

\begin{assumption}\label{assumption_h_10}
	(i). $\sum_{l=1}^{L}{\tilde{g}^2(\tau;\tilde{w}_l)}>0$ for any $\tau \neq \tau_{0}$; (ii). $\partial \tilde{g}(\tau_{0};\tilde{w}_l)/\partial\tau \neq 0$ for some $l\leq L$.
\end{assumption}
Define $a_l=-[\partial \tilde{g}(\tau_{0};\tilde{w}_l)/\partial\tau]/\sum_{l\in \mathcal{A}}{[\partial \tilde{g}(\tau_{0};\tilde{w}_l)/\partial\tau]^2}$ and
$$\tilde{h}_{t}(s)=\sum_{l\in \mathcal{A}}{a_l[\tilde{w}_l(s,\tilde{Y}^{\top}_{t-1}(s))-b^{\top}(\tau_{0};\tilde{w}_l)\Sigma^{-1}(\tau_{0}) w(s,Y^{\top}_{t-1}(s))Z_{t-1}(s)  ]},$$
where  $\mathcal{A}=\{l: \partial \tilde{g}(\tau_{0};\tilde{w}_l)/\partial\tau \neq  0\}$. The asymptotic normalities of $\hat{\tau}_n$ and $\hat{\theta}_n$ are shown as follows.
\begin{theorem}\label{theorem_h_3}
	Suppose that the conditions in Theorem \ref{theorem_h_2} and Assumptions \ref{assumption_h_9}-\ref{assumption_h_10} hold. Then, it follows that
	\begin{align*}
		\sqrt{n}(\hat{\tau}_n-\tau_{0}) \overset{d}{\longrightarrow} N(0,\gamma^2_1),
	\end{align*}
	as $n \to \infty$, where $\gamma^2_1=\tau_{0}(1-\tau_{0})\int_{0}^{1}{E\tilde{h}^2_t(s)ds   }$.
\end{theorem}

\begin{theorem}\label{theorem_h_4}
	Suppose that the conditions in Theorem \ref{theorem_h_3} hold. Then we have
	\begin{align*}
		\sqrt{n}(\hat{\theta}_n-\theta_0)\overset{d}{\longrightarrow} N(0, \Gamma_1),
	\end{align*}
	as $n \to \infty$, where $\Gamma_1=\tau_{0}(1-\tau_{0})\int_{0}^{1}{E\Big\{\tilde{H}_t(s)\tilde{H}^{\top}_t(s)\Big\}ds}$ with $$\tilde{H}_t(s)=\Sigma^{-1}(\tau_{0})w(s,Y^{\top}_{t-1}(s))Z_{t-1}(s)+\frac{\partial \delta_0(\tau_{0})}{\partial \tau}\tilde{h}_t(s).$$
\end{theorem}

\subsection{\bf The bootstrapped approximation}
\begin{assumption}\label{assumption_h_11}
	$\{w^*_t\}$ and $\{y_t\}$ are independent.
\end{assumption}
\begin{theorem}\label{theorem_h_5}
	Suppose that all the conditions in Theorem \ref{theorem_h_2} and Assumption \ref{assumption_h_11} hold. Conditional on $\{y_1,\cdots,y_n\}$, it follows that
	\begin{align*}
		\sqrt{n}(\hat{\theta}^{*}_n(\tau)-\hat{\theta}_n(\tau)) \rightsquigarrow \Sigma^{-1}(\tau) T^{*}_0(\tau)
	\end{align*}
	in probability, where $T^{*}_0(\tau)$ is a centered Gaussian process with  $Cov(T^{*}_0(\tau_{1}),T^{*}_0(\tau_{2}))=\int_{0}^{1}{\mbox{E}[\xi_{0}(\tau_{1},s)\xi^{\top}_{0}(\tau_{2},s)] ds}$, where $\xi_{0}(\tau,s)$ is defined in Lemma \ref{lemma_h_2}.
\end{theorem}
\begin{theorem}\label{theorem_h_6}
	Suppose that all the conditions in Theorem \ref{theorem_h_3} and Assumption \ref{assumption_h_11} hold. Conditional on $\{y_1,\cdots,y_n\}$,  we have
	\begin{align*}
		\sqrt{n}(\hat{\tau}^{*}_n-\hat{\tau}_n) \overset{d}{\longrightarrow}N(0,\gamma^2_1)\mbox{ and }\sqrt{n}(\hat{\theta}^{*}_n-\hat{\theta}_n)\overset{d}{\longrightarrow} N(0,\Gamma_1),
	\end{align*}
	in probability.
\end{theorem}

\subsection{\bf Proofs of the results to the general noise}
For the GARCH-type noise in (2.1), we have simple closed forms:
\begin{align}\label{eq_h_3_new}
	F(y,x;\mathcal{F}_{t-1}) & =F_{\eta}(y/\sigma_{t}(x))\nonumber\\
	F_1(y,x;\mathcal{F}_{t-1})& =f_{\eta}(y/\sigma_{t}(x))/\sigma_{t}(x)\nonumber\\
	F_2(y,x;\mathcal{F}_{t-1})& =f^{(1)}_{\eta}(y/\sigma_{t}(x))/[\sigma_{t}(x)]^2,
\end{align}
where $F_{\eta}(\cdot)$ is the distribution function of $\eta_{t}$, and $f_{\eta}(\cdot)$ is the density function with $f_{\eta}(\cdot)$ being its derivative. 
Based on this, we can adopt the same proof strategy and framework, with the sole modification being the replacement of the GARCH-specific technical assumptions (i.e., Assumptions 2.4-2.5, 5.1) by general ones (i.e., Assumptions \ref{assumption_h_4}-\ref{assumption_h_6}, \ref{assumption_h_8}).
Therefore, owing to the similarity of the proof structures to those in the main manuscript, we only present the key distinguishing steps as follows; the full details are provided upon request.
\begin{proof}[\bf Proof of Lemma \ref{lemma_h_1}]
	(i). Without loss of generality, we only focus on the case when $\tau>\tau_{0}$. Suppose that there exists a $\delta\in\mathbb{R}^{p+1}$ such that   $P(Z^{\top}_{t-1}\delta=F_t^{-1}(\tau))=1$. Denote $\delta_{i}$ as the $i$-th element in the vector $\delta$. Since $F_t(\cdot)=F(\cdot,t/n;\mathcal{F}_{t-1})$, by the non-degeneracy of $F_t^{-1}(\cdot)$ in Assumption \ref{assumption_h_6} (ii), there exists one smallest $j \in \{1,\cdots,p\}$ such that the $\delta_{j+1}\neq 0$.  So, by the fact that $F_t^{-1}(\tau)\geq F_t^{-1}(\tau_{0})=0$ in Assumption \ref{assumption_h_6} (i), we get that
	\begin{align}\label{eq_h_4}
		P\bigg(\delta_{j+1}y_{t-j}+\delta_1+\sum_{k=j+1}^{p}{\delta_{k+1}y_{t-k}}\geq 0\bigg)=1.
	\end{align}
	Since $y_{t-j}=Z^{\top}_{t-j-1}\theta_0+\varepsilon_{t-j}$, (\ref{eq_h_4}) implies that
	\begin{align}\label{eq_h_5}
		\begin{cases}
			& P(\varepsilon_{t-j}\geq g_1(Z_{t-j-1},\delta) )=1,\,\,\mbox{ if }\delta_{j+1}>0;\\
			& P(\varepsilon_{t-j}\leq g_1(Z_{t-j-1},\delta) )=1,\,\,\mbox{ if }\delta_{j+1}<0,\\
		\end{cases}
	\end{align}
	where $g_1$ is a deterministic function.
	Notice that $Z_{t-j-1}\in \mathcal{F}_{t-j-1}$.  Thus, using the fact that the conditional density of $\varepsilon_{t-j}$ given $\mathcal{F}_{t-j-1}$ is positive on the whole real line in Assumption \ref{assumption_h_5} (i), we know that (\ref{eq_h_5}) can not hold and hence (i) in Lemma \ref{lemma_h_1} holds.
	
	(ii). We only need notice that if $P(Z^{\top}_{t-1}\delta=0)=1$ for some vector $\delta\neq 0$, then for some $j \in \{1,\cdots,p\}$, we have $\delta_{j+1}\neq 0$ and 
	\begin{align}\label{eq_h_6}
		P\bigg(\delta_{j+1}y_{t-j}+\delta_1+\sum_{k=j+1}^{p}{\delta_{k+1}y_{t-k}}=0\bigg)=1.
	\end{align}
	Similarly, this is in contradiction with Assumption \ref{assumption_h_5} (i).
\end{proof}

\begin{proof}[\bf Proof of Theorem \ref{theorem_h_1}]
	(i). By the law of total expectation and Assumption \ref{assumption_h_7}, we have
	\begin{align}\label{eq_h_7}
		\frac{\partial E\tilde{L}_n(\theta;\tau)}{\partial \theta}=-\frac{1}{n} \sum_{t=p+1}^{n}{E\Big(w_t Z_{t-1}\big[\tau-F_t\big( {(\theta-\theta_0)^{\top}Z_{t-1}} \big)\big]     \Big)},
	\end{align}
	{where $F_t(\cdot):=F(\cdot,t/n;\mathcal{F}_{t-1})$ is the conditional distribution function of $\varepsilon_t$. Therefore, by Assumption \ref{assumption_h_5} (ii) and Assumption \ref{assumption_h_7}, for any $s_1 \in [0,1]$, we can derive that}
	\begin{flalign}\label{eq_h_8}
		& \quad
		\sup_{\tau \in \mathcal{T}, \theta \in \Theta}\bigg\vert  w_tZ_{t-1} \big[\tau-F_t\big( { (\theta-\theta_0)^{\top} Z_{t-1}} \big)\big]\nonumber\\
		&\qquad\qquad\qquad- w(s_1,Y^{\top}_{t-1}(s_1))Z_{t-1}(s_1)\big[\tau-F\big({(\theta-\theta_0)^{\top} Z_{t-1}(s_1)},s_1;\mathcal{F}_{t-1} \big) \big] \bigg\vert\nonumber\\
		&\leq \vert w(s_1,Y^{\top}_{t-1}(s_1))Z_{t-1}(s_1) \vert \sup_{\theta\in \Theta}{\Big\vert F\big({(\theta-\theta_0)^{\top} Z_{t-1}},\frac{t}{n};\mathcal{F}_{t-1} \big)-F\big({(\theta-\theta_0)^{\top} Z_{t-1}(s_1)},\frac{t}{n};\mathcal{F}_{t-1} \big)  \Big\vert}\nonumber\\
		&\qquad + \vert w(s_1,Y^{\top}_{t-1}(s_1))Z_{t-1}(s_1) \vert\sup_{\theta\in \Theta}{\Big\vert F\big({(\theta-\theta_0)^{\top} Z_{t-1}(s_1)},\frac{t}{n};\mathcal{F}_{t-1} \big)-F\big({(\theta-\theta_0)^{\top} Z_{t-1}(s_1)},s_1;\mathcal{F}_{t-1} \big)  \Big\vert}\nonumber\\
		& \qquad + \vert w_t Z_{t-1}-w(s_1,Y^{\top}_{t-1}(s_1))Z_{t-1}(s_1)\vert\nonumber \\
		& \leq C \Big\{ \vert t/n-s_1\vert^{\alpha_0}+\sum_{i=1}^{p}{\vert y_{t-i}-y_{t-i}(s_1) \vert^{\alpha_0}}+\sup_{y}{\big[  {\vert F(y,t/n;\mathcal{F}_{t-1})-F(y,s_1;\mathcal{F}_{t-1})\vert}/{(\vert y \vert+1)} \big]}\Big\},
	\end{flalign}
	where $C$ is a constant only relying on $\sup_{\theta\in \Theta}{\vert \theta-\theta_0\vert}$, $w(\cdot)$, and $C_0$.
	
	Similarly, we can also prove that, for any $s_1,s_2 \in [0,1]$,
	\begin{align}\label{eq_h_9}
		& \sup_{\tau \in \mathcal{T},\theta\in \Theta}  \Big \vert w(s_1,Y^{\top}_{t-1}(s_1))Z_{t-1}(s_1)\big[\tau-F\big({(\theta-\theta_0)^{\top} Z_{t-1}(s_1)},s_1;\mathcal{F}_{t-1} \big) \big]\nonumber\\
		& \qquad \qquad \qquad-w(s_2,Y^{\top}_{t-1}(s_2))Z_{t-1}(s_2)\big[\tau-F\big({(\theta-\theta_0)^{\top} Z_{t-1}(s_2)},s_2;\mathcal{F}_{t-1} \big) \big] \Big \vert\nonumber\\
		& \leq C \bigg\{\vert s_1-s_2\vert^{\alpha_0}+ \sum_{i=1}^{p}{\vert y_{t-i}(s_1)-y_{t-i}(s_2) \vert^{\alpha_0}}+\sup_{y}{\Big(  \frac{\vert F(y,s_1;\mathcal{F}_{t-1})-F(y,s_2;\mathcal{F}_{t-1})\vert}{\vert y \vert+1} \Big)}   \bigg\}.
	\end{align}
	On the other hand, by Assumption \ref{assumption_h_1} and using the standard transformation in \citet{hamilton1994}, we can get that
	\begin{align}\label{eq_h_10}
		y_t=\sum_{j=0}^{t-p-1}{\pi_{j0} \varepsilon_{t-j}\Big(\frac{t-j}{n}\Big) }+\sum_{j=0}^{t-p-1}{\pi_{j0} \mu_0}+\Pi^{\top}_{t-p} Y_{p},
	\end{align}
	where $t\geq p+1$ and $\Pi_k=O(c^k_0)$. Thus, by Assumptions \ref{assumption_h_1}--\ref{assumption_h_3}, (\ref{eq_h_8}) implies that
	\begin{align}\label{eq_h_11}
		& \sup_{\tau \in \mathcal{T},\theta\in \Theta}  \bigg \vert   \frac{1}{n}\sum_{t=2p+1}^{n}{E\Big(w_t Z_{t-1}\big[\tau-F_t\big( {(\theta-\theta_0)^{\top}Z_{t-1}} \big)\big]     \Big)}\\
		& \qquad \qquad-\frac{1}{n} \sum_{t=2p+1}^{n}{E\Big(w(t/n,Y^{\top}_{t-1}(t/n))Z_{t-1}(t/n)\big[\tau-F\big({(\theta-\theta_0)^{\top} Z_{t-1}({t}/{n})},{t}/{n};\mathcal{F}_{t-1} \big) \big]\Big)} \bigg \vert  \nonumber\\
		& \qquad \leq \frac{C_1}{n}\sum_{i=1}^{p}{\sum_{t=2p+1}^{n}{ \Vert y_{t-i}-y_{t-i}(t/n)  \Vert^{\alpha_0}_{\alpha_0} }}\nonumber\\
		& \qquad \leq \frac{C_1}{n}\sum_{i=1}^{p}{\sum_{t=2p+1}^{n}{ \bigg\{  \sum_{j=0}^{t-i-p-1}{\tilde{c}^j_0 (\frac{i+j}{n})^{\alpha_0}}+\sum_{j=t-i-p}^{\infty}{\tilde{c}^j_0} \bigg\} }}=O(n^{-\alpha_0}),\nonumber
	\end{align}
	where $C_1$ is a constant relying on $\Vert Y_p\Vert_{\alpha_0}$, $\Vert \sup_{x}\vert \varepsilon_t(x)\vert \Vert_{\alpha_0}$ and the constant $C$ in (\ref{eq_h_8}), and  $\tilde{c}_0=c^{\alpha_0}_0$. Meanwhile, by (\ref{eq_h_9}) and Assumptions \ref{assumption_h_1}-\ref{assumption_h_4}, it is easy to show that
	\begin{eqnarray}\label{eq_h_12}
		&& \sup_{\tau \in \mathcal{T},\theta\in\Theta} \bigg  \vert   \frac{1}{n}\sum_{t=2p+1}^{n}{E\Big(w(t/n,Y^{\top}_{t-1}(t/n))Z_{t-1}(t/n) \big[\tau-F\big({(\theta-\theta_0)^{\top} Z_{t-1}({t}/{n})},{t}/{n};\mathcal{F}_{t-1} \big) \big]\Big)}\nonumber\\
		&& \qquad\qquad\quad - \int_{0}^{1}{E\Big\{w(s,Y^{\top}_{t-1}(s))Z_{t-1}(s)\big[\tau-F\big({(\theta-\theta_0)^{\top} Z_{t-1}(s)},s;\mathcal{F}_{t-1} \big) \big]   \Big\} ds} \bigg \vert \nonumber \\  
		&& \qquad\qquad=O(n^{-\alpha_0}). 
	\end{eqnarray}
	Therefore, {Theorem \ref{theorem_h_1} (i)  follows from (\ref{eq_h_7}) and (\ref{eq_h_11})-(\ref{eq_h_12}).}
	
	For (ii), by Assumption \ref{assumption_h_5} (ii) and Assumption \ref{assumption_h_7} and utilizing the dominated convergence theorem, for the limit $g(x,\tau)$ in (i), we have
	\begin{align}\label{eq_h_13}
		\frac{\partial g(x,\tau)}{\partial \tau} & =-\int_{0}^{1}{E[w(s,Y^{\top}_{t-1}(s))Z_{t-1}(s)]ds},\nonumber \\
		\frac{\partial g(x,\tau)}{\partial x}& =\int_{0}^{1}{E\Big[w(s,Y^{\top}_{t-1}(s))Z_{t-1}(s)Z^{\top}_{t-1}(s)F_1(x^{\top}Z_{t-1}(s),s;\mathcal{F}_{t-1})  \Big]ds}.
	\end{align}
In addition, we define
\begin{align*}
	\tilde{g}(r,\phi, \tau)=-\int_{0}^{1}{E\bigg\{w(s,Y^{\top}_{t-1}(s))\phi^{\top} Z_{t-1}(s)\big[\tau-F\big({r\phi^{\top} Z_{t-1}(s)},s;\mathcal{F}_{t-1} \big) \big]   \bigg\} ds}.
\end{align*}
Then the conclusion (ii) can be easily established by a similar proof procedure as that in the Appendix A.
\end{proof}
\begin{proof}[\bf Proofs of Lemma \ref{lemma_h_2} and Theorem \ref{theorem_h_1}]
	The most important point is to show the generality of Lemma \ref{lemma_b_3}, Lemmas \ref{lemma_b_1}--\ref{lemma_b_2}. In this general case, Equations (\ref{eq_b_5}) and (\ref{eq_b_7}) are changed to be 
	\begin{align}\label{eq_h_14}
		\Vert d^2_t \Vert^{q/2}_{q/2} & \leq E\Big \{ \vert w_t Z_{t-1,i}\vert^q E\big ( \vert  \eta_{t}(x,h)-\eta_{t}(\tilde{x},\tilde{h})-E\{[\eta_{t}(x,h)-\eta_{t}(\tilde{x},\tilde{h})]\vert\mathcal{F}_{t-1}  \}  \vert^2   \vert \mathcal{F}_{t-1}\big) \Big \}\nonumber\\
		& \leq E\Big \{ \vert w_t Z_{t-1,i}\vert^q E\big ( \vert  \eta_{t}(x,h)-\eta_{t}(\tilde{x},\tilde{h})\vert^2   \vert \mathcal{F}_{t-1}\big) \Big \}\nonumber\\
		& \leq E\Big \{ \vert w_t Z_{t-1,i}\vert^q \times \vert F_t(x^{\top} Z_{t-1}+h \vert Z_{t-1}\vert)-F_t(\tilde{x}^{\top} Z_{t-1}+\tilde{h} \vert Z_{t-1}\vert) \vert \Big \}\nonumber\\
		& \leq {C}_q\times (\vert x-\tilde{x}\vert+\vert h-\tilde{h}\vert),
	\end{align}
and
\begin{eqnarray}
	&& \Vert L_n \Vert^{q/2}_{q/2} \leq n^{q/2} \times \max_{t}{\{ \Vert E(d^2_t \vert \mathcal{F}_{t-1})\Vert^{q/2}_{q/2} \}}\nonumber\\
	&&\qquad \leq n^{q/2} \max_{t}\Big\{\Big  \Vert w^2_t Z^2_{t-1,i} \vert F_t(x^{\top} Z_{t-1}+h \vert Z_{t-1}\vert)-F_t(\tilde{x}^{\top} Z_{t-1}+\tilde{h} \vert Z_{t-1}\vert) \vert \Big \Vert^{q/2}_{q/2} \Big \}\nonumber \\
	&&\qquad \leq n^{q/2} C_q \times (\vert x-\tilde{x}\vert^{q/2}+\vert h-\tilde{h}\vert^{q/2}).\label{eq_h_15}
\end{eqnarray}
In addition, we have
\begin{align}\label{eq_h_16}
	Q'_t(\tau)=\frac{1}{\sqrt{n}}w_t Z_{t-1,i}\Big[1-F_1(\delta^{\top}_0(\tau)Z_{t-1},t/n;\mathcal{F}_{t-1})Z^{\top}_{t-1} \frac{\partial \delta_0(\tau)}{\partial \tau} \Big].
\end{align}
\begin{align}\label{eq_h_17}
	Q''_t(\tau)& =-\frac{1}{\sqrt{n}}w_t Z_{t-1,i}\Big \{ F_1(\delta^{\top}_0(\tau)Z_{t-1},t/n;\mathcal{F}_{t-1})Z^{\top}_{t-1} \frac{\partial^2 \delta_0(\tau)}{\partial \tau^2}  \nonumber\\
	& \qquad +F_2(\delta^{\top}_0(\tau)Z_{t-1},t/n;\mathcal{F}_{t-1})\big[Z^{\top}_{t-1} \frac{\partial \delta_0(\tau)}{\partial \tau}\big]^2 \Big\}.
\end{align}
and 
\begin{align}\label{eq_h_18}
	\Vert \mathcal{P}_{t-m}[Q'_t(\tau)] \Vert^2 & \leq C_1/n \times \Big \{  \Vert Y^{*(t-m)}_{t-1}-Y_{t-1} \Vert^{\alpha_0}_{\alpha_0}\nonumber\\
	& +\sup_{x\in [0,1]}{\big\Vert \sup_{y \in \mathbb{R}}{\big[\vert F_1(y,x;\mathcal{F}_{t-1})-F_1(y,x;\mathcal{F}^{*(t-m)}_{t-1})\vert/(\vert y \vert+1)\big]} \big \Vert^2} \Big\}.
\end{align}
Based on the above differences, the conclusions can be proved by using the same proof framework for Lemma 5.1 and Theorem 5.1 in the main manuscript.\\
\end{proof}

\begin{proof}[\bf Proofs of the results from Lemma \ref{lemma_h_3} to Theorem \ref{theorem_h_6}]
The proofs of Lemma \ref{lemma_h_3} through Theorem \ref{theorem_h_6} follow the same lines as those in Sections 6–7—merely replacing the GARCH-specific inequalities with the general ones as stated above, so we omit the details.\\
\end{proof}

